# Electrically Tunable Excitonic-Hyperbolicity in Chirality-Pure Carbon Nanotubes

Jason Lynch[1], Pavel Shapturenka[2, 3], Mohammad Mojtaba Sadafi[4], Zoey Liu[1], Tobia Ruth[5], Kritika Jha[3], Zahra Fakhraai[3], Hossein Mosallaei[4], Nader Engheta[1, 6, 7], Jeffrey A. Fagan[2], Deep Jariwala[1, *]

[1]Electrical and Systems Engineering, University of Pennsylvania, Philadelphia, PA 19104, USA

[2]Materials Science and Engineering Division, National Institute of Standards and Technology, Gaithersburg, Maryland 20899, United States

[3]Department of Chemical Engineering, University of Pennsylvania, Philadelphia, PA 19104, USA

[4]Metamaterials Laboratory, Electrical and Computer Engineering Department, Northeastern University, Boston, Massachusetts 02115, USA

[5]Materials Science and Engineering, University of Pennsylvania, Philadelphia, PA 19104, USA

[6]Department of Bioengineering, University of Pennsylvania, Philadelphia, Pennsylvania 19104, United States

[7]Department of Physics and Astronomy, University of Pennsylvania, Philadelphia, Pennsylvania, 19104, United States

*Corresponding Author: dmj@seas.upenn.edu

## Abstract

Metamaterials exhibiting hyperbolic dispersion enable unprecedented control over light-matter interactions, from sub-diffraction imaging to enhanced spontaneous emission. However, conventional plasmonic hyperbolic metamaterials suffer from limited tunability and lack intrinsic emission capabilities, constraining their utility for active photonic devices. Here, we demonstrate the first room-temperature, electrically tunable, excitonic hyperbolic metamaterial using aligned films of chirality-pure semiconducting carbon nanotubes. Unlike plasmonic systems, these excitonic metamaterials of aligned nanotubes combine strong optical anisotropy with dynamic electrostatic tunability. Spectroscopic ellipsometry reveals that the hyperbolic dispersion window can be electrically shifted by 53 meV, enabling real-time switching between hyperbolic and elliptical regimes. Theory predicts that this tunability translates to the propagation angle being modulated by 34°, driven by a momentum enhancement 3.11 times that of free space, limited primarily by material losses that can be mitigated through improved alignment. In addition, simulations of the system exhibit a high Purcell factor of 1550 and a modulation of 37 % without an optical cavity for a dipole placed 5 nm above the aligned nanotubes. These findings establish excitonic carbon nanotubes as a versatile platform for dynamically reconfigurable photonic metamaterials, opening pathways for adaptive optical devices, electrically-controlled spontaneous emission, and tunable hyper-lenses operating at room temperature.

**Introduction**

Hyperbolic metamaterials present a paradigm shift for photonics, offering exotic light propagation regimes rarely in natural materials. These artificially engineered media enable phenomena ranging from super-resolution imaging and cloaking to enhanced nonlinear optical processes and directional spontaneous emission[1–3]. The key requirement—simultaneous optical anisotropy and negative real permittivity ($\varepsilon_1 < 0$) along one axis—has traditionally been achieved through plasmonic architectures comprised of patterned metals or heavily-doped semiconductors[4,5]. Although these systems are excellent at confining micron-wavelength-scale light to the nanometer-scale, such plasmonic media are neither emissive nor highly tunable to external stimuli such as electrostatic doping. Therefore, these systems normally benefit from the integration of an active layer to boost their performance for emissive[6] and electro-optical applications[7,8]. Despite this drawback, plasmonic media have dominated the field of hyperbolic photonics for visible and near infrared wavelength light, as other resonance phenomena in this energy range lack sufficiently strong optical responses needed for negative permittivities.

Excitons, tightly-bound electron-hole pairs near the band gap of a semiconductor, are another optical resonance phenomenon in the visible to near infrared range[9], but they also do not normally possess strong enough optical responses for negative permittivity. Recently, several semiconductor materials have exhibited negative permittivities near their exciton resonances including organics[10–12] and 2D semiconductors[13–19]. Excitons must possess a strong oscillator strength and narrow linewidth to exhibit negative real permittivities. Unfortunately, for most excitonic media this only occurs at low temperatures. However, the phenomena of excitonic hyperbolicity has the potential to augment hyperbolic photonics and metamaterials, as these media can be easily switched between the hyperbolic and elliptical regimes, and they can enable emission directly from the hyperbolic medium. Achieving such a useful applications has proven challenging due to the mentioned frequent requirement of cryogenic temperatures for sufficiently narrow exciton linewidths[13,17,20], and room-temperature excitonic systems typically lacking the negative permittivity[10–12].

Of specific interest in this work are the isolation and ordered thin-film assembly of chirality-pure single-walled carbon nanotubes (SWCNTs), which can be made using either a sorting process[21–23] or by direct growth[24]. Previous work on aligned, unsorted films of SWCNTs have exhibited hyperbolic windows that can be tuned through chemical doping[25]. However, the hyperbolic windows are the result of plasmonic and free-carrier effects. Chemical doping is also much slower than electrostatic doping and it can potentially introduce defects in addition to free-carriers. A negative excitonic permittivity has been observed with multiple different purified SWCNT chiralities[21,26], demonstrating that this phenomenon can occur in SWCNTs from the visible to near infrared since the band gap of SWCNTs is highly dependent on the nanotube diameter and carbon lattice orientation (chiral index ($n,m$))[27]. Combining this with our recent observation of gate-tunable, in-plane optical anisotropy in non-chirality-pure aligned SWCNTs[28], chirality-pure SWCNTs should allow for actively tunable hyperbolicity to occur in aligned films.

Based on our extensive review of current literature, we report the first demonstration of electrically-tunable, in-plane excitonic hyperbolicity at room temperature under ambient conditions using locally-aligned films of chirality-pure semiconducting carbon nanotubes. Our approach leverages the exceptional optical properties of small-diameter SWCNTs combined with controlled alignment to achieve simultaneous requirements for hyperbolicity: strong optical anisotropy and negative permittivity. Crucially, electrostatic doping enables real-time switching between hyperbolic and elliptical dispersion regimes, providing unprecedented control over light propagation and emission. This work establishes excitonic metamaterials as a viable platform for next-generation active photonic devices operating under ambient conditions. In this work, we present the first demonstration of electrically-tunable excitonic hyperbolicity at room temperature with a hyperbolic window that can be modulated by 53 meV. Theory predicts that hyperbolic SWCNTs can enhance the momentum of by a factor of up to 3.11 compared to vacuum and the propagation angle can be modulated by up to 34°. Further, simulations show that the systems support Purcell factors as large as 1550 without optical engineering, and the Purcell factor can be modulated by 37 %. Although this work is performed on aligned areas on the scale of 10's of microns, wafer-scale, globally-aligned films of SWCNTs can be fabricated using highly-optimized vacuum filtration[29,30] or other solution-based approaches[31]. Therefore, our work introduces chirality-pure SWCNTs as a highly-tunable hyperbolic medium. Future works can scale this medium up to the wafer-scale and focus on further improving its hyperbolic performance.

**Results and Discussion**

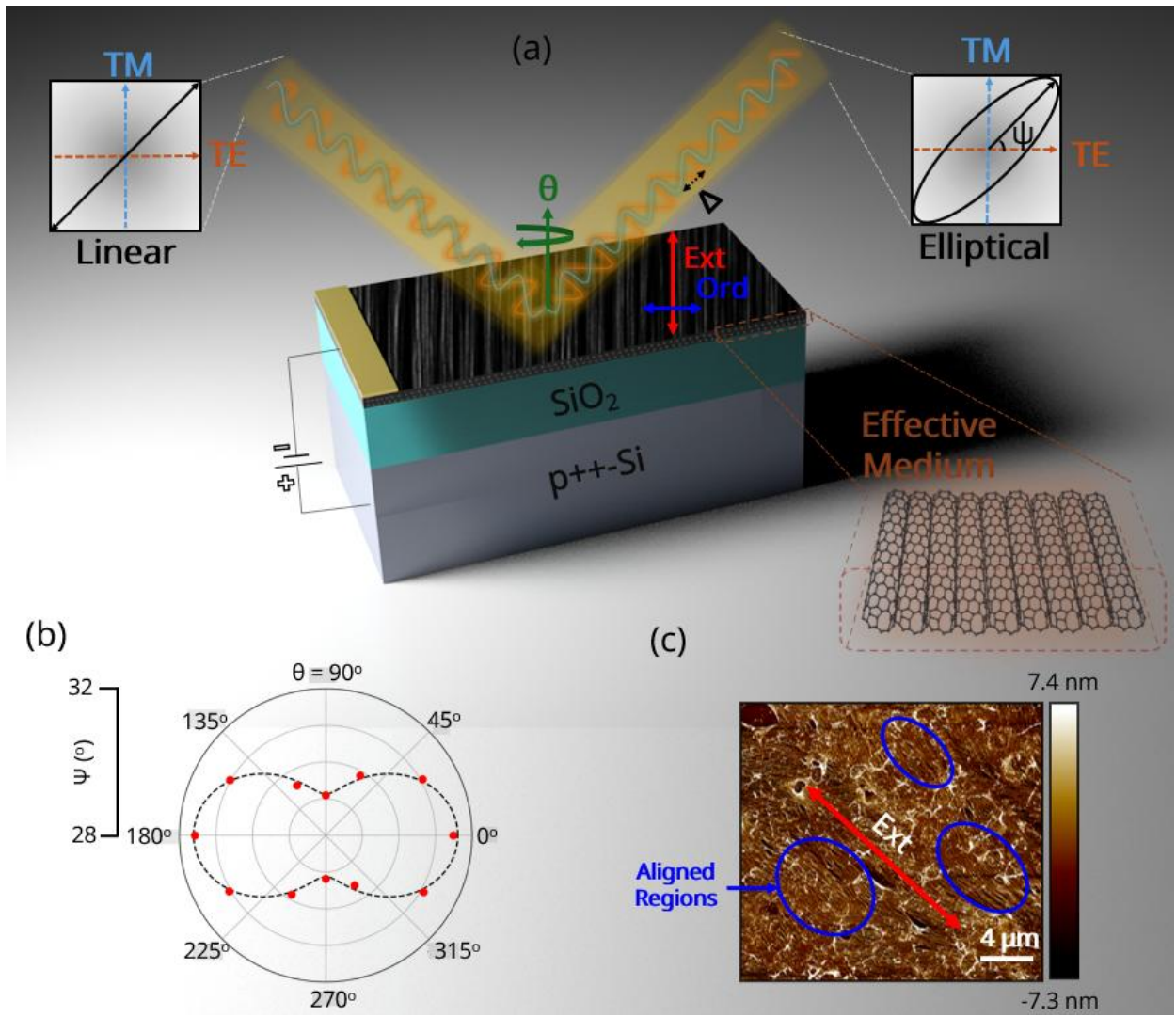

**Figure 1. Aligned films of chirality-pure SWCNTs. (a)** Schematic of spectroscopic ellipsometry where 45° polarized light becomes elliptically polarized upon reflection from the sample due to the polarization-dependent reflection coefficients. The in-plane alignment angle of the SWCNTs (θ) is shown along with the experimentally measured values characterizing the relative intensity ($\psi = \arctan\left(\frac{|r_{TM}|}{|r_{TE}|}\right)$) and phase difference ($\Delta = \arg(r_{TM}) - \arg(r_{TE})$). The effective medium of the collection of SWCNTs is shown in the cross-section on the right. **(b)** θ-dependent ψ values of the (6,5) SWCNT film. The varying value with θ indicates the in-plane optical anisotropy that is induced by nanotube alignment. **(c)** Atomic force microscopy (AFM) scan of the film further confirming the alignment of the nanotubes. The nanotubes are found to be generally aligned from the upper left to bottom right of the scan.

The small-diameter SWCNTs used herein are commercially sourced, with the raw powder containing a distribution of SWCNT chiralities, lengths and impurities (see Methods). A set of processing methods is sequentially applied to a dispersion of these SWCNTs to remove impurities and refine the (*n,m*) distribution. Approximately single chirality populations (> 90 % pure, see Methods)[23] are produced. In its processing the (6,5) fraction primarily used in this contribution is also sorted by length as this improves eventual film alignment[32]. Previous work determined the average length to be 605 nm ± 150 nm[23]. The chirality purity of dispersions, and thus the resulting films are confirmed using absorbance and photoluminescence spectroscopies, and Raman scattering demonstrates that the nanotubes have the expected diameter for (6,5) SWCNTs (Figure S1). The aligned films are produced from SWCNT dispersions *via* a slow vacuum filtration process that allows for micron-scale domains of very well aligned SWCNTs, with domains overall aligned along a general director[29,32,33].

Spectroscopic ellipsometry (SE) is a powerful tool in characterizing the optical properties of thin films since it collects information on both the amplitude and phase of reflected light at non-normal incident angles as shown in Figure 1a. The reflected relative amplitude ($\psi = \arctan(|r_{TM}|/|r_{TE}|)$ where $r_{TM}$ and $r_{TE}$ are the complex reflection coefficients of transverse magnetic (TM) and transverse electric (TE) polarized light) and relative phase difference ($\Delta = \arg(r_{TM}) - \arg(r_{TE})$) allow for the complex permittivity of a sample to be determined with a high level of certainty (Section S1). The optical properties of SWCNT films depend on both the properties of individual nanotubes, such as their chiral indices *(n,m)*[34–36], and on global properties such as packing density, degree of alignment, and the dielectric environment[21,37,38]. Therefore, the optical properties reported for SWCNTs are best interpreted as the effective medium of the full nanotube-dielectric network as shown in the cross-section of Figure 1a. Importantly, the alignment of the nanotubes induces the optical anisotropy that is necessary for hyperbolicity. The azimuthal direction of the nanotubes is called the extraordinary axis since it has its own unique permittivity while the other two directions are the ordinary axes. The alignment is observed optically by performing SE while rotating the in-plane orientation, characterized by the angle θ, of the sample (Figures 1b and S2). The SE parameter ψ is found to exhibit the two-fold symmetry that is expected for aligned SWCNT films. The film alignment is further confirmed using atomic force microscopy (AFM) as shown in Figure 1c. AFM shows that the nanotubes are generally aligned in the same direction (upper left to bottom right).

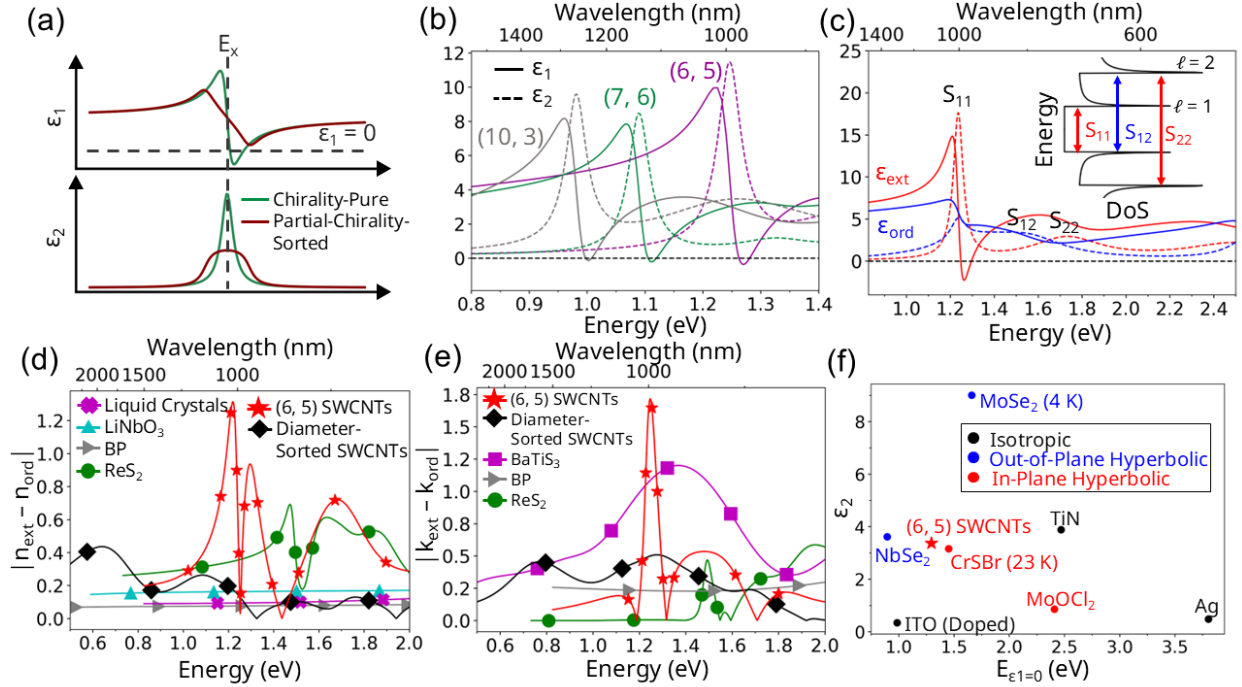

**Figure 2. Static permittivity of chirality-pure SWCNT films.** **(a)** Permittivity of chirality-pure (red) and partial-chirality-sorted (blue) SWCNT films. The chirality-pure SWCNT is modelled as a single Lorentz oscillator, while the partial-chirality-enriched SWCNT is treated as a set of equally-spaced Lorentz oscillators with the same total oscillator strength as the chirality-pure SWCNT film. The partial-chirality-sorted SWCNT films appear to have a broader exciton-like resonance than the chirality-pure ones and do not exhibit negative permittivity. **(b)** The complex permittivity of isotropic films of (10,3), (7,6), and (6,5) chiralities showing that excitons in chiralities-pure networks of SWCNTs are strong and narrow enough to support negative real permittivities at energies just above their primary resonance. **(c)** Anisotropic permittivity of (6,5) SWCNTs that is extracted from measuring the Mueller matrix with a micro-ellipsometer. The measurement range is 400 nm to 1,000 nm, but the permittivity is extrapolated to lower energies to fully show the $S_{11}$ resonance. The hyperbolic region of interest falls entirely within the measurement range. Comparison of **(d)** birefringence and **(e)** dichroism of the (6,5) SWCNTs with other semiconductors and insulators taken from literature[28,39–43]. **(f)** Comparison of the loss at the $\varepsilon_1 = 0$ energy for other anisotropic excitonic media, layered metals, and bulk plasmonic media[13,17,42,44–47].

In addition to nanotube alignment, chirality-purity is essential for observing hyperbolicity in SWCNTs. First, the chiral indices studied in this work all satisfy the semiconducting nanotube condition of $(n - m)$ mod $3 \neq 0$. Therefore, the nanotubes studied here all possess both band gaps and excitons which are needed to host a strong enough susceptibility for negative $\varepsilon_1$ values to occur. Second, the diameter of the nanotubes is determined by the indices, and the band gap of the nanotubes is highly dependent on their diameters[35]. As a result, chirality-pure films will have a narrow resonance near the band gap whereas partial-chirality-sorted films will have a convolution of narrow peaks from the present chiralities which widens the optical response. This mechanism is shown in Figure 2a where the narrow chirality-pure permittivity is strong enough to exhibit $\varepsilon_1 < 0$ while the partial-chirality-sorted one does not.

The complex permittivity of excitonic media is commonly modelled using the multi-Lorentz oscillator model[48]:

$$\varepsilon(E) = \varepsilon_\infty + \sum_i \frac{f_i \Gamma_i E_i}{E_i^2 - E^2 - iE\Gamma_i} \qquad (1)$$

Where the sum is over all the resonances near the energies measured, the sum is over all the resonances near the energies measured, E is the energy of light, $\varepsilon_\infty$ is the background permittivity of the medium, and $E_i$, $f_i$, and $\Gamma_i$ are the resonant energy, oscillator strength, and linewidth of the i[th] resonance, respectively. As discussed above, the permittivity of a nanotube film is the effective value of the full nanotube-dielectric nanotube network since optical modelling assumes the film is a rectangular prism. Therefore, for SWCNTs, the parameters in Eq. 1 are effective values that depend on the properties of individual nanotubes, the surrounding dielectric, and intertube coupling. The background permittivity is inversely related to diameter[34]. This results in our SWCNTs have a relatively small background permittivity compared to other low-dimensional excitonic systems such as transition metal dichalcogenides. The oscillator strength of an exciton is interpreted as the probability of a transition absorbing an incident photon, and it is proportional to the dipole moment[9]. As defined in Eq. 1, the oscillator strength is equal to $\varepsilon_2$ at the resonant energy. For SWCNTs, the oscillator strength is also inversely related to the diameter because the exciton radius decreases with diameter which increases the electron-hole interaction strength of the exciton[36]. As for exciton linewidth, an isolated SWCNT can have a linewidth of ≈ 30 meV at room temperature[49]. However, in a network of SWCNTs, intertube interactions lead to exciton tunneling between nanotubes, increased Coulomb screening between electron-hole pairs, and increased exciton quenching, which broadens the exciton linewidth[38]. In our films, these mechanisms yield and exciton linewidth of ≈ 55 meV.

By studying a single exciton system, the ideal properties of an exciton for negative permittivity can be determined (Section S2). This model gives the negative real permittivity criterion as:

$$\varepsilon_\infty < \frac{f}{2 + (\Gamma/E)} \qquad (2)$$

When this criterion is met, a window of negative real permittivity will open just above the resonant energy of the exciton. From Eq. 2, the ideal exciton has a large oscillator strength and a narrow linewidth compared to its energy while the background permittivity should be small. As discussed above, small-diameter excitons have all of these properties making them excellent candidates to support negative real permittivities. This is confirmed by measuring the isotropic complex permittivity of multiple chiralities of SWCNTs (Figures 2b and S4-S6). The strong, narrow excitons in all three chiralities show negative real permittivities which is consistent with previous work on chirality-pure SWCNTs[21]. Importantly, these measurements are performed under ambient conditions at room temperature unlike many other excitonic media that require cryogenic temperatures to observe negative real permittivities[13,17].

The measurements in Figure 2b are performed using an ellipsometer with a spot size of ≈ 100 µm. The non-normal angle of incidence causes the beam to elongate into an ellipse with major and minor axes around 292 µm and 100 µm, respectively, at an incident angle of 70°. Therefore, the sample area probed is approximately 29,000 µm² which is much larger than the aligned domain size, and optical anisotropy cannot be observed using this measurement. This is confirmed by performing θ-dependent ellipsometry on the films to show that they do not exhibit optical anisotropy at this scale (Figure S7). To investigate the effects of alignment, an imaging micro-ellipsometer with micron-sized resolution is used to perform Mueller matrix ellipsometry (see Methods). After the extraordinary axis is identified by measuring the θ-dependence of ψ (Figure 1b), the nanotubes are aligned such that the extraordinary axis is at a 45° angle of the in-plane momentum of incident light. Mueller matrix ellipsometry provides a complete description of the polarization of reflected light which allows for the determination of the anisotropic, complex permittivity without the need to perform an in-plane rotation scan (Section S3). The results of this measurement on aligned (6,5) SWCNTs are shown in Figure 2c. The measurement range of the micro-ellipsometer is λ = 400 nm to 1,000 nm (E = 1.24 eV to 3.1 eV). For clarity, the multi-Lorentz oscillator is used to extrapolate the results to energies below 1.24 eV to show the primary exciton, but our focus in this work, the hyperbolic window, falls entirely within the measurement range.

The one-dimensional nature results in the appearance of van Hove singularities in the density of states (DoS) in SWCNTs as shown as sharp spikes in the inset in Figure 2c, and each singularity corresponds to a sub-band in SWCNTs[50]. This gives SWCNTs a strong optical response when incident light is resonant with a transition between sub-bands. However, the circular cross-section of SWCNTs also results in each sub-band having its own angular momentum which limits certain transitions from occurring. The angular momentum at the conduction band minimum is labelled as $\ell$ = 1, the next higher energy sub-band is $\ell$ = 2, and the trend continues. The same trend occurs in the valence band where the valence band maximum is labelled as $\ell$ = 1. The interband transitions in semiconducting SWCNTs are then labelled using angular momentum numbers as the subscripts ($S_{12}$ represents a transition between an $\ell$ = 1 and $\ell$ = 2 sub-band). The optical selection rules of SWCNTs state that the angular momentum number is conserved for light polarized along the extraordinary axis and $\ell$ changes by ± 1. Despite the optical selections rules predicting that the fundamental exciton ($S_{11}$) would only appear in the extraordinary direction, a weak $S_{11}$ resonance is observed in the ordinary direction because the SWCNT networks are not perfectly aligned. Therefore, the extraordinary axis some individual nanotubes will be misaligned to the experimental extraordinary axis, and their $S_{11}$ resonance can be excited by light polarized along the experimental ordinary axis. Since the nanotubes are all aligned in-plane, but with some variation in the in-plane orientation, the $S_{11}$ resonance should not contribute to the out-of-plane permittivity of the film making the system biaxial in nature. However, similar to few-layer graphene, the film is too thin to accurately measure the out-of-plane permittivity so the system is treated as uniaxial. Using the relative oscillator strengths of the $S_{11}$ resonance along the two axes, the in-plane angular variation is estimated as 22° (Section S4). Assuming a normal distribution of the alignment of nanotubes, this corresponds to a 2D nematic order parameter ($S_{2D}$) of 0.74, which agrees well with the measured values for similarly prepared films reported to have $S_{2D}$ values of up to 0.66 over a 625 µm$^2$ area[32].

Despite not being a perfectly aligned network of nanotubes, our (6,5) SWCNTs exhibit excellent birefringence and dichroism compared to both bulk and low-dimensional media (Figures 2d and 2e). We directly observe a birefringence of up to 0.93 for SWCNTs within our measurement region (1.30 eV), and the Lorentz oscillator model predicts a birefringence of 1.31 at 1.22 eV, which is just 20 meV below our measurement region. Similarly, we observe giant dichroism near the exciton that is as large as 1.72 at 1.25 eV. Both the birefringence and dichroism are record values for low-dimensional semiconductors in the near infrared and combined with the facts that SWCNTs are actively tunable and can be fabricated on the wafer-scale, demonstrates that SWCNTs are an excellent material for active polarization optics.

The extreme optical anisotropy in chirality-pure SWCNTs is also seen as the appearance of a hyperbolic window at energies just above the $S_{11}$ resonance where $\varepsilon_{1,ext} < 0$ and $\varepsilon_{1,ord} > 0$. Typically, excitonic hyperbolicity is only observed in the near infrared using a plasmonic medium which lacks its own tunability or by cooling down an excitonic system to reduce its linewidth to satisfy Eq. 2[13,17]. However, SWCNTs are able to display hyperbolicity under ambient conditions at room temperature which confirms previous theoretical prediction[37]. The main drawback for excitonic hyperbolicity is that it will only occur at energies above the resonance as predicted by the Lorentz model which increases the loss. In the case of excitonic hyperbolicity, two hyperbolic transition points are observed: one near the resonance that we will refer to as the low energy transition (LET), and one at a higher energy one called the high energy transition (HET). The proximity of the LET to the $S_{11}$ resonance makes it more absorptive than the HET. However, we can compare $\varepsilon_2$ at the HET to other negative real permittivity media at their $\varepsilon_1 = 0$ to quantify how absorptive these media are (Figure 2f). The lowest loss media are the noble metal Ag, heavily doped ITO, and metallic $MoOCl_2$. Of these, $MoOCl_2$ is the only one that is intrinsically anisotropic with its in-plane hyperbolicity (the extraordinary axis lays in-plane). (6,5) SWCNTs have comparable absorptivity to both cryogenic excitons in CrSBr and metallic TiN which is a popular plasmonic medium[47]. This suggests that although $\varepsilon_2$ is relatively large in SWCNTs, it may be used in similar ways to TiN in hyperbolic systems with the added feature that it is electrically tunable as discussed below. It is also important to mention that the media in Figure 2f with $\varepsilon_2 > 1$ are not epsilon-near-zero (ENZ) media since $|\varepsilon| = \varepsilon_2$ when $\varepsilon_1 = 0$. In the narrow linewidth limit ($\Gamma << E_x$ where $E_x$ is the exciton energy) of the Lorentz oscillator model, an exciton will have an ENZ region when $\frac{\varepsilon_\infty^2}{f} + \frac{\varepsilon_\infty \Gamma}{2} < 1$ (Section S2).

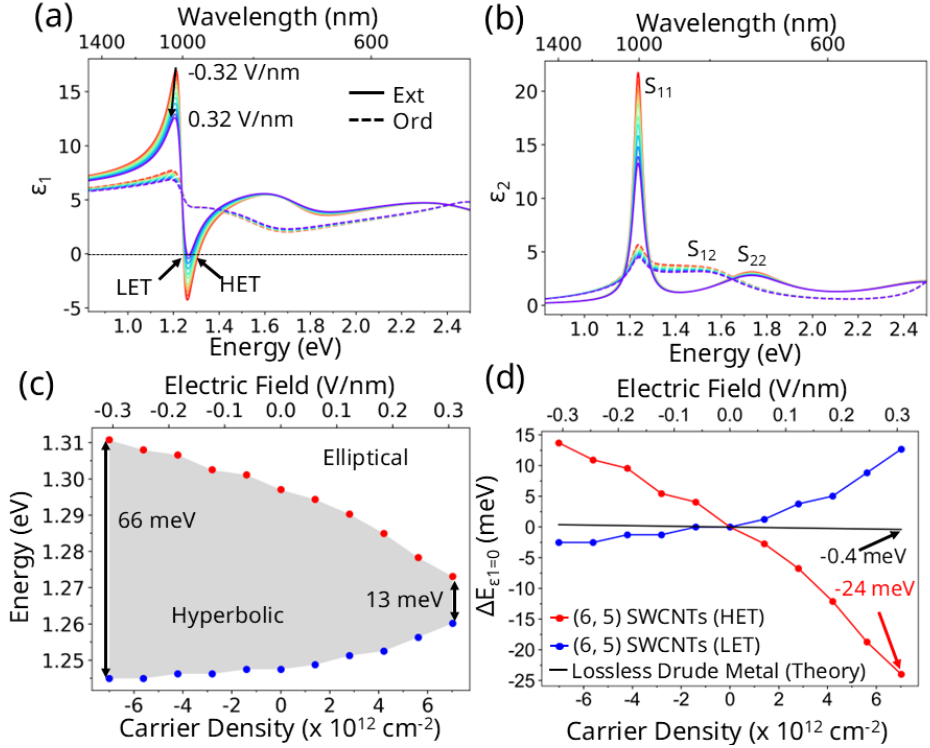

**Figure 3. Electrically tunable permittivity of (6,5) SWCNTs.** The gate-dependent **(a)** real and **(b)** imaginary parts of the anisotropic permittivity of (6,5) SWCNTs that are extracted by measuring the Mueller matrix with a micro-ellipsometer. The permittivity has been extrapolated to wavelengths longer than 1,000 nm to show the full exciton. The applied voltage ranged from -100 V to 100 V which produces electric fields ranging from -0.32 V/nm to 0.32 V/nm across the 307 nm SiO$_2$ layer. **(c)** The injected carrier density dependent energy of the HET (red) and LET (blue). **(d)** The change in energy of the HET (red), LET (blue), and a lossless Drude metal (green) as a function of injected carrier densities showing that excitonic hyperbolic transitions are far more tunable than plasmonic ones.

A key difference between excitons and plasmons is that excitons are much more tunable by electrostatic doping making them ideal for active applications. We inject free carriers into (6,5) SWCNTs using a lateral gate geometry as shown in Figure 1a. This approach injects electrons and holes when a positive and negative voltage is applied to the substrate, respectively. The injected carriers can be calculated by modelling the film as a parallel plate capacitor ($N = \frac{\varepsilon_0 \varepsilon_{SiO2} V}{e * t_{SiO2}} \approx (7x10^{10}\ cm^{-2}V^{-1}) * V$ where N is the injected carrier density, $t_{SiO2}$ is the thickness of the oxide layer (307 nm), $\varepsilon_{SiO2}$ is the relative static permittivity of SiO$_2$ (3.9), $\varepsilon_0$ is permittivity of free space, and $e$ is the charge of an electron). Ellipsometry determined the thickness of the (6,5) SWCNTs to be 4 nm which is around 5 nanotubes thick. However, the Debye screening length in SWCNTs is ~1 nm so only the bottom most nanotubes will have free carriers injected into them while the rest remain unchanged[51]. Measurements are performed 10 μm away from the top contact since the contact is not transparent. The injection of charge modulates the complex permittivity through the plasma dispersion effect[52]. The injected electrons counteract the p-type nature of SWCNTs, making them more intrinsic[53]. This results in a decrease in the Pauli blocking and exciton-carrier scattering, which leads to a larger oscillator strength[54] and narrower exciton[55]. Inversely, injected holes make the film more p-type, which leads to a weaker and broader exciton. The gate-tunable anisotropic permittivity of (6,5) SWCNTs are shown in Figures 3a and 3b. Over the voltage range studied, the oscillator strength and linewidth are found to increase and decrease with the gate voltage, respectively [28,56,57]. The permittivity is found to be highly tunable near the $S_{11}$ which is consistent with the gate-dependent isotropic permittivity of chirality-pure films (Figure S9). The HET and LET are modulated by 38 meV and 15 meV, respectively, which adds to a total modulation of the hyperbolic window by 53 meV (Figure 3c). This modulation allows the medium to be switched between the hyperbolic and elliptical regimes, and the hyperbolic window can be further modulated by injecting more carriers in future works. Additionally, the tunability of the SWCNTs can be compared to the ideal plasmonic medium using the lossless Drude model[48]. In this case, the carrier density of the lossless Drude model is chosen such that its plasmon energy is equal to the HET, and we find that it is orders of magnitude less tunable than the excitonic system (Figure 3d). The same injected carrier density only modulates the $\varepsilon_1 = 0$ energy of the Drude model by a total of 0.8 meV which is only 2% the tunability of the HET.

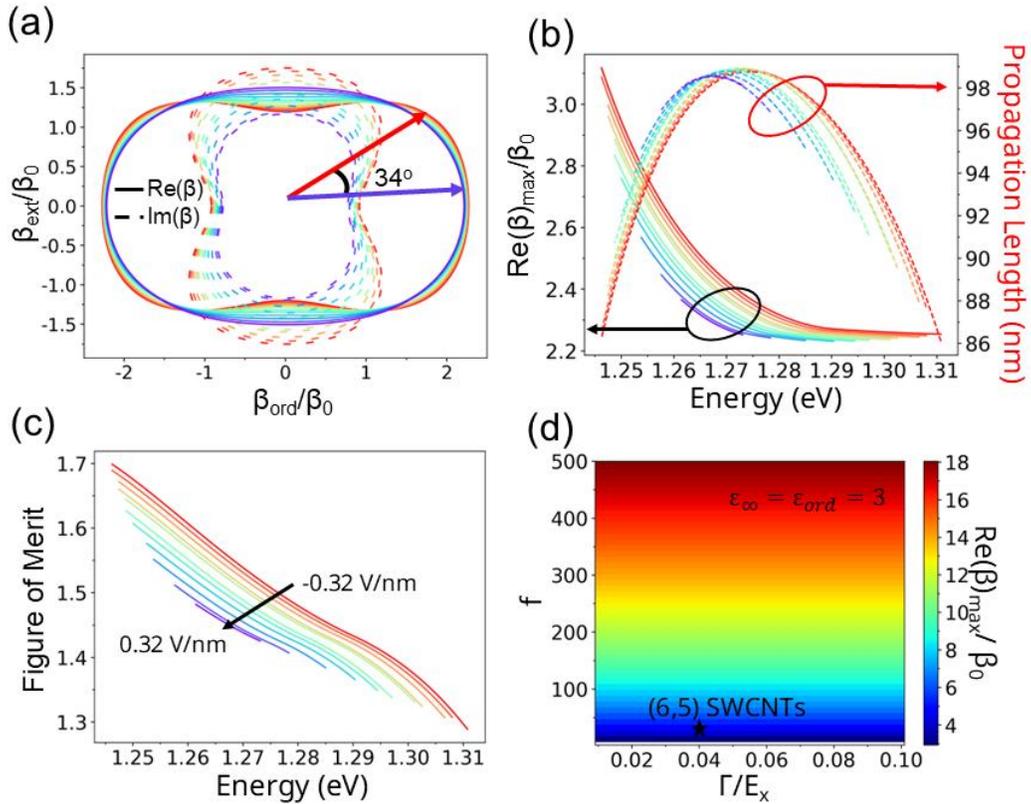

**Figure 4. Tunable bulk dispersion in SWCNTs.** The gate-dependent **(a)** dispersion at E = 1.27 eV, **(b)** momentum enhancement (left, solid lines) and propagation length (right, dashed lines), and **(c)** FoM of (6,5) SWCNTs that are calculated using the dispersion relation and the experimental complex permittivity. Although SWCNTs allows for the switching between the elliptical and hyperbolic regime, the maximum

momentum enhancement remains limited. As such, we calculated the maximum momentum enhancement of a single anisotropic Lorentz oscillator as a function of the oscillator strength (f) and normalized linewidth (Γ/E$_x$). The momentum enhancement is found to scale with $\sqrt{f}$ and depend little on the normalized linewidth.

For TM polarized light, its propagation through a medium is described by the dispersion relation $\frac{\beta_{ord}^2}{\varepsilon_{ext}} + \frac{\beta_{ext}^2}{\varepsilon_{ord}} = \frac{\omega^2}{c^2}$, where β, ω, and c are the momentum, angular frequency, and speed of light, respectively. This relation determines both the propagation of energy in the system which occurs in the direction where the real part of the momentum is maximum, and how quickly the light is absorbed by the medium which is related to the imaginary part of the momentum. The relation also allows us to model the dispersion of light in (6,5) SWCNTs using the complex permittivity that we measured with spectroscopic ellipsometry. When both ε$_1$ values are positive, the dispersion is elliptical in shape, as seen in Figure 4a for electric fields > 0.06 V/nm. In this elliptical regime, the maximum momentum for light occurs along either the extraordinary or ordinary axes depending on their refractive index. However, when ε$_1$ becomes negative along only one axis, the dispersion relation becomes an open hyperbolic curve in the lossless case, and the maximum momentum occurs away from the optical axes. This would mean that potential momentum enhancement is unbound, but loss closes the isofrequency curve and reduces the maximum momentum of light[58,59]. For (6,5) SWCNTs, the isofrequency curve resembles a pinched ellipse as seen in Figure 4a for electric fields ≤ 0.06 V/nm. Additionally, the maximum momentum occurs 34° away from the ordinary axis meaning that the direction of propagation can be electrically modulated by up to 34°. The maximum momentum of light, normalized to the momentum of free space, reaches 2.8 in the unbiased case and can be increased to 3.1 with electrostatic doping (Figure 4b). However, the increased momentum enhancement as you go from the HET to the LET comes at the expense of reduced propagation length. In the unbiased case, the propagation length varies from 86 nm to 98 nm. However, the momentum enhancement also means that the wavelength of light within the medium is also reduced in the hyperbolic regime which means that the light may travel multiple wavelengths before it decays. A natural figure of merit (FoM) is then to calculate the number of wavelengths the light travels before the electric field decays by 1/e. The FoM is calculated as FoM = Re(β)/Im(β). The gate-dependent FoM of (6,5) SWCNTs reaches as large as 1.7 at the maximum momentum propagation angle, and the FoM reaches its maximum value at the LET for all electric fields studied. This occurs despite the LET being more absorptive in the extraordinary direction. Further, the FoM can be larger at other propagation angles, but this occurs at the expense momentum enhancement[59]. For comparison, a superlattice of alumina and Au can host a momentum enhancement of 9.6 at an energy of 1.73 eV, and at this propagation angle, the propagation length and FoM are 26 nm and 1.8, respectively[60]. Therefore, the loss is regularly substantial at the maximum momentum propagation angle, but the momentum enhancement in SWCNTs is still well behind plasmonic metamaterials.

To further investigate the effects of loss on the momentum enhancement, the enhancement is calculated when the loss along one of the optical axes is set to 0 (Figure S12). The maximum momentum enhancements are calculated as 3.4 and 2.7 when the ordinary and extraordinary axes, respectively, are treated as lossless. This means that the absorption along the ordinary axes of the SWCNTs is the limiting factor. As discussed above, the absorptivity of the ordinary axes is increased due to the imperfect alignment of the nanotubes (Section S4). Therefore, the momentum can be further enhanced by better alignment of the nanotubes, and our theory predicts that the enhancement can be up to 3.5 which is a 25% improvement over the current film (Figure S13). However, this is still well below the enhancement of plasmonic metamaterials that can reach up to around 15[59], but excitonic hyperbolicity remains an emerging subfield with a growing library of media the exhibit excitonic hyperbolicity. Therefore, there is the opportunity to improve the quality of excitons to improve the momentum enhancement of light. In this vein, we study the effects that the parameters of a single Lorentz oscillator have on momentum enhancement (Figure 4d). The momentum enhancement reaches 18.1 at an oscillator strength of 500. For reference, hBN-encapsulated monolayer $WS_2$ has exhibited an oscillator strength of around 500 at cryogenic temperatures[61]. The momentum enhancement is found to depend on the oscillator strength more than the normalized linewidth. We attribute this to the magnitude of a Lorentz oscillator increasing linearly with the oscillator strength (Section S5) Further, the maximum momentum scales with the square root of $|\varepsilon|$. Therefore, $Re(\beta)_{max}/\beta_0$ should be proportional to $\sqrt{f}$, and this is confirmed by fitting the oscillator-strength-dependent momentum enhancement to a square root curve (Figure S13). The fit gives the relation $Re(\beta)_{max}/\beta_0 = 0.8\sqrt{f}$. However, the momentum enhancement maximum is unaffected by the linewidth since it does not affect peak amplitude of the permittivity (Figure S15).

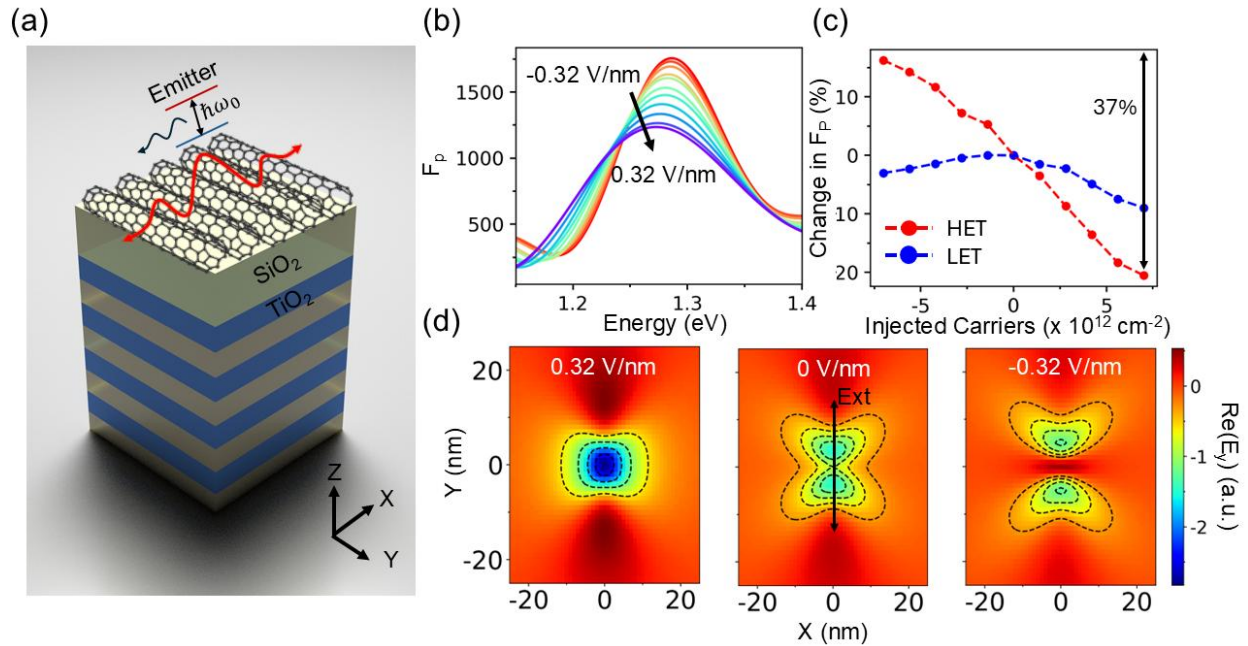

**Figure 5. Gate-dependent Purcell Factor. (a)** Schematic of aligned chiral SWCNTs on top of 300 nm $SiO_2$ and a $TiO_2/SiO_2$ distributed Bregg reflector (DBR), and a dipole emitter placed above it with the x, y, and z directions labelled. **(b)** Gate-dependent $F_p$ for an emitter placed 5 nm above the SWCNTs as a function of emitter energy. **(c)** Percent change in the Purcell factor as a function of injected carriers at the unbiased LET (blue) and HET (red) energies. **(d)** The real part of the electric field in the y direction (parallel to the extraordinary axis) distribution in the middle of the SWCNT layer at E = 1.26 eV (λ = 983 nm) with electric fields of 0.32 V/nm, 0 V/nm, and -0.32 V/nm from left to right showing the gate-tunable transition from an elliptical system at 0.32 V/nm to a hyperbolic one at 0 V/nm and -0.32 V/nm. The contour lines show the propagation of waves away from the point dipole which is located at the (0 nm, 0 nm, 5 nm) point.

In addition to enhanced momentum, a feature of hyperbolic media is a sharp increase in the systems' photonic density of optical states (PDoS). The large PDoS concentrates the electric field near the hyperbolic medium, and it increases the spontaneous emission rate of a nearby emitter[62]. The improvement in spontaneous emission rate is characterized by the ratio of the emitter's spontaneous emission in a specific system and in a vacuum. This ratio is called the Purcell factor ($F_p$). We have chosen to study the system of aligned SWCNTs on top of $SiO_2$ (300 nm) and a $TiO_2/SiO_2$ distributed Bragg reflector (DBR) with an emitter placed above the SWCNTs and aligned along the extraordinary direction (Figure 5a). The system in Figure 5a is modelled using Lumerical to calculate $F_p$ (see Methods). Since the PDOS and electric field intensity are maximum at the surface of the hyperbolic medium, the Purcell factor increases as the emitter approaches the SWCNTs (Figure S16). At a height of 5 nm, $F_p$ is as large as 1520 within the hyperbolic region. The large growth of $F_p$ agrees with previous theoretical work on SWCNTs[63]. Despite seeing a Purcell enhancement in the hyperbolic region, the cusp in $F_p$ near the hyperbolic transition is not observed[64]. Hyperbolic media tend to see a sharp increase near their ENZ, with a faster decrease in $F_p$ occurring for $\varepsilon_1 > 0$ than for $\varepsilon_1 < 0$. However, we do not observe this feature since the SWCNTs do not have an ENZ point, but the Purcell enhancement is still seen in the hyperbolic regime. Additionally, the Purcell factor is found to be highly tunable to carrier injection (Figure 5b). $F_p$ is found to increase with the excitonic oscillator strength, which reduces $\varepsilon_1$. The peak in $F_p$ is found to be modulated by 14 meV despite the exciton energy being constant. This indicates that the peak is due to hyperbolic effects since the exciton's energy remains unchanged. Further, $F_p$ is also found to be modulated by as much as 37 % at the unbiased HET (Figure 5c). Importantly, this is done without coupling the SWCNTs to a high-Q cavity, which would further increase $F_p$ and its sensitivity to electrostatic doping. Therefore, the modulation of SWCNTs allows for a drastic modulation in the PDOS and $F_p$, which is useful for active emission applications. Further, by studying the spatial distribution of $Re(E_y)$ within the SWCNT layer, the switching between a hyperbolic and elliptical system is observed (Figure 5d). At an electric field of 0.32 V/nm, the SWCNTs are elliptical at E = 1.26 eV. The contour lines show that the waves are propagating along the ordinary axis since this direction has a larger refractive index than the extraordinary. However, at electric fields of 0.32 V/nm and -0.32 V/nm, the SWCNTs become hyperbolic, and the contours of show that the waves propagate away from the two optical axes.

## Conclusion

We demonstrate electrically-tunable, room-temperature excitonic hyperbolicity in chirality-pure carbon nanotube films under ambient conditions, representing a fundamental paradigm shift from conventional plasmonic metamaterials toward dynamic, emissive excitonic systems. We have demonstrated this on regions of aligned, chirality-pure SWCNTs that are on the scale of 10's of microns. However, wafer-scale films of aligned SWCNTs can be fabricated using highly-optimized vacuum filtration[29,30] or other solution-based approaches[31]. By establishing excitons as a viable alternative to plasmons for engineering hyperbolic dispersion, we have achieved unprecedented capabilities, real-time electrical switching between hyperbolic and elliptical regimes (53 meV tunability measured experimentally), and significant emission enhancement (Purcell factors exceeding 1550 in simulations) without requiring optical cavities or cryogenic operation. This breakthrough directly addresses the central limitation of metamaterial physics - creating active, reconfigurable systems that maintain exotic properties under practical conditions, and further improvements are achievable through the engineering of excitons in the systems. The intrinsic compatibility with semiconductor architectures opens pathways to transformative applications ranging from adaptive hyperlenses and reconfigurable antennas to quantum light sources with voltage-controlled emission. Beyond carbon nanotubes, the design principles demonstrated here—combining strong, narrow excitons with controllable optical anisotropy—provide a roadmap for engineering hyperbolicity across diverse excitonic materials, fundamentally reshaping our approach to light manipulation using low-dimensional semiconductors.

**Methods**

Certain equipment, instruments, software, or materials, commercial or non-commercial, are identified in this paper in order to specify the experimental procedure adequately. Such identification is not intended to imply recommendation or endorsement of any product or service by the National Institute of Standards and Technology (NIST), nor is it intended to imply that the materials or equipment identified are necessarily the best available for the purpose.

*Processing and Assembly of Chirality-pure SWCNTs*

SWCNTs were commercially sourced from Chasm Nanotechnologies (SG65i-L64 and CG400-P0630). Briefly, SWCNT dispersions were generated *via* tip sonication and purified by both differential centrifugation and rate-zonal centrifugation as reported elsewhere[65]. Single (*n,m*) populations were then generated through a multistep aqueous two-polymer extraction (ATPE) methodology[65]. The (6,5) fraction was additionally sorted by length prior to ATPE (605 nm ± 150 nm). To generate aligned films, aqueous SWCNT dispersions (in 10 g/L sodium deoxycholate) were diluted with water and filtered within a hydrophilic treated glass funnel onto polycarbonate track-etched membranes (Whatman Nuclepore, 80 nm pores) *via* an automated, flow rate-controlled process. This process yielded a near-monolayer (4 nm) film of aligned SWCNTs on the membrane surface[32]. The membrane-bound SWCNT films were then adhered (SWCNT side down) onto Si substrates with 307 nm of surface $SiO_2$ using water, and the membrane was slowly rinsed away with chloroform. Further details of the filtration and film transfer process have been reported previously[32,33].

*Deposition of Electrical Contacts*

The 0.4 cm x 0.2 cm metal contacts were deposited under vacuum (< 6.7 x $10^{-4}$ Pa) using an E-beam evaporator (Kurt J. Lesker PVD 75). The 10 nm Pd and 40 nm Au were deposited sequentially with a deposition rate of 0.2 nm $s^{-1}$, and the wires were connected using an Ag paste. The SWCNTs were grounded while the voltage was applied to the $p^{++}$-Si substrate. The ellipsometry measurements were made approximately 10 μm away from the contact.

*Absorbance, Raman, and Photoluminescence Spectroscopy*

Raman spectroscopy and normal incidence photoluminescence (from 600 nm to 1,050 nm) were performed using a Horiba Scientific confocal microscope (LabRAM HR Evolution). This instrument used is an Olympus objective lens (×100) and a 600 grating-based spectrometer which is coupled to a Si focal plane array detector. A continuous-wave excitation source with excitation wavelength at 633 nm was used with the ×100 objective lens at 3.2 % laser power, which corresponds to a power of ≈ 7 μW. Absorbance was measured with a Varian Cary 5000 spectrophotometer with a 1 nm step and a 2 nm bandpass through a 1 mm path quartz cell. A reference spectrum of the dispersing surfactant in water was collected and subtracted from the sample spectra during data analysis.

*Ellipsometry*

Millimeter-scale spectroscopic ellipsometry was used to measure the isotropic permittivities of several chiralities shown in Figure 2b. An M-2000 ellipsometer (J. A. Woollam) was used with a spot size ≈100 μm, which covers the visible to near-infrared range (371 nm to 1687 nm). The permittivities were extracted using an isotropic, multi-Lorentz oscillator model in the CompleteEase software. The anisotropic results were measured using an Accurion EP4 (Park Systems) imaging spectroscopic ellipsometer. This ellipsometer performed Mueller matrix ellipsometer from 400 nm to 1,000 nm with a resolution of ≈ 1 μm and a 50 x lens. The permittivity was extracted using a uniaxial, multi-Lorentz oscillator model in the EP4Model software.

*Simulations*

All numerical simulations were carried out based on the finite-difference time-domain (FDTD) method. A commercial software package (Lumerical) was used to perform the FDTD calculations. Perfectly matched layer (PML) absorbing boundary conditions were applied in order to terminate the simulation region and prevent spurious reflections. SWCNTs were treated as an effective anisotropic slab with a height of 4 nm, and a causal multi-coefficient model was fitted to the experimentally measured refractive index data to account for dispersion in the FDTD simulations. The emitter was modeled as a Hertzian point dipole oriented along the extraordinary axis of the SWCNTs, and the Purcell factor was calculated by measuring the dipole's output power. Besides, a non-uniform Cartesian mesh was utilized over the SWCNTs and source regions, to effectively capture the subwavelength attributes of the structure.

## Acknowledgements

D. J. and J.L. acknowledge primary support from the Office of Naval Research Young Investigator Award, Metamaterials Program (N00014-23-1-203). This work was carried out in part at the Singh Center for Nanotechnology, which is supported by the NSF National Nanotechnology Coordinated Infrastructure Program under Grant NNCI-2025608. Additional support to the Nanoscale Characterization facility by the NSF through the University of Pennsylvania Materials Research Science and Engineering Center (MRSEC) (DMR-2309043) is acknowledged. T. R. acknowledges support from the Vagelos Institute for Energy Science and Technology. Z. F. and K. J. acknowledge support from NSF CHE-2318105. K. J. acknowledges support from the Vagelos Institute for Energy Science and Technology (VIEST) Graduate Fellowship. J. F. was supported by internal National Institute of Standards and Technology (NIST) funding. P. S. acknowledges funding support from the National Academies Postdoctoral Research Fellowship.

# Supplementary Information: Electrically Tunable Excitonic-Hyperbolicity in Chirality-Pure Carbon Nanotubes

Jason Lynch[1], Pavel Shapturenka[2, 3], Mohammad Mojtaba Sadafi[4], Zoey Liu[1], Tobia Ruth[5], Kritika Jha[3], Zahra Fakhraai[3], Hossein Mosallaei[4], Nader Engheta[1, 6, 7], Jeffrey A. Fagan[2], Deep Jariwala[1, *]

[1]Electrical and Systems Engineering, University of Pennsylvania, Philadelphia, PA 19104, USA

[2]Materials Science and Engineering Division, National Institute of Standards and Technology, Gaithersburg, Maryland 20899, United States

[3]Department of Chemical Engineering, University of Pennsylvania, Philadelphia, PA 19104, USA

[4]Metamaterials Laboratory, Electrical and Computer Engineering Department, Northeastern University, Boston, Massachusetts 02115, USA

[5]Materials Science and Engineering, University of Pennsylvania, Philadelphia, PA 19104, USA

[6]Department of Bioengineering, University of Pennsylvania, Philadelphia, Pennsylvania 19104, United States

[7]Department of Physics and Astronomy, University of Pennsylvania, Philadelphia, Pennsylvania, 19104, United States

*Corresponding Author: dmj@seas.upenn.edu

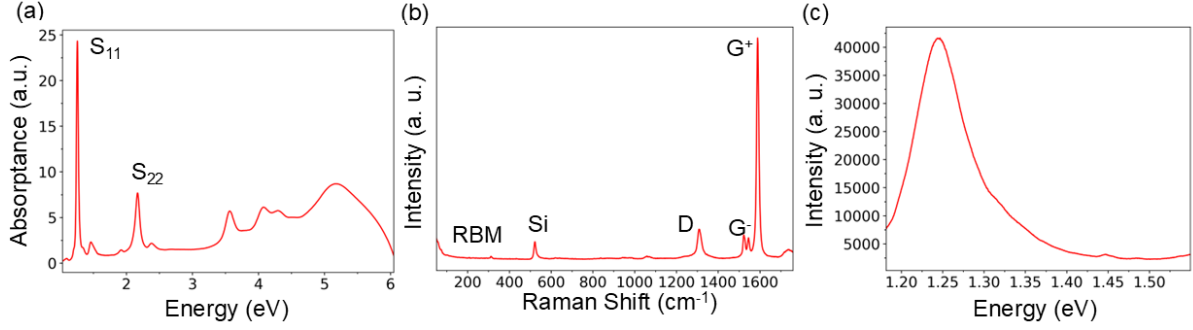

**Figure S1. Characterization of chirality-pure (6, 5) SWCNT film. (a)** Absorptance in solution, **(b)** Raman, and **(c)** photoluminescence (PL) spectroscopy for (6, 5) SWCNTs with the two fundamental excitons along the extraordinary axis and the Raman modes labelled. The radial breathing mode (RBM) depends on the nanotube diameter, and our observed value of 310 cm$^{-1}$ agrees well with the literature value of 306 cm$^{-1}$ [1]. The PL energy of 1.24 eV also agrees well with the literature value of 1.26 eV[2]. The Raman and PL spectra are measured using the film on a SiO$_2$/Si substrate.

## Section S1: Isotropic Spectroscopic Ellipsometry and the Lorentz Oscillator Model

Spectroscopic ellipsometry (SE) uses the difference in complex reflection coefficients ($\tilde{r} = |r|e^{i\phi}$) between transverse magnetic (TM) and transverse electric (TE) polarized light at non-normal angles of incidence to determine the complex permittivity of films[3]. SE does this by characterizing the ratio of reflected amplitude (ψ = arctan(|r$_{TM}$|/|r$_{TE}$|)) and the phase difference between the two polarizations (Δ = Φ$_{TM}$ − Φ$_{TE}$). For the most accurate results, these values are typically collected at multiple angles of incidence. The complex permittivity is then extracted by modelling the system and varying the permittivity of the layer of interest until the experimental data is replicated by the model. The goodness-of-fit is determined by the weighted root-mean-square-error (RMSE):

$$RMSE = \sqrt{\frac{1}{2p-q}\sum_i\left(\left(\psi_i^{mod} - \psi_i^{exp}\right)^2 + \left(\Delta_i^{mod} - \Delta_i^{exp}\right)^2\right)}$$

where p and q refer to the number of wavelengths measured and the number of fit parameters, respectively, the sum is over the wavelengths, and the superscripts "mod" and "exp" denote the modeled and experimental values, respectively. Typically, an RMSE less than 5 is considered accurate.

The contribution to the permittivity of a single exciton resonance is modelled using the Lorentz oscillator model[3]:

$$\varepsilon_i^{Lorentz}(E) = \frac{f_i E_i \Gamma_i}{E_i^2 - E^2 - iE\Gamma_i}$$

where $f_i$, $E_i$, and $\Gamma_i$ are the oscillator strength, resonant energy, and the damping factor of the $i^{th}$ transition, respectively, and all three are fit parameters. Using this model, the permittivity of the SWCNTs is modelled as the sum of a series of Lorentz oscillators, each representing an interband transition, and the static permittivity ($\varepsilon_\infty$) which is also a fit parameter:

$$\varepsilon(E) = \varepsilon_\infty + \sum_i \varepsilon_i^{Lorentz}(E)$$

The CompleteEase software then minimizes the RMSE to determine the permittivity, and an RMSE < 5 is normally considered accurate[3]. The fitted parameters also include the thickness of the film which is found to be approximately 5 nanotubes thick (4 nm) for the (6,5) SWCNT film. To improve the accuracy of the results, we also performed ellipsometry on the bare $SiO_2$/Si substrate to determine the refractive indices and thicknesses of the $SiO_2$, Si, and the interlayer oxide. This approach allows for the substrate effects to be accounted for in the SWCNT/$SiO_2$/Si sample, and therefore, the model will more accurately model the SWCNT layer.

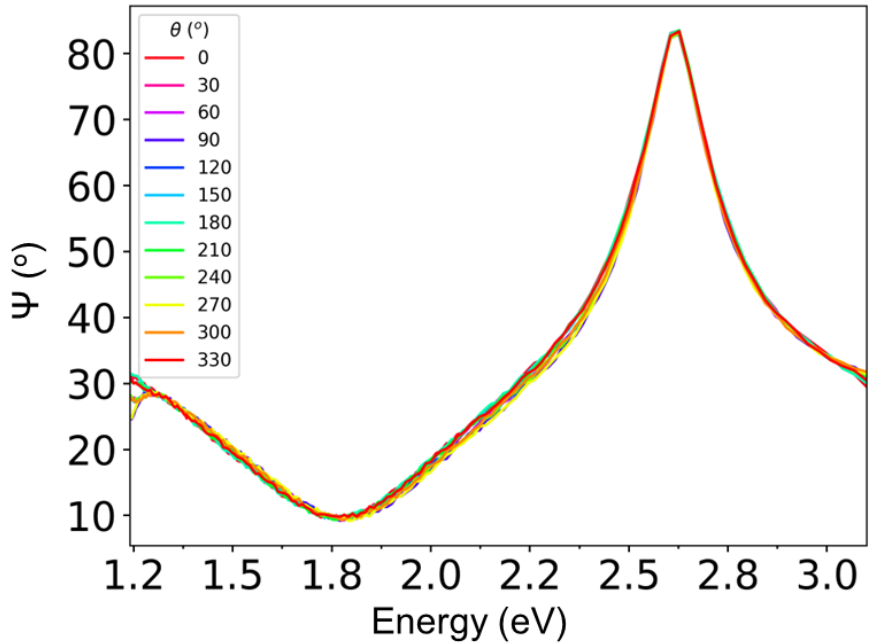

**Figure S2. In-plane-rotation-dependent ellipsometry.** $\Psi$ ($\Psi$ = arctan($|r_{TM}/r_{TE}|$)) of the (6, 5) SWCNTs as a function of the in-plane rotation ($\theta$). $\theta = 0°$ means that the light is propagating along the extraordinary direction and the in-plane electric field points along the ordinary direction. The large modulation near 1.2 eV is due to the optical anisotropy of the $S_{11}$ resonance of (6,5) SWCNTs, and it indicates that the SWCNTs are aligned.

## Section S2: Negative permittivities and epsilon-near-zero points in excitonic systems

The Lorentz oscillator model is commonly used to model the optical response of exciton. Therefore, it can provide insights into what is needed for excitons to exhibit negative permittivities and epsilon-near-zero (ENZ) points. For simplicity, we will focus on a single isolated exciton. In this case the permittivity is:

$$\varepsilon(E) = \varepsilon_\infty + \frac{fE_x\Gamma_x}{E_x^2 - E^2 - iE\Gamma_x}$$

Where $\varepsilon_\infty$ is the static permittivity and $f$, $E_x$, and $\Gamma_x$ are the oscillator strength, resonant energy, and linewidth of the exciton, respectively. This model can also be broken down into real and imaginary parts by multiplying the numerator and denominator by the complex conjugate of the denominator:

$$\varepsilon(E) = \varepsilon_\infty + \frac{fE_x\Gamma_x(E_x^2 - E^2)}{(E_x^2 - E^2)^2 + (E\Gamma_x)^2} + i\frac{fE_x\Gamma_x(E\Gamma_x)}{(E_x^2 - E^2)^2 + (E\Gamma_x)^2}$$

Negative permittivity then occurs when the real part is less than zero. To do this, we must first find the minimum value of the real part of the permittivity. We first find the energy where the minimum value occurs. First, we will normalize the parameters to the resonant energy.

$$\varepsilon_1(\tilde{E}) = \varepsilon_\infty + \frac{f\tilde{\Gamma}_x(1 - \tilde{E}^2)}{(1 - \tilde{E}^2)^2 + (\tilde{E}\tilde{\Gamma})^2}$$

Where the ~ denotes that the parameter has been normalized to the resonant energy. The derivative of the real part with respect to normalized energy is then:

$$\frac{d\varepsilon_1}{d\tilde{E}} = f\tilde{\Gamma}\left[\frac{(-2\tilde{E})\left((1 - \tilde{E}^2)^2 + \tilde{E}^2\tilde{\Gamma}^2\right) - (1 - \tilde{E}^2)(2(1 - \tilde{E}^2)(-2\tilde{E}) + 2\tilde{E}\tilde{\Gamma}^2)}{\left((1 - \tilde{E}^2)^2 + (\tilde{E}\tilde{\Gamma})^2\right)^2}\right]$$

The extrema of the function are then found by setting the derivative equal to 0:

$$0 = (1 - \tilde{E}^2)^2 + \tilde{E}^2\tilde{\Gamma}^2 - 2(1 - \tilde{E}^2)^2 + \tilde{\Gamma}^2(1 - \tilde{E}^2)$$

$$0 = (1 - \tilde{E}^2)^2 - \tilde{\Gamma}^2(1 - \tilde{E}^2) - \tilde{E}^2\tilde{\Gamma}^2$$

$$0 = (1 - \tilde{E}^2)^2 - \tilde{\Gamma}^2$$

$$\tilde{E} = \sqrt{1 \pm \tilde{\Gamma}}$$

The two extrema solutions correspond to the maximum and minimum on either side of the resonance. However, since the oscillator strength and linewidth are both positive, we know that the minimum of the permittivity occurs above the resonance while the maximum occurs below it. Therefore, we can focus on the high energy solution. The minimum permittivity can then be found by plugging in the high energy solution into the normalized permittivity equation:

$$\varepsilon_{1,min}\left(\sqrt{1+\tilde{\Gamma}}\right) = \varepsilon_\infty - \frac{f\tilde{\Gamma}^2}{\tilde{\Gamma}^2 + \tilde{\Gamma}^2(1+\tilde{\Gamma})}$$

$$\varepsilon_{1,min}\left(\sqrt{1+\tilde{\Gamma}}\right) = \varepsilon_\infty - \frac{f}{2+\tilde{\Gamma}}$$

Therefore, the real part of the permittivity has a negative region when:

$$\varepsilon_\infty - \frac{f}{2+\tilde{\Gamma}} < 0$$

$$\frac{f}{2+\tilde{\Gamma}} > \varepsilon_\infty$$

$$\frac{f}{2+\left(\frac{\Gamma}{E_x}\right)} > \varepsilon_\infty$$

Although $\varepsilon_1 = 0$ at some points when this criterion is met, these are not epsilon-near-zero (ENZ) points since $\varepsilon_2 = |\varepsilon| > 1$ at these points. The lower loss $\varepsilon_1 = 0$ point is the higher energy one since the lower energy point is nearly resonant with the lossy exciton. We can analytically study when the higher energy point exhibits ENZ behavior by making a narrow exciton approximation ($\Gamma_x << E_x$). In this regime, the real part of the permittivity is approximately

$$\varepsilon_1(\tilde{E}) = \varepsilon_\infty + \frac{f\tilde{\Gamma}_x}{(1-\tilde{E}^2)}$$

We can see that this approximation is accurate near the higher energy $\varepsilon_1 = 0$ energy below using values comparable to the $S_{11}$ exciton in SWCNTs (Figure S6). The approximation then breaks down near the resonance. By setting $\varepsilon_1 = 0$ and solving for $\tilde{E}$, the energy of the $\varepsilon_1 = 0$ point is found to be:

$$\tilde{E} = \sqrt{1 + \frac{f\tilde{\Gamma}_x}{\varepsilon_\infty}}$$

In this approximation, the imaginary part of the permittivity is

$$\varepsilon_2(\tilde{E}) = \frac{f\tilde{\Gamma}_x^2\tilde{E}}{(1-\tilde{E}^2)^2}$$

Plugging the $\varepsilon_1 = 0$ energy into this equation gives

$$\varepsilon_2 = |\varepsilon| = \frac{\varepsilon_\infty^2}{f}\sqrt{1 + \frac{f\tilde{\Gamma}_x}{\varepsilon_\infty}}$$

Further, the square root can be Taylor expanded since $\tilde{\Gamma}_x \ll 1$

$$|\varepsilon| = \frac{\varepsilon_\infty^2}{f} + \frac{\varepsilon_\infty \tilde{\Gamma}}{2}$$

Therefore, an ENZ point occurs at the HET when

$$\frac{\varepsilon_\infty^2}{f} + \frac{\varepsilon_\infty \tilde{\Gamma}}{2} < 1$$

Here, we can see that the same excitonic properties are desired for negative permittivities and ENZ points as expected. However, an important difference is that the ENZ criterion has an $\varepsilon_\infty^2$ term indicating that static permittivity plays a larger role in ENZ excitons than in the negative permittivity case. Therefore, low static permittivity semiconductors are more likely to exhibit ENZ points. Alternatively, another approach for achieving ENZ excitons would be to use semiconductors with highly negative permittivities. These semiconductors can either be patterned, or layered with dielectrics, to reduce the effective static permittivity of the system while maintaining $\varepsilon_1 < 0$ regions.

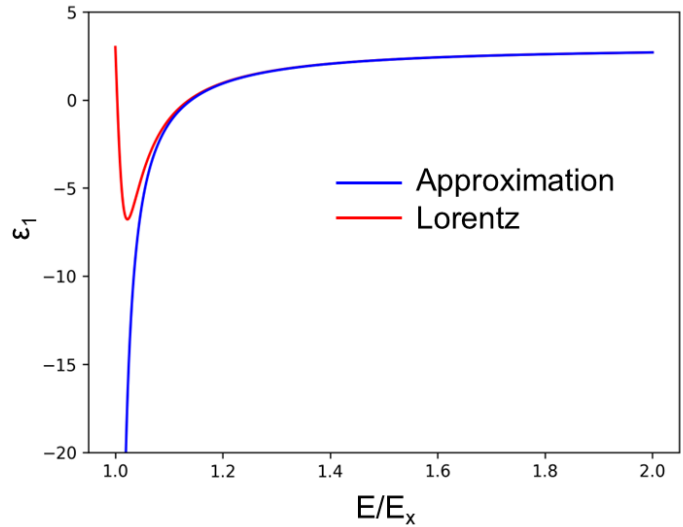

**Figure S3. Narrow exciton approximation**. The real part of the permittivity calculated using the exact Lorentz oscillator model (red) and the narrow exciton approximation (blue) as discussed in Section S2. Comparable values to the measured $S_{11}$ resonance in aligned (6, 5) SWCNTs were used (f = 20, $\tilde{\Gamma}_x$=0.045, and $\varepsilon_\infty$ = 3). The Lorentz model and narrow exciton approximation are found to agree well at the higher energy $\varepsilon_1 = 0$ point.

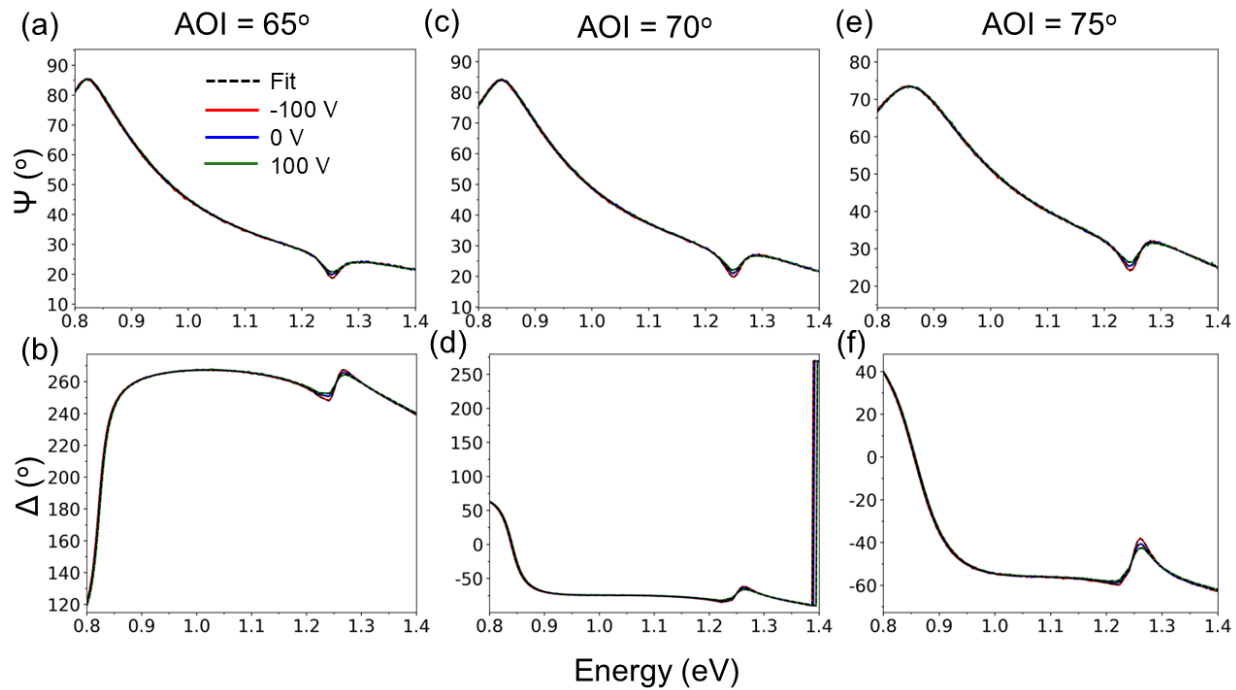

**Figure S4. Ellipsometry of Isotropic (6, 5) SWCNTs.** Gate-dependent **(a, c, e)** ψ and **(b, d, f)** Δ of isotropic (6, 5) SWCNTs at angles of incidence (AOI) of 65° (left), 70° (middle), 75° (right). The largest tunability is observed near the $S_{11}$ resonance. The fitted permittivity is shown in Figure 2a and b. The RMSE in these fits are 3.05, 2.87, and 2.98 for V = -100 V, 0 V, and 100 V, respectively.

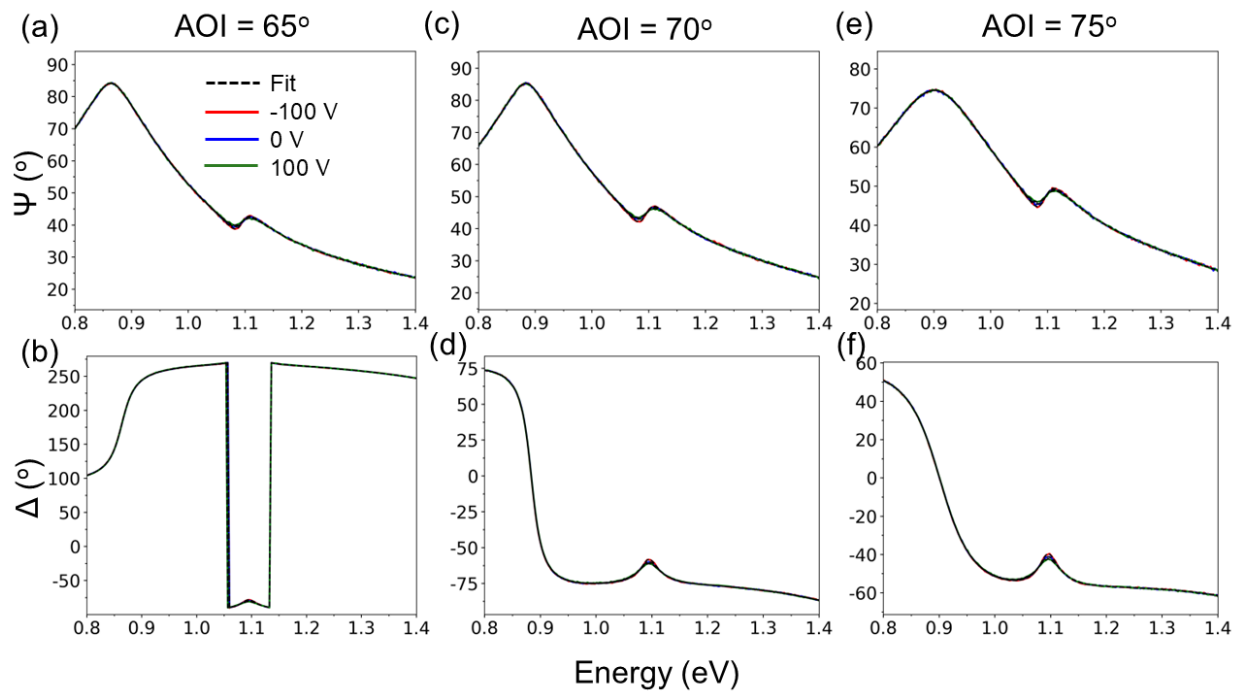

**Figure S5. Ellipsometry of Isotropic (7, 6) SWCNTs.** Gate-dependent **(a, c, e)** ψ and **(b, d, f)** Δ of isotropic (7, 6) SWCNTs at angles of incidence (AOI) of 65° (left), 70° (middle), 75° (right). The largest tunability is observed near the $S_{11}$ resonance. The fitted permittivity is shown in Figure 2a and b. The RMSE in these fits are 2.76, 2.40, and 2.82 for V = -10 V, 0 V, and 100 V, respectively.

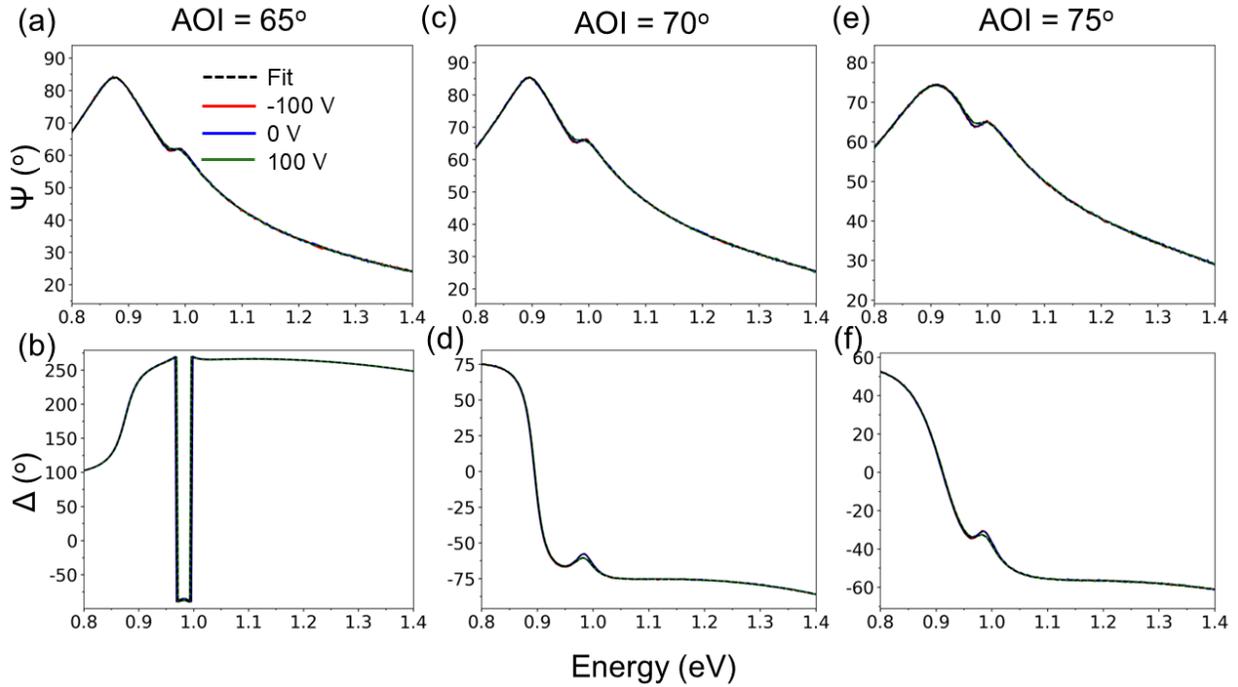

**Figure S6. Ellipsometry of Isotropic (10, 3) SWCNTs.** Gate-dependent **(a, c, e)** ψ and **(b, d, f)** Δ of isotropic (10, 3) SWCNTs at angles of incidence (AOI) of 65° (left), 70° (middle), 75° (right). The largest tunability is observed near the $S_{11}$ resonance. The fitted permittivity is shown in Figure 2a and b. The RMSE in these fits were 2.61, 2.50, and 2.54 for V = -100 V, 0 V, and 100 V, respectively.

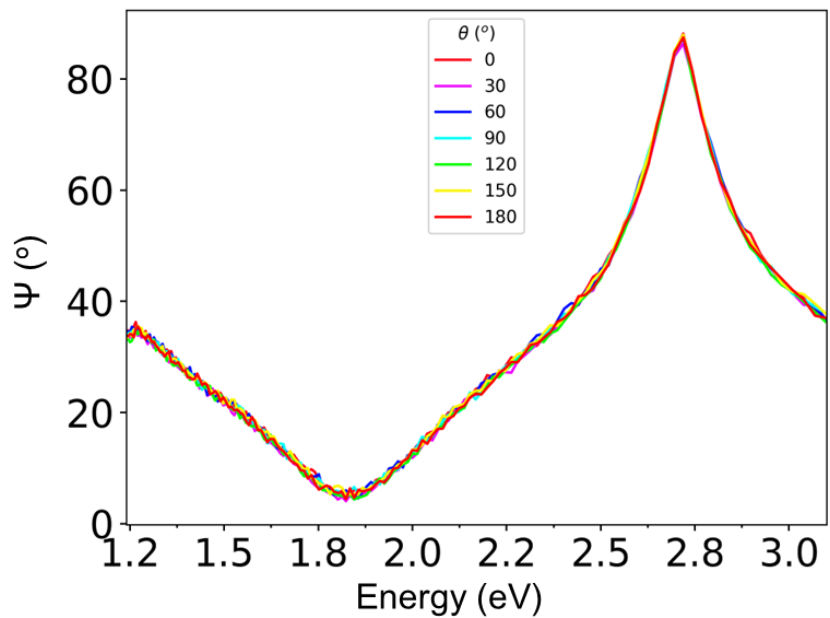

**Figure S7. Large-area, in-plane-rotation-dependent ellipsometry.** Ψ (Ψ = arctan(|r_TM/r_TE|)) of the (6, 5) SWCNTs as a function of the in-plane rotation (θ). θ = 0° means that the light is propagating along the extraordinary direction and the in-plane electric field points along the ordinary direction. The measurement is performed over an area (100 µm x 300 µm) that is much larger than the domain size of aligned regions. The lack of modulation with in-plane rotation demonstrates that the film is effectively isotropic at this scale.

## Section S3: Measuring Optical Anisotropy using Mueller Matrix Ellipsometry

Spectroscopic ellipsometry (SE), as discussed in section S1, is excellent at determining the permittivity of a thin film since it captures information on both the reflected intensity and phase. However, a single SE measurement is insufficient for systems with optical anisotropy. For aligned SWCNTs, there is only a single axis of broken symmetry (along the nanotube) which is called the extraordinary axis. This extraordinary axis has its own unique permittivity while the ordinary axes have the same permittivity. This system is said to be uniaxial since it has a single unique optical axis, and the permittivity tensor is:

$$\varepsilon(E) = \begin{pmatrix} \varepsilon_{ext} & 0 & 0 \\ 0 & \varepsilon_{ord} & 0 \\ 0 & 0 & \varepsilon_{ord} \end{pmatrix}$$

Since $\varepsilon_{ext}$ and $\varepsilon_{ord}$ have real and imaginary parts, there are four unknown optical constants at every energy probed. Therefore, we can see that a single SE scan is insufficient since only two values are measured[3]. One approach around this is to perform multiple SE scans where the momentum of light along the extraordinary axis is changed. This would mean using multiple angles of incidence for systems where the extraordinary axis is out-of-plane and rotating the sample in-plane for systems where the extraordinary axis lays in-plane. We used this in-plane rotation approach to observe the optical anisotropy of aligned SWCNTs in Figure 1e.

However, another approach to measure optical anisotropy is to perform Mueller Matrix (MM) Spectroscopic Ellipsometry. The MM describes the transformation of the Stokes vector of incident light upon reflection. The Stokes vector has 4 elements where the first element is commonly normalized to 1, and the second, third, and fourth elements are the difference in reflected intensity of TM and TE, 45° and -45°, and circular left-hand and right-hand polarized light, respectively. The second through fourth elements are also the principle axes of the Poincare Sphere so the Stokes vector presents a complete description of the polarization of light. Therefore, the MM provides a complete description of the transformation of the polarization of light, and it can be used to accurately measure the permittivity tensor of a thin film. Although this can be done using a single angle of incidence, we performed MM Spectroscopic ellipsometry at 65°, 70°, and 75° for improved accuracy as shown below. The fitting process is similar to the process described in Section S! with the extraordinary and ordinary axes having their own permittivity, the in-plane angle of the extraordinary axis is a fit parameter, and the weighted RMSE equation becomes:

$$RMSE = \sqrt{\frac{1}{15p - q} \sum_{i,j} \left(M_{ij}^{mod} - M_{ij}^{exp}\right)^2}$$

Where $M_{ij}$ is an element of the MM, the sum is over all of the 15 independent elements of the Mueller matrix ($M_{11}$ has been normalized to 1).

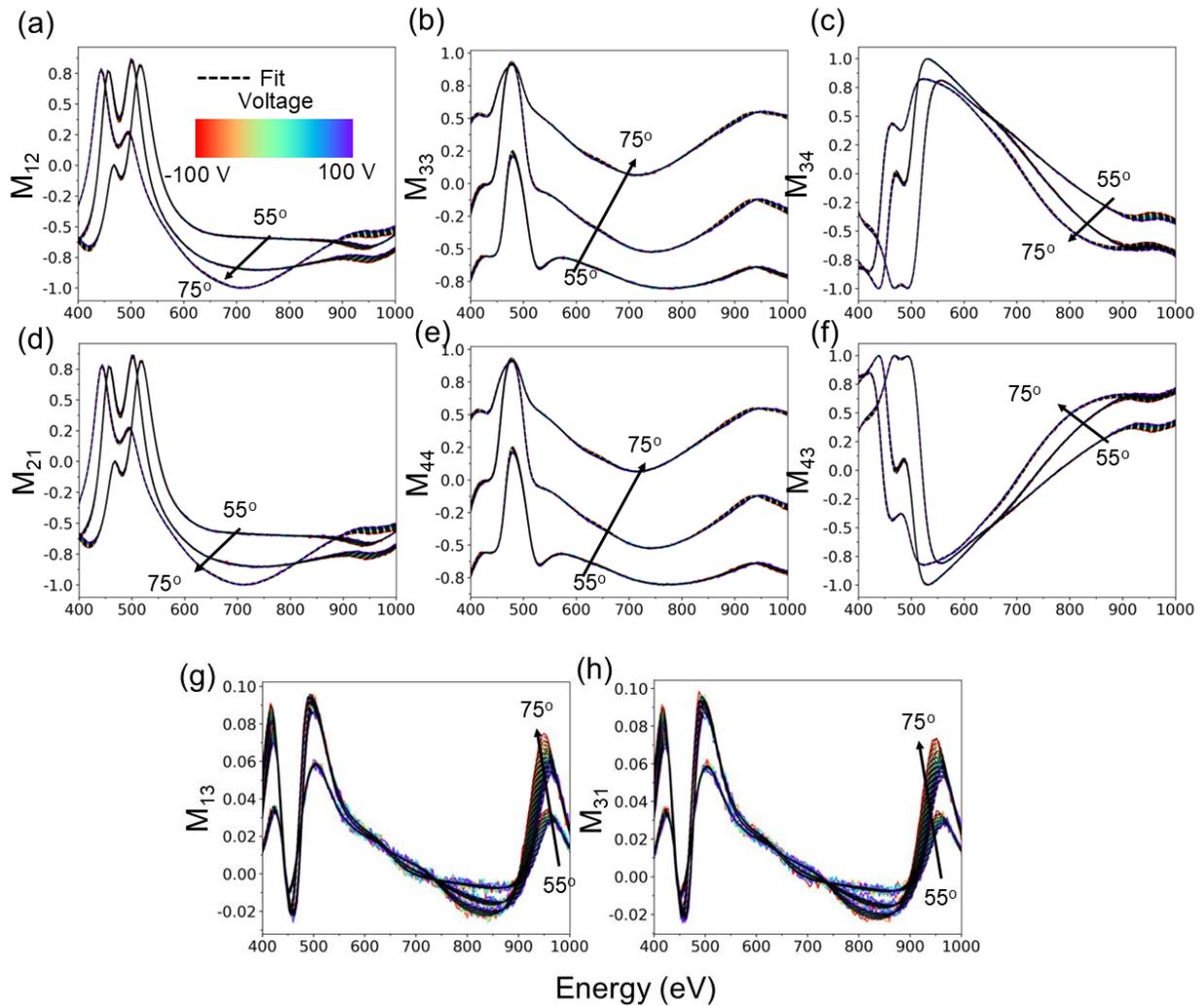

**Figure S8. Gate-dependent Mueller matrix elements of aligned (6, 5) SWCNTs.** Select Mueller matrix (MM) measured using imaging spectroscopic ellipsometer and the fitted values. The angles of incidences are labelled from 55° to 75° (10° increments). The fitted permittivity is shown in Figure 2c and d. The RMSE ranges from 3.81 to 4.66 for the models.

## Section S4: Effects of Alignment on the Permittivity and Dispersion

The persistence of the $S_{11}$ resonance along the ordinary axis indicates that the SWCNT film is not perfectly aligned which is consistent with previous observations of similarly prepared films[4]. To estimate the degree of alignment, we will assume that the orientation of individual nanotubes are normally distributed along a single direction. Therefore, the probability of finding a nanotube aligned along the angle θ is $P(\theta) = \frac{1}{\sigma\sqrt{2\pi}}\exp\left(-\frac{\theta^2}{2\sigma^2}\right)$ where σ is the standard deviation of angular alignment. Additionally, it is well know that the interaction strength of the nanotubes with polarized light is proportional to $\cos^2(\theta)$[4,5]. As a result, we will model the effective oscillator strength ($f_\theta$) for a nanotube aligned at the angle θ as $f_\theta = f_{CNT}*\cos^2(\theta)$ where $f_{CNT}$ is the oscillator strength of a nanotube. We can then model the experimental oscillator strength ($f_{exp}$) as:

$$f_{exp} = \frac{1}{\sigma\sqrt{2\pi}} \int_{-\frac{\pi}{2}}^{\frac{\pi}{2}} \exp\left(-\frac{\theta^2}{2\sigma^2}\right) * f_{CNT} \cos^2(\theta) d\theta$$

Next, we can estimate σ by comparing the experimental $S_{11}$ oscillator strengths along the two optical axes

$$\frac{f_{exp,ext}}{f_{exp,ord}} = \frac{\int_{-\frac{\pi}{2}}^{\frac{\pi}{2}} \exp\left(-\frac{\theta^2}{2\sigma^2}\right) \cos^2(\theta) d\theta}{\int_{-\frac{\pi}{2}}^{\frac{\pi}{2}} \exp\left(-\frac{\theta^2}{2\sigma^2}\right) \sin^2(\theta) d\theta}$$

Numerically solving this equation gives σ = 22.2° which corresponds to a 2D nematic order parameter ($S_{2D}$) of 0.74. This value is slightly larger than measured on similar films that had order up to $S_{2D}$ = 0.66 (σ = 26° for a normal distribution).

Further, we can use this model to estimate the permittivity and dispersion of perfectly aligned (6, 5) SWCNTs. This is done by solving $f_{CNT}$ for the $S_{11}$ and $S_{22}$ resonances and using those values (along with the $S_{12}$ oscillator strength being set to 0) in the multi-Lorentz oscillator to calculate the extraordinary permittivity. The process was repeated for the ordinary direction and $S_{12}$ resonance while the $S_{11}$ and $S_{22}$ oscillator strengths were set to 0. Additionally, we assume that the linewidths and energies are unchanged. The resulting permittivity is shown in Figure S11. This model predicts an improvement in the hyperbolicity of the nanotubes as $\varepsilon_{1,ext}$ becomes more negative and $\varepsilon_{2,ord}$ is reduced along the ordinary direction. However, nanotubes cannot be perfectly aligned. To the best of our knowledge, the current record for wafer-scale alignment is σ = 8.3°. In this case, we can use our model to determine the effective oscillator strength using our known values for $f_{CNT}$, σ = 8.3°, and replacing $\cos^2(\theta)$ with $\sin^2(\theta)$ for the ordinary direction. This gives a similar improvement in the permittivity, and the maximum momentum enhancement becomes 3.41 as shown below in Figure S12.

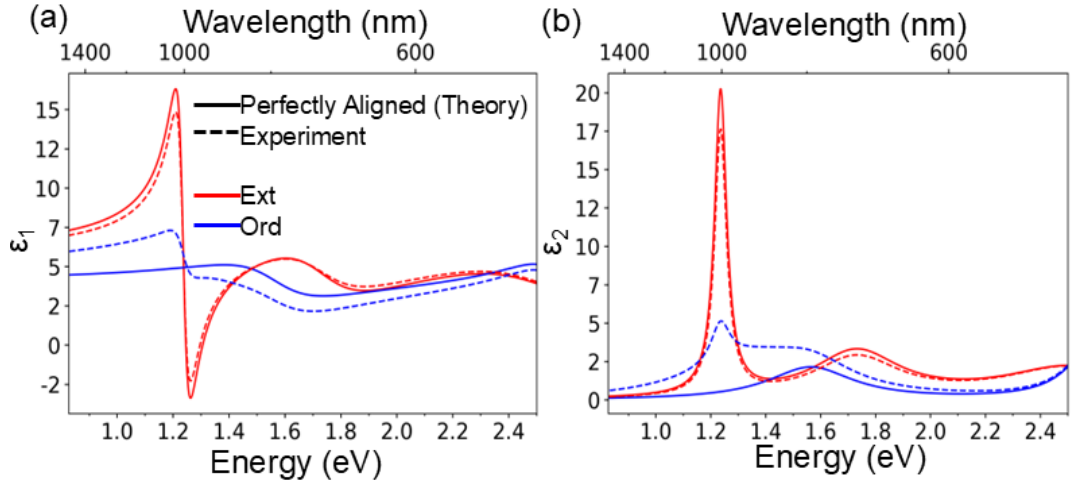

**Figure S9. Theoretical hyperbolicity in perfectly aligned films.** The **(a)** real and **(b)** imaginary parts of the permittivity along the extraordinary (red) and ordinary (blue) axes for experimental (dashed) and theoretical, perfectly aligned (solid) films of (6, 5) SWCNTs.

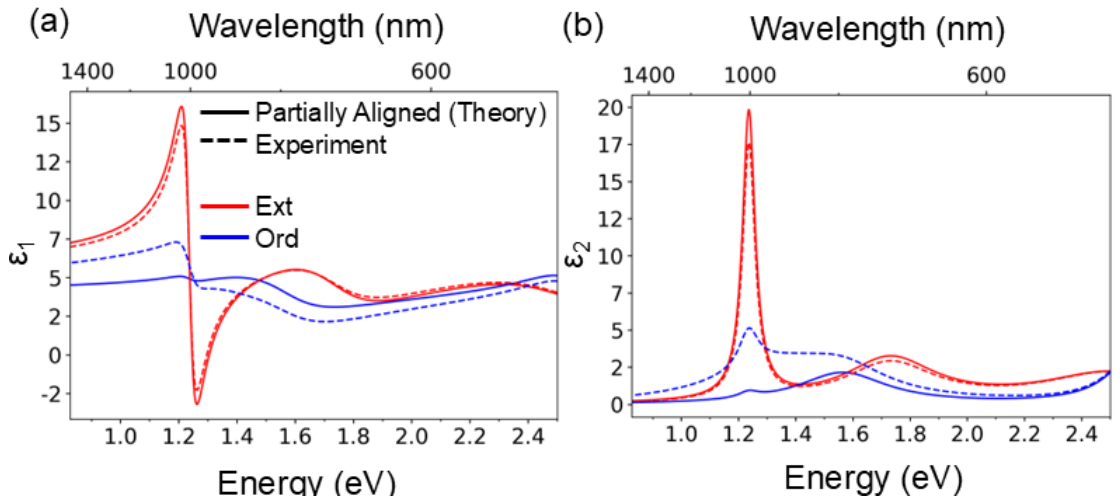

**Figure S10. Theoretical hyperbolicity in highly aligned films.** The **(a)** real and **(b)** imaginary parts of the permittivity along the extraordinary (red) and ordinary (blue) axes for experimental (dashed) and theoretical, highly aligned (solid, σ = 8.3°) films of (6, 5) SWCNTs.

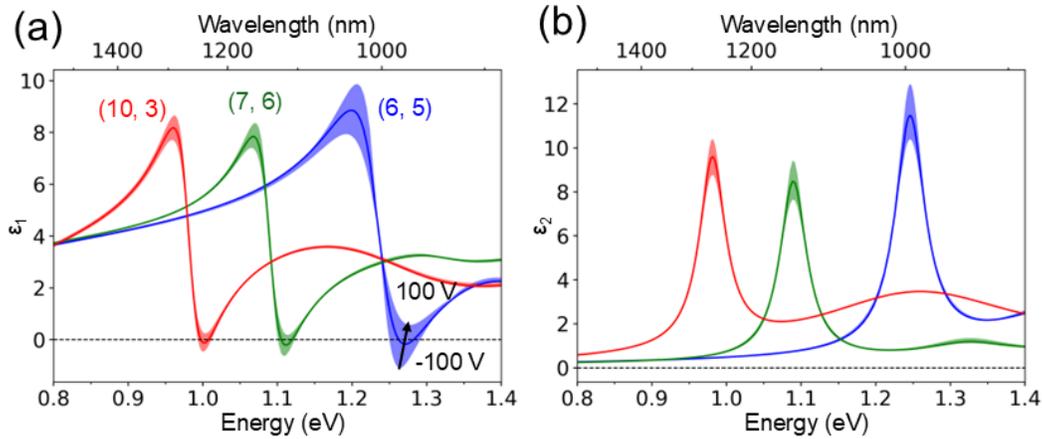

**Figure S11. Tunable permittivity of isotropic films.** The gate-dependent **(a)** real and **(b)** imaginary parts of the permittivity of isotropic films of (10,3) (red), (7,6) (green), and (6,5) (blue) SWCNTs over a voltage range of -100 V to 100 V.

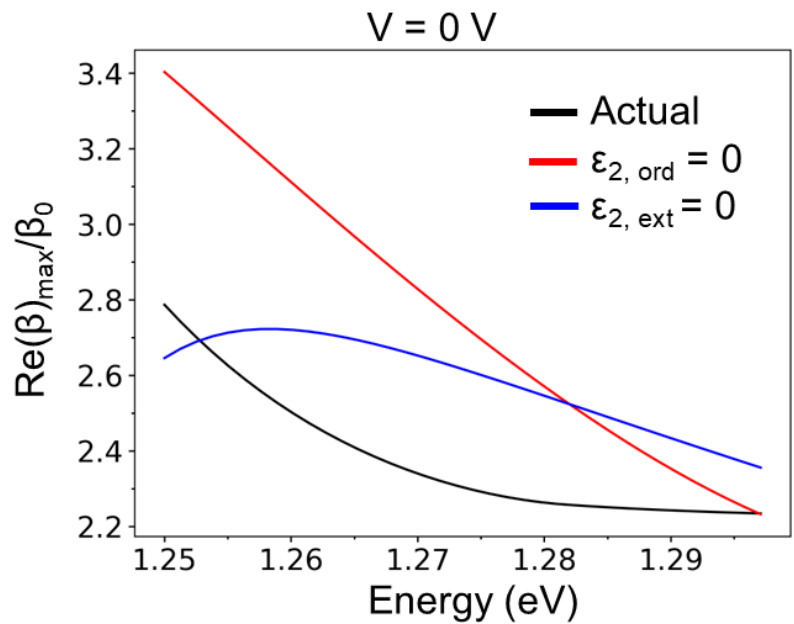

**Figure S12. Effects of loss on momentum enhancement.** The momentum enhancement according to the bulk dispersion relation for the actual SWCNTs, and when $\varepsilon_2$ is set to 0 along the ordinary (red) and extraordinary (blue) direction. A larger improvement is observed when loss is ignored along the extraordinary axis showing that it is more important to reduce its loss.

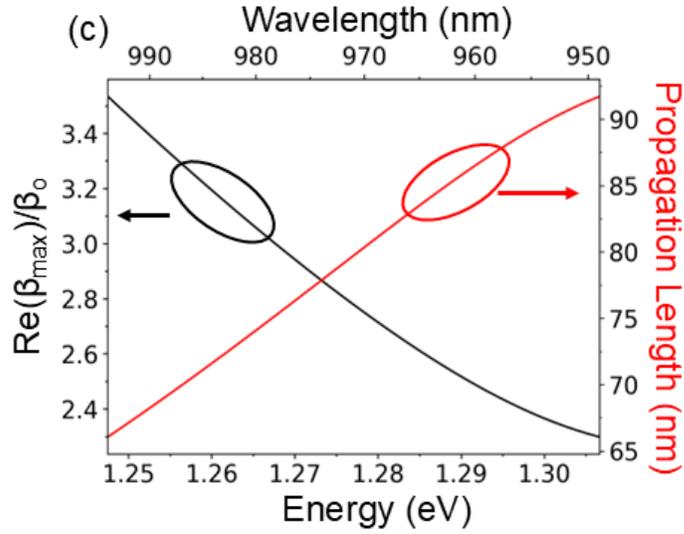

**Figure S13 Momentum enhancement of perfectly aligned SWCNTs.** The momentum enhancement (black) and propagation length in perfectly aligned films of (6,5) SWCNTs using the theory in Section S4 and the dispersion relation.

## Section S5: Magnitude of Lorentz Oscillator

Building off the Lorentz oscillator model discussed in section S2, the peak amplitude of the oscillator occurs at the resonant energy of the transition. Therefore, the normalized permittivity at this energy is

$$\varepsilon(\tilde{E} = 1) = \varepsilon_\infty + if$$

And the magnitude is

$$|\varepsilon(\tilde{E} = 1)| = \sqrt{f^2 + \varepsilon_\infty^2}$$

Therefore, the maximum amplitude of the permittivity increases linearly with the oscillator strength, and it does not depend on the linewidth. Although this maximum value occurs in the hyperbolic regime, the LET will occur near it in strong exciton systems.

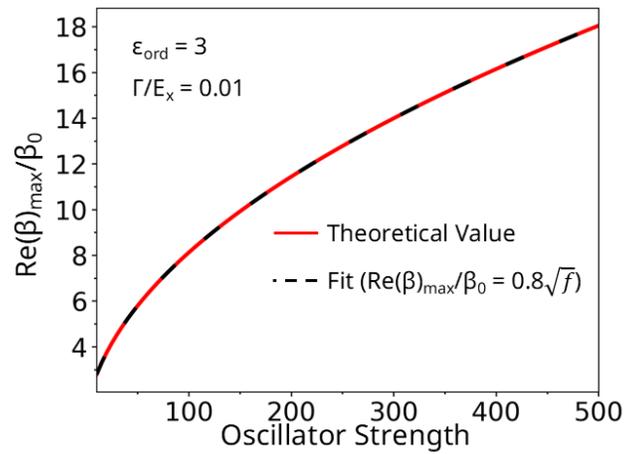

**Figure S14. Theoretical momentum enhancement of strong excitons.** The calculated maximum momentum enhancement of an anisotropic exciton with a normalized linewidth of 0.01 and $\varepsilon_{ord}$ = 3 as a function of oscillator strength (f). The enhancement is found to follow a square root curve which is predicted by theory[6].

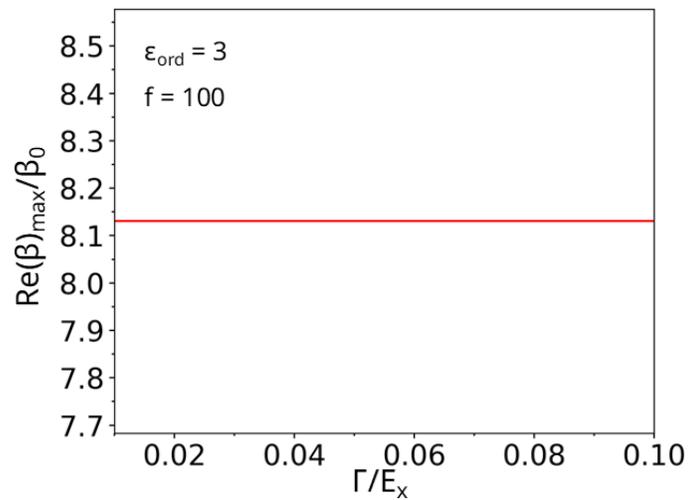

**Figure S15. Theoretical momentum enhancement of strong excitons.** The calculated maximum momentum enhancement of an anisotropic exciton with an oscillator strength of 100 and $\varepsilon_{ord}$ = 3 as a function of the normalized linewidth.

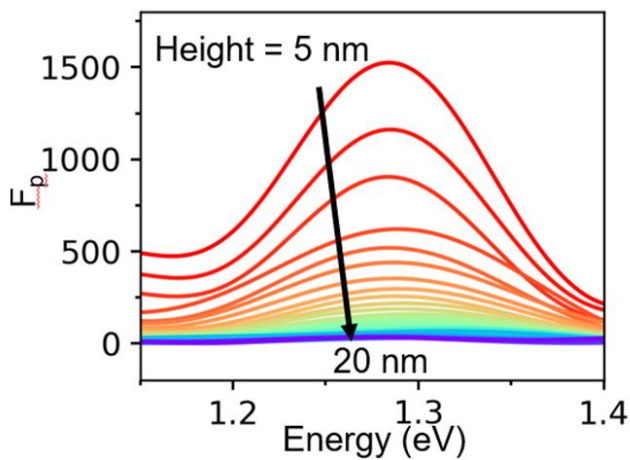

**Figure S16. Dipole height dependent Purcell factor.** The Purcell factor ($F_p$) of a dipole in the system shown in Figure 4a for heights from 5 nm to 20 nm showing a large enhancement as the dipole approaches the hyperbolic SWCNTs.

**Table S1. Gate-dependent Permittivity of (6, 5) SWCNTs).** The tabulated anisotropic, complex permittivity of (6, 5) SWCNTs at V = -100 V, 0 V, and 100 V. The other voltages are available upon reasonable request to the corresponding author. Extrapolated values are in red.

|   | -100 | | | | 0 | | | | 100 | | | |
|---|---|---|---|---|---|---|---|---|---|---|---|---|
| Wavelength (nm) | $\varepsilon_{1,ext}$ | $\varepsilon_{2,ext}$ | $\varepsilon_{1,ord}$ | $\varepsilon_{2,ord}$ | $\varepsilon_{1,ext}$ | $\varepsilon_{2,ext}$ | $\varepsilon_{1,ord}$ | $\varepsilon_{2,ord}$ | $\varepsilon_{1,ext}$ | $\varepsilon_{2,ext}$ | $\varepsilon_{1,ord}$ | $\varepsilon_{2,ord}$ |
| 496 | 4.00 | 2.29 | 4.78 | 2.30 | 4.04 | 2.23 | 4.78 | 2.26 | 4.08 | 2.18 | 4.80 | 2.22 |
| 497 | 4.03 | 2.29 | 4.79 | 2.21 | 4.07 | 2.23 | 4.79 | 2.17 | 4.11 | 2.18 | 4.81 | 2.13 |
| 498 | 4.06 | 2.29 | 4.79 | 2.12 | 4.10 | 2.22 | 4.79 | 2.09 | 4.14 | 2.17 | 4.81 | 2.05 |
| 499 | 4.09 | 2.29 | 4.79 | 2.04 | 4.13 | 2.22 | 4.79 | 2.01 | 4.17 | 2.17 | 4.81 | 1.97 |
| 500 | 4.12 | 2.28 | 4.79 | 1.97 | 4.15 | 2.22 | 4.78 | 1.93 | 4.19 | 2.17 | 4.81 | 1.89 |
| 501 | 4.15 | 2.28 | 4.78 | 1.89 | 4.18 | 2.21 | 4.78 | 1.86 | 4.22 | 2.17 | 4.80 | 1.82 |
| 502 | 4.18 | 2.27 | 4.77 | 1.82 | 4.21 | 2.21 | 4.76 | 1.79 | 4.25 | 2.16 | 4.79 | 1.75 |
| 503 | 4.21 | 2.26 | 4.75 | 1.75 | 4.24 | 2.20 | 4.75 | 1.72 | 4.27 | 2.15 | 4.78 | 1.69 |
| 504 | 4.23 | 2.26 | 4.74 | 1.69 | 4.26 | 2.19 | 4.74 | 1.66 | 4.30 | 2.15 | 4.76 | 1.63 |
| 505 | 4.26 | 2.25 | 4.72 | 1.63 | 4.29 | 2.19 | 4.72 | 1.60 | 4.32 | 2.14 | 4.75 | 1.57 |
| 506 | 4.29 | 2.24 | 4.70 | 1.58 | 4.31 | 2.18 | 4.70 | 1.55 | 4.35 | 2.13 | 4.73 | 1.52 |
| 507 | 4.31 | 2.23 | 4.68 | 1.52 | 4.34 | 2.17 | 4.68 | 1.50 | 4.37 | 2.12 | 4.71 | 1.46 |
| 508 | 4.34 | 2.21 | 4.66 | 1.47 | 4.36 | 2.16 | 4.66 | 1.45 | 4.39 | 2.11 | 4.69 | 1.41 |
| 509 | 4.36 | 2.20 | 4.64 | 1.42 | 4.38 | 2.14 | 4.64 | 1.40 | 4.42 | 2.10 | 4.67 | 1.37 |
| 510 | 4.38 | 2.19 | 4.62 | 1.38 | 4.41 | 2.13 | 4.62 | 1.36 | 4.44 | 2.09 | 4.65 | 1.32 |
| 511 | 4.40 | 2.17 | 4.59 | 1.33 | 4.43 | 2.12 | 4.60 | 1.31 | 4.46 | 2.08 | 4.63 | 1.28 |
| 512 | 4.42 | 2.16 | 4.57 | 1.29 | 4.45 | 2.11 | 4.57 | 1.27 | 4.48 | 2.06 | 4.61 | 1.24 |
| 513 | 4.44 | 2.15 | 4.55 | 1.26 | 4.47 | 2.09 | 4.55 | 1.24 | 4.50 | 2.05 | 4.58 | 1.21 |

| | | | | | | | | | | | |
|---|---|---|---|---|---|---|---|---|---|---|---|
| 514 | 4.46 | 2.13 | 4.52 | 1.22 | 4.49 | 2.08 | 4.53 | 1.20 | 4.52 | 2.04 | 4.56 | 1.17 |
| 515 | 4.48 | 2.11 | 4.50 | 1.19 | 4.50 | 2.06 | 4.50 | 1.17 | 4.53 | 2.02 | 4.54 | 1.14 |
| 516 | 4.50 | 2.10 | 4.48 | 1.15 | 4.52 | 2.05 | 4.48 | 1.14 | 4.55 | 2.01 | 4.52 | 1.11 |
| 517 | 4.52 | 2.08 | 4.45 | 1.12 | 4.54 | 2.03 | 4.46 | 1.10 | 4.57 | 1.99 | 4.49 | 1.08 |
| 518 | 4.53 | 2.06 | 4.43 | 1.09 | 4.55 | 2.02 | 4.43 | 1.08 | 4.58 | 1.98 | 4.47 | 1.05 |
| 519 | 4.55 | 2.05 | 4.41 | 1.07 | 4.57 | 2.00 | 4.41 | 1.05 | 4.60 | 1.96 | 4.45 | 1.02 |
| 520 | 4.56 | 2.03 | 4.38 | 1.04 | 4.58 | 1.98 | 4.39 | 1.02 | 4.61 | 1.95 | 4.43 | 1.00 |
| 521 | 4.57 | 2.01 | 4.36 | 1.02 | 4.59 | 1.97 | 4.37 | 1.00 | 4.62 | 1.93 | 4.40 | 0.97 |
| 522 | 4.58 | 1.99 | 4.34 | 0.99 | 4.60 | 1.95 | 4.34 | 0.98 | 4.63 | 1.92 | 4.38 | 0.95 |
| 523 | 4.60 | 1.98 | 4.31 | 0.97 | 4.62 | 1.93 | 4.32 | 0.96 | 4.64 | 1.90 | 4.36 | 0.93 |
| 524 | 4.61 | 1.96 | 4.29 | 0.95 | 4.63 | 1.92 | 4.30 | 0.93 | 4.65 | 1.88 | 4.34 | 0.91 |
| 525 | 4.62 | 1.94 | 4.27 | 0.93 | 4.63 | 1.90 | 4.28 | 0.92 | 4.66 | 1.87 | 4.32 | 0.89 |
| 526 | 4.62 | 1.92 | 4.25 | 0.91 | 4.64 | 1.88 | 4.26 | 0.90 | 4.67 | 1.85 | 4.30 | 0.87 |
| 527 | 4.63 | 1.90 | 4.22 | 0.89 | 4.65 | 1.86 | 4.24 | 0.88 | 4.68 | 1.83 | 4.28 | 0.86 |
| 528 | 4.64 | 1.88 | 4.20 | 0.88 | 4.66 | 1.85 | 4.21 | 0.86 | 4.69 | 1.82 | 4.26 | 0.84 |
| 529 | 4.65 | 1.87 | 4.18 | 0.86 | 4.67 | 1.83 | 4.19 | 0.85 | 4.70 | 1.80 | 4.24 | 0.82 |
| 530 | 4.65 | 1.85 | 4.16 | 0.84 | 4.67 | 1.81 | 4.17 | 0.83 | 4.70 | 1.79 | 4.22 | 0.81 |
| 531 | 4.66 | 1.83 | 4.14 | 0.83 | 4.68 | 1.80 | 4.15 | 0.82 | 4.71 | 1.77 | 4.20 | 0.79 |
| 532 | 4.66 | 1.81 | 4.12 | 0.82 | 4.68 | 1.78 | 4.13 | 0.80 | 4.71 | 1.75 | 4.18 | 0.78 |
| 533 | 4.66 | 1.80 | 4.10 | 0.80 | 4.68 | 1.76 | 4.11 | 0.79 | 4.72 | 1.74 | 4.16 | 0.77 |
| 534 | 4.67 | 1.78 | 4.08 | 0.79 | 4.69 | 1.75 | 4.09 | 0.78 | 4.72 | 1.72 | 4.14 | 0.76 |
| 535 | 4.67 | 1.76 | 4.06 | 0.78 | 4.69 | 1.73 | 4.07 | 0.77 | 4.72 | 1.71 | 4.12 | 0.74 |
| 536 | 4.67 | 1.74 | 4.04 | 0.77 | 4.69 | 1.71 | 4.06 | 0.76 | 4.72 | 1.69 | 4.10 | 0.73 |
| 537 | 4.67 | 1.73 | 4.02 | 0.76 | 4.69 | 1.70 | 4.04 | 0.75 | 4.73 | 1.68 | 4.09 | 0.72 |
| 538 | 4.67 | 1.71 | 4.00 | 0.75 | 4.69 | 1.68 | 4.02 | 0.74 | 4.73 | 1.66 | 4.07 | 0.71 |
| 539 | 4.67 | 1.70 | 3.98 | 0.74 | 4.69 | 1.67 | 4.00 | 0.73 | 4.73 | 1.65 | 4.05 | 0.70 |
| 540 | 4.67 | 1.68 | 3.96 | 0.73 | 4.69 | 1.65 | 3.98 | 0.72 | 4.73 | 1.63 | 4.03 | 0.70 |
| 541 | 4.67 | 1.66 | 3.95 | 0.72 | 4.69 | 1.64 | 3.97 | 0.71 | 4.73 | 1.62 | 4.02 | 0.69 |
| 542 | 4.67 | 1.65 | 3.93 | 0.71 | 4.69 | 1.62 | 3.95 | 0.70 | 4.73 | 1.60 | 4.00 | 0.68 |
| 543 | 4.67 | 1.63 | 3.91 | 0.70 | 4.69 | 1.61 | 3.93 | 0.70 | 4.72 | 1.59 | 3.98 | 0.67 |
| 544 | 4.66 | 1.62 | 3.89 | 0.70 | 4.69 | 1.59 | 3.91 | 0.69 | 4.72 | 1.58 | 3.97 | 0.66 |
| 545 | 4.66 | 1.60 | 3.88 | 0.69 | 4.68 | 1.58 | 3.90 | 0.68 | 4.72 | 1.56 | 3.95 | 0.66 |
| 546 | 4.66 | 1.59 | 3.86 | 0.68 | 4.68 | 1.57 | 3.88 | 0.68 | 4.72 | 1.55 | 3.93 | 0.65 |
| 547 | 4.65 | 1.58 | 3.84 | 0.68 | 4.68 | 1.55 | 3.86 | 0.67 | 4.72 | 1.54 | 3.92 | 0.65 |
| 548 | 4.65 | 1.56 | 3.83 | 0.67 | 4.67 | 1.54 | 3.85 | 0.66 | 4.71 | 1.53 | 3.90 | 0.64 |
| 549 | 4.64 | 1.55 | 3.81 | 0.67 | 4.67 | 1.53 | 3.83 | 0.66 | 4.71 | 1.51 | 3.89 | 0.63 |
| 550 | 4.64 | 1.54 | 3.79 | 0.66 | 4.67 | 1.52 | 3.82 | 0.65 | 4.70 | 1.50 | 3.87 | 0.63 |
| 551 | 4.63 | 1.52 | 3.78 | 0.66 | 4.66 | 1.51 | 3.80 | 0.65 | 4.70 | 1.49 | 3.86 | 0.62 |

| | | | | | | | | | | | |
|---|---|---|---|---|---|---|---|---|---|---|---|
| 552 | 4.63 | 1.51 | 3.76 | 0.65 | 4.66 | 1.49 | 3.79 | 0.64 | 4.70 | 1.48 | 3.84 | 0.62 |
| 553 | 4.62 | 1.50 | 3.74 | 0.65 | 4.65 | 1.48 | 3.77 | 0.64 | 4.69 | 1.47 | 3.83 | 0.62 |
| 554 | 4.61 | 1.49 | 3.73 | 0.64 | 4.64 | 1.47 | 3.75 | 0.64 | 4.69 | 1.46 | 3.81 | 0.61 |
| 555 | 4.61 | 1.48 | 3.71 | 0.64 | 4.64 | 1.46 | 3.74 | 0.63 | 4.68 | 1.45 | 3.80 | 0.61 |
| 556 | 4.60 | 1.47 | 3.70 | 0.64 | 4.63 | 1.45 | 3.73 | 0.63 | 4.67 | 1.44 | 3.78 | 0.61 |
| 557 | 4.59 | 1.46 | 3.68 | 0.63 | 4.62 | 1.44 | 3.71 | 0.63 | 4.67 | 1.43 | 3.77 | 0.60 |
| 558 | 4.58 | 1.45 | 3.67 | 0.63 | 4.62 | 1.43 | 3.70 | 0.62 | 4.66 | 1.42 | 3.76 | 0.60 |
| 559 | 4.58 | 1.44 | 3.65 | 0.63 | 4.61 | 1.42 | 3.68 | 0.62 | 4.66 | 1.41 | 3.74 | 0.60 |
| 560 | 4.57 | 1.43 | 3.64 | 0.63 | 4.60 | 1.41 | 3.67 | 0.62 | 4.65 | 1.40 | 3.73 | 0.59 |
| 561 | 4.56 | 1.42 | 3.62 | 0.62 | 4.59 | 1.40 | 3.65 | 0.62 | 4.64 | 1.39 | 3.72 | 0.59 |
| 562 | 4.55 | 1.41 | 3.61 | 0.62 | 4.59 | 1.40 | 3.64 | 0.61 | 4.63 | 1.39 | 3.70 | 0.59 |
| 563 | 4.54 | 1.40 | 3.60 | 0.62 | 4.58 | 1.39 | 3.63 | 0.61 | 4.63 | 1.38 | 3.69 | 0.59 |
| 564 | 4.53 | 1.39 | 3.58 | 0.62 | 4.57 | 1.38 | 3.61 | 0.61 | 4.62 | 1.37 | 3.68 | 0.58 |
| 565 | 4.52 | 1.38 | 3.57 | 0.62 | 4.56 | 1.37 | 3.60 | 0.61 | 4.61 | 1.36 | 3.66 | 0.58 |
| 566 | 4.51 | 1.38 | 3.55 | 0.62 | 4.55 | 1.37 | 3.59 | 0.61 | 4.60 | 1.36 | 3.65 | 0.58 |
| 567 | 4.50 | 1.37 | 3.54 | 0.61 | 4.54 | 1.36 | 3.57 | 0.61 | 4.60 | 1.35 | 3.64 | 0.58 |
| 568 | 4.49 | 1.36 | 3.53 | 0.61 | 4.54 | 1.35 | 3.56 | 0.61 | 4.59 | 1.34 | 3.63 | 0.58 |
| 569 | 4.48 | 1.36 | 3.51 | 0.61 | 4.53 | 1.35 | 3.55 | 0.60 | 4.58 | 1.34 | 3.61 | 0.58 |
| 570 | 4.47 | 1.35 | 3.50 | 0.61 | 4.52 | 1.34 | 3.53 | 0.60 | 4.57 | 1.33 | 3.60 | 0.58 |
| 571 | 4.46 | 1.34 | 3.49 | 0.61 | 4.51 | 1.34 | 3.52 | 0.60 | 4.56 | 1.33 | 3.59 | 0.58 |
| 572 | 4.45 | 1.34 | 3.47 | 0.61 | 4.50 | 1.33 | 3.51 | 0.60 | 4.55 | 1.32 | 3.58 | 0.58 |
| 573 | 4.44 | 1.33 | 3.46 | 0.61 | 4.49 | 1.33 | 3.50 | 0.60 | 4.54 | 1.32 | 3.57 | 0.58 |
| 574 | 4.43 | 1.33 | 3.45 | 0.61 | 4.48 | 1.32 | 3.48 | 0.60 | 4.54 | 1.31 | 3.55 | 0.58 |
| 575 | 4.42 | 1.32 | 3.43 | 0.61 | 4.47 | 1.32 | 3.47 | 0.60 | 4.53 | 1.31 | 3.54 | 0.58 |
| 576 | 4.41 | 1.32 | 3.42 | 0.61 | 4.46 | 1.31 | 3.46 | 0.60 | 4.52 | 1.31 | 3.53 | 0.58 |
| 577 | 4.40 | 1.32 | 3.41 | 0.61 | 4.45 | 1.31 | 3.45 | 0.60 | 4.51 | 1.30 | 3.52 | 0.58 |
| 578 | 4.39 | 1.31 | 3.40 | 0.61 | 4.44 | 1.31 | 3.43 | 0.60 | 4.50 | 1.30 | 3.51 | 0.58 |
| 579 | 4.38 | 1.31 | 3.38 | 0.61 | 4.43 | 1.30 | 3.42 | 0.60 | 4.49 | 1.30 | 3.50 | 0.58 |
| 580 | 4.36 | 1.31 | 3.37 | 0.61 | 4.42 | 1.30 | 3.41 | 0.61 | 4.48 | 1.30 | 3.48 | 0.58 |
| 581 | 4.35 | 1.30 | 3.36 | 0.61 | 4.41 | 1.30 | 3.40 | 0.61 | 4.47 | 1.29 | 3.47 | 0.58 |
| 582 | 4.34 | 1.30 | 3.35 | 0.62 | 4.40 | 1.30 | 3.39 | 0.61 | 4.46 | 1.29 | 3.46 | 0.58 |
| 583 | 4.33 | 1.30 | 3.33 | 0.62 | 4.38 | 1.29 | 3.37 | 0.61 | 4.45 | 1.29 | 3.45 | 0.58 |
| 584 | 4.32 | 1.30 | 3.32 | 0.62 | 4.37 | 1.29 | 3.36 | 0.61 | 4.44 | 1.29 | 3.44 | 0.58 |
| 585 | 4.31 | 1.29 | 3.31 | 0.62 | 4.36 | 1.29 | 3.35 | 0.61 | 4.43 | 1.29 | 3.43 | 0.58 |
| 586 | 4.29 | 1.29 | 3.30 | 0.62 | 4.35 | 1.29 | 3.34 | 0.61 | 4.42 | 1.29 | 3.42 | 0.58 |
| 587 | 4.28 | 1.29 | 3.29 | 0.62 | 4.34 | 1.29 | 3.33 | 0.61 | 4.41 | 1.29 | 3.41 | 0.58 |
| 588 | 4.27 | 1.29 | 3.27 | 0.62 | 4.33 | 1.29 | 3.32 | 0.62 | 4.40 | 1.29 | 3.39 | 0.59 |
| 589 | 4.26 | 1.29 | 3.26 | 0.63 | 4.32 | 1.29 | 3.30 | 0.62 | 4.39 | 1.29 | 3.38 | 0.59 |

| | | | | | | | | | | | |
|---|---|---|---|---|---|---|---|---|---|---|---|
| 590 | 4.25 | 1.29 | 3.25 | 0.63 | 4.31 | 1.29 | 3.29 | 0.62 | 4.38 | 1.29 | 3.37 | 0.59 |
| 591 | 4.23 | 1.29 | 3.24 | 0.63 | 4.30 | 1.29 | 3.28 | 0.62 | 4.37 | 1.29 | 3.36 | 0.59 |
| 592 | 4.22 | 1.29 | 3.23 | 0.63 | 4.29 | 1.29 | 3.27 | 0.62 | 4.36 | 1.29 | 3.35 | 0.59 |
| 593 | 4.21 | 1.29 | 3.22 | 0.64 | 4.27 | 1.29 | 3.26 | 0.63 | 4.35 | 1.29 | 3.34 | 0.59 |
| 594 | 4.20 | 1.30 | 3.20 | 0.64 | 4.26 | 1.29 | 3.25 | 0.63 | 4.34 | 1.29 | 3.33 | 0.60 |
| 595 | 4.18 | 1.30 | 3.19 | 0.64 | 4.25 | 1.29 | 3.24 | 0.63 | 4.33 | 1.29 | 3.32 | 0.60 |
| 596 | 4.17 | 1.30 | 3.18 | 0.64 | 4.24 | 1.30 | 3.23 | 0.63 | 4.32 | 1.29 | 3.31 | 0.60 |
| 597 | 4.16 | 1.30 | 3.17 | 0.65 | 4.23 | 1.30 | 3.22 | 0.64 | 4.31 | 1.30 | 3.30 | 0.60 |
| 598 | 4.15 | 1.30 | 3.16 | 0.65 | 4.22 | 1.30 | 3.20 | 0.64 | 4.30 | 1.30 | 3.29 | 0.61 |
| 599 | 4.13 | 1.31 | 3.15 | 0.65 | 4.21 | 1.30 | 3.19 | 0.64 | 4.29 | 1.30 | 3.28 | 0.61 |
| 600 | 4.12 | 1.31 | 3.13 | 0.65 | 4.20 | 1.31 | 3.18 | 0.65 | 4.28 | 1.30 | 3.27 | 0.61 |
| 601 | 4.11 | 1.31 | 3.12 | 0.66 | 4.18 | 1.31 | 3.17 | 0.65 | 4.27 | 1.31 | 3.26 | 0.61 |
| 602 | 4.10 | 1.32 | 3.11 | 0.66 | 4.17 | 1.31 | 3.16 | 0.65 | 4.26 | 1.31 | 3.25 | 0.62 |
| 603 | 4.08 | 1.32 | 3.10 | 0.66 | 4.16 | 1.32 | 3.15 | 0.65 | 4.25 | 1.32 | 3.24 | 0.62 |
| 604 | 4.07 | 1.33 | 3.09 | 0.67 | 4.15 | 1.32 | 3.14 | 0.66 | 4.24 | 1.32 | 3.23 | 0.62 |
| 605 | 4.06 | 1.33 | 3.08 | 0.67 | 4.14 | 1.33 | 3.13 | 0.66 | 4.23 | 1.32 | 3.22 | 0.63 |
| 606 | 4.05 | 1.33 | 3.07 | 0.68 | 4.13 | 1.33 | 3.12 | 0.67 | 4.22 | 1.33 | 3.21 | 0.63 |
| 607 | 4.03 | 1.34 | 3.06 | 0.68 | 4.12 | 1.34 | 3.11 | 0.67 | 4.21 | 1.33 | 3.20 | 0.63 |
| 608 | 4.02 | 1.35 | 3.05 | 0.68 | 4.11 | 1.34 | 3.10 | 0.67 | 4.20 | 1.34 | 3.19 | 0.64 |
| 609 | 4.01 | 1.35 | 3.03 | 0.69 | 4.09 | 1.35 | 3.09 | 0.68 | 4.19 | 1.35 | 3.18 | 0.64 |
| 610 | 4.00 | 1.36 | 3.02 | 0.69 | 4.08 | 1.36 | 3.08 | 0.68 | 4.18 | 1.35 | 3.17 | 0.64 |
| 611 | 3.98 | 1.36 | 3.01 | 0.70 | 4.07 | 1.36 | 3.07 | 0.68 | 4.17 | 1.36 | 3.16 | 0.65 |
| 612 | 3.97 | 1.37 | 3.00 | 0.70 | 4.06 | 1.37 | 3.06 | 0.69 | 4.16 | 1.36 | 3.15 | 0.65 |
| 613 | 3.96 | 1.38 | 2.99 | 0.70 | 4.05 | 1.38 | 3.05 | 0.69 | 4.15 | 1.37 | 3.14 | 0.65 |
| 614 | 3.95 | 1.39 | 2.98 | 0.71 | 4.04 | 1.38 | 3.04 | 0.70 | 4.14 | 1.38 | 3.13 | 0.66 |
| 615 | 3.94 | 1.39 | 2.97 | 0.71 | 4.03 | 1.39 | 3.02 | 0.70 | 4.13 | 1.39 | 3.12 | 0.66 |
| 616 | 3.92 | 1.40 | 2.96 | 0.72 | 4.02 | 1.40 | 3.01 | 0.71 | 4.12 | 1.39 | 3.11 | 0.67 |
| 617 | 3.91 | 1.41 | 2.95 | 0.72 | 4.01 | 1.41 | 3.00 | 0.71 | 4.11 | 1.40 | 3.10 | 0.67 |
| 618 | 3.90 | 1.42 | 2.94 | 0.73 | 4.00 | 1.42 | 2.99 | 0.72 | 4.10 | 1.41 | 3.09 | 0.67 |
| 619 | 3.89 | 1.43 | 2.93 | 0.73 | 3.99 | 1.42 | 2.98 | 0.72 | 4.09 | 1.42 | 3.08 | 0.68 |
| 620 | 3.88 | 1.44 | 2.91 | 0.74 | 3.98 | 1.43 | 2.97 | 0.73 | 4.08 | 1.43 | 3.07 | 0.68 |
| 621 | 3.86 | 1.45 | 2.90 | 0.74 | 3.97 | 1.44 | 2.96 | 0.73 | 4.07 | 1.44 | 3.06 | 0.69 |
| 622 | 3.85 | 1.46 | 2.89 | 0.75 | 3.96 | 1.45 | 2.95 | 0.74 | 4.06 | 1.45 | 3.05 | 0.69 |
| 623 | 3.84 | 1.47 | 2.88 | 0.75 | 3.95 | 1.46 | 2.94 | 0.74 | 4.06 | 1.45 | 3.04 | 0.70 |
| 624 | 3.83 | 1.48 | 2.87 | 0.76 | 3.94 | 1.47 | 2.93 | 0.75 | 4.05 | 1.46 | 3.03 | 0.70 |
| 625 | 3.82 | 1.49 | 2.86 | 0.77 | 3.93 | 1.48 | 2.92 | 0.75 | 4.04 | 1.47 | 3.02 | 0.71 |
| 626 | 3.81 | 1.50 | 2.85 | 0.77 | 3.92 | 1.50 | 2.91 | 0.76 | 4.03 | 1.49 | 3.01 | 0.71 |
| 627 | 3.80 | 1.52 | 2.84 | 0.78 | 3.91 | 1.51 | 2.90 | 0.76 | 4.02 | 1.50 | 3.01 | 0.72 |

| | | | | | | | | | | | |
|---|---|---|---|---|---|---|---|---|---|---|---|
| 628 | 3.78 | 1.53 | 2.83 | 0.78 | 3.90 | 1.52 | 2.89 | 0.77 | 4.01 | 1.51 | 3.00 | 0.72 |
| 629 | 3.77 | 1.54 | 2.82 | 0.79 | 3.89 | 1.53 | 2.88 | 0.78 | 4.01 | 1.52 | 2.99 | 0.73 |
| 630 | 3.76 | 1.55 | 2.81 | 0.80 | 3.88 | 1.54 | 2.87 | 0.78 | 4.00 | 1.53 | 2.98 | 0.73 |
| 631 | 3.75 | 1.57 | 2.80 | 0.80 | 3.87 | 1.56 | 2.86 | 0.79 | 3.99 | 1.54 | 2.97 | 0.74 |
| 632 | 3.74 | 1.58 | 2.79 | 0.81 | 3.86 | 1.57 | 2.85 | 0.79 | 3.98 | 1.55 | 2.96 | 0.75 |
| 633 | 3.73 | 1.59 | 2.78 | 0.82 | 3.85 | 1.58 | 2.84 | 0.80 | 3.98 | 1.57 | 2.95 | 0.75 |
| 634 | 3.72 | 1.61 | 2.77 | 0.82 | 3.84 | 1.60 | 2.83 | 0.81 | 3.97 | 1.58 | 2.94 | 0.76 |
| 635 | 3.71 | 1.62 | 2.76 | 0.83 | 3.84 | 1.61 | 2.82 | 0.81 | 3.96 | 1.59 | 2.93 | 0.76 |
| 636 | 3.70 | 1.64 | 2.74 | 0.84 | 3.83 | 1.62 | 2.81 | 0.82 | 3.96 | 1.61 | 2.92 | 0.77 |
| 637 | 3.69 | 1.66 | 2.73 | 0.84 | 3.82 | 1.64 | 2.80 | 0.83 | 3.95 | 1.62 | 2.91 | 0.78 |
| 638 | 3.68 | 1.67 | 2.72 | 0.85 | 3.81 | 1.65 | 2.79 | 0.83 | 3.94 | 1.63 | 2.90 | 0.78 |
| 639 | 3.68 | 1.69 | 2.71 | 0.86 | 3.81 | 1.67 | 2.78 | 0.84 | 3.94 | 1.65 | 2.89 | 0.79 |
| 640 | 3.67 | 1.70 | 2.70 | 0.86 | 3.80 | 1.68 | 2.77 | 0.85 | 3.93 | 1.66 | 2.88 | 0.79 |
| 641 | 3.66 | 1.72 | 2.69 | 0.87 | 3.79 | 1.70 | 2.76 | 0.85 | 3.93 | 1.68 | 2.88 | 0.80 |
| 642 | 3.65 | 1.74 | 2.68 | 0.88 | 3.79 | 1.72 | 2.75 | 0.86 | 3.92 | 1.69 | 2.87 | 0.81 |
| 643 | 3.64 | 1.76 | 2.67 | 0.89 | 3.78 | 1.73 | 2.74 | 0.87 | 3.92 | 1.71 | 2.86 | 0.81 |
| 644 | 3.64 | 1.77 | 2.66 | 0.90 | 3.77 | 1.75 | 2.74 | 0.88 | 3.91 | 1.72 | 2.85 | 0.82 |
| 645 | 3.63 | 1.79 | 2.65 | 0.90 | 3.77 | 1.77 | 2.73 | 0.88 | 3.91 | 1.74 | 2.84 | 0.83 |
| 646 | 3.62 | 1.81 | 2.64 | 0.91 | 3.76 | 1.78 | 2.72 | 0.89 | 3.90 | 1.76 | 2.83 | 0.84 |
| 647 | 3.61 | 1.83 | 2.63 | 0.92 | 3.76 | 1.80 | 2.71 | 0.90 | 3.90 | 1.77 | 2.82 | 0.84 |
| 648 | 3.61 | 1.85 | 2.62 | 0.93 | 3.75 | 1.82 | 2.70 | 0.91 | 3.90 | 1.79 | 2.81 | 0.85 |
| 649 | 3.60 | 1.87 | 2.61 | 0.94 | 3.75 | 1.84 | 2.69 | 0.92 | 3.89 | 1.81 | 2.80 | 0.86 |
| 650 | 3.60 | 1.89 | 2.60 | 0.95 | 3.74 | 1.86 | 2.68 | 0.92 | 3.89 | 1.82 | 2.79 | 0.87 |
| 651 | 3.59 | 1.91 | 2.59 | 0.95 | 3.74 | 1.88 | 2.67 | 0.93 | 3.89 | 1.84 | 2.79 | 0.87 |
| 652 | 3.59 | 1.93 | 2.58 | 0.96 | 3.74 | 1.90 | 2.66 | 0.94 | 3.89 | 1.86 | 2.78 | 0.88 |
| 653 | 3.58 | 1.95 | 2.57 | 0.97 | 3.74 | 1.91 | 2.65 | 0.95 | 3.89 | 1.88 | 2.77 | 0.89 |
| 654 | 3.58 | 1.97 | 2.56 | 0.98 | 3.73 | 1.93 | 2.64 | 0.96 | 3.89 | 1.89 | 2.76 | 0.90 |
| 655 | 3.58 | 2.00 | 2.55 | 0.99 | 3.73 | 1.95 | 2.63 | 0.97 | 3.89 | 1.91 | 2.75 | 0.91 |
| 656 | 3.57 | 2.02 | 2.54 | 1.00 | 3.73 | 1.97 | 2.62 | 0.98 | 3.89 | 1.93 | 2.74 | 0.91 |
| 657 | 3.57 | 2.04 | 2.53 | 1.01 | 3.73 | 1.99 | 2.61 | 0.99 | 3.89 | 1.95 | 2.73 | 0.92 |
| 658 | 3.57 | 2.06 | 2.52 | 1.02 | 3.73 | 2.02 | 2.60 | 1.00 | 3.89 | 1.97 | 2.72 | 0.93 |
| 659 | 3.57 | 2.09 | 2.51 | 1.03 | 3.73 | 2.04 | 2.59 | 1.01 | 3.89 | 1.99 | 2.71 | 0.94 |
| 660 | 3.57 | 2.11 | 2.50 | 1.04 | 3.73 | 2.06 | 2.58 | 1.01 | 3.89 | 2.00 | 2.71 | 0.95 |
| 661 | 3.57 | 2.13 | 2.49 | 1.05 | 3.73 | 2.08 | 2.57 | 1.02 | 3.89 | 2.02 | 2.70 | 0.96 |
| 662 | 3.57 | 2.16 | 2.48 | 1.06 | 3.73 | 2.10 | 2.56 | 1.03 | 3.89 | 2.04 | 2.69 | 0.97 |
| 663 | 3.57 | 2.18 | 2.47 | 1.07 | 3.73 | 2.12 | 2.56 | 1.04 | 3.90 | 2.06 | 2.68 | 0.98 |
| 664 | 3.57 | 2.20 | 2.46 | 1.08 | 3.74 | 2.14 | 2.55 | 1.05 | 3.90 | 2.08 | 2.67 | 0.98 |
| 665 | 3.57 | 2.23 | 2.45 | 1.09 | 3.74 | 2.16 | 2.54 | 1.06 | 3.90 | 2.10 | 2.66 | 0.99 |

| | | | | | | | | | | | |
|---|---|---|---|---|---|---|---|---|---|---|---|
| 666 | 3.58 | 2.25 | 2.44 | 1.10 | 3.74 | 2.19 | 2.53 | 1.08 | 3.91 | 2.12 | 2.65 | 1.00 |
| 667 | 3.58 | 2.28 | 2.43 | 1.11 | 3.75 | 2.21 | 2.52 | 1.09 | 3.91 | 2.14 | 2.65 | 1.01 |
| 668 | 3.58 | 2.30 | 2.42 | 1.12 | 3.75 | 2.23 | 2.51 | 1.10 | 3.92 | 2.16 | 2.64 | 1.02 |
| 669 | 3.59 | 2.33 | 2.41 | 1.14 | 3.76 | 2.25 | 2.50 | 1.11 | 3.92 | 2.18 | 2.63 | 1.03 |
| 670 | 3.59 | 2.35 | 2.40 | 1.15 | 3.76 | 2.27 | 2.49 | 1.12 | 3.93 | 2.20 | 2.62 | 1.04 |
| 671 | 3.60 | 2.38 | 2.39 | 1.16 | 3.77 | 2.30 | 2.48 | 1.13 | 3.94 | 2.22 | 2.61 | 1.05 |
| 672 | 3.61 | 2.40 | 2.38 | 1.17 | 3.78 | 2.32 | 2.47 | 1.14 | 3.95 | 2.24 | 2.60 | 1.06 |
| 673 | 3.62 | 2.43 | 2.37 | 1.18 | 3.79 | 2.34 | 2.47 | 1.15 | 3.95 | 2.26 | 2.60 | 1.08 |
| 674 | 3.62 | 2.45 | 2.36 | 1.20 | 3.80 | 2.36 | 2.46 | 1.16 | 3.96 | 2.28 | 2.59 | 1.09 |
| 675 | 3.63 | 2.48 | 2.36 | 1.21 | 3.80 | 2.38 | 2.45 | 1.18 | 3.97 | 2.30 | 2.58 | 1.10 |
| 676 | 3.64 | 2.50 | 2.35 | 1.22 | 3.81 | 2.41 | 2.44 | 1.19 | 3.98 | 2.32 | 2.57 | 1.11 |
| 677 | 3.65 | 2.53 | 2.34 | 1.23 | 3.82 | 2.43 | 2.43 | 1.20 | 3.99 | 2.34 | 2.56 | 1.12 |
| 678 | 3.66 | 2.55 | 2.33 | 1.25 | 3.84 | 2.45 | 2.42 | 1.21 | 4.00 | 2.36 | 2.56 | 1.13 |
| 679 | 3.68 | 2.58 | 2.32 | 1.26 | 3.85 | 2.47 | 2.42 | 1.22 | 4.01 | 2.37 | 2.55 | 1.14 |
| 680 | 3.69 | 2.60 | 2.31 | 1.27 | 3.86 | 2.49 | 2.41 | 1.24 | 4.03 | 2.39 | 2.54 | 1.15 |
| 681 | 3.70 | 2.63 | 2.30 | 1.29 | 3.87 | 2.51 | 2.40 | 1.25 | 4.04 | 2.41 | 2.53 | 1.17 |
| 682 | 3.72 | 2.65 | 2.29 | 1.30 | 3.89 | 2.54 | 2.39 | 1.26 | 4.05 | 2.43 | 2.53 | 1.18 |
| 683 | 3.73 | 2.67 | 2.28 | 1.31 | 3.90 | 2.56 | 2.38 | 1.28 | 4.07 | 2.45 | 2.52 | 1.19 |
| 684 | 3.75 | 2.70 | 2.27 | 1.33 | 3.92 | 2.58 | 2.37 | 1.29 | 4.08 | 2.47 | 2.51 | 1.20 |
| 685 | 3.77 | 2.72 | 2.27 | 1.34 | 3.93 | 2.60 | 2.37 | 1.30 | 4.10 | 2.48 | 2.50 | 1.21 |
| 686 | 3.78 | 2.74 | 2.26 | 1.36 | 3.95 | 2.62 | 2.36 | 1.32 | 4.11 | 2.50 | 2.50 | 1.23 |
| 687 | 3.80 | 2.77 | 2.25 | 1.37 | 3.97 | 2.64 | 2.35 | 1.33 | 4.13 | 2.52 | 2.49 | 1.24 |
| 688 | 3.82 | 2.79 | 2.24 | 1.38 | 3.98 | 2.65 | 2.34 | 1.34 | 4.14 | 2.53 | 2.48 | 1.25 |
| 689 | 3.84 | 2.81 | 2.23 | 1.40 | 4.00 | 2.67 | 2.34 | 1.36 | 4.16 | 2.55 | 2.47 | 1.27 |
| 690 | 3.86 | 2.83 | 2.22 | 1.41 | 4.02 | 2.69 | 2.33 | 1.37 | 4.18 | 2.57 | 2.47 | 1.28 |
| 691 | 3.88 | 2.85 | 2.22 | 1.43 | 4.04 | 2.71 | 2.32 | 1.39 | 4.20 | 2.58 | 2.46 | 1.29 |
| 692 | 3.90 | 2.87 | 2.21 | 1.45 | 4.06 | 2.73 | 2.31 | 1.40 | 4.21 | 2.60 | 2.45 | 1.31 |
| 693 | 3.93 | 2.89 | 2.20 | 1.46 | 4.08 | 2.74 | 2.31 | 1.42 | 4.23 | 2.61 | 2.45 | 1.32 |
| 694 | 3.95 | 2.91 | 2.19 | 1.48 | 4.10 | 2.76 | 2.30 | 1.43 | 4.25 | 2.63 | 2.44 | 1.33 |
| 695 | 3.97 | 2.93 | 2.19 | 1.49 | 4.12 | 2.77 | 2.29 | 1.45 | 4.27 | 2.64 | 2.43 | 1.35 |
| 696 | 4.00 | 2.94 | 2.18 | 1.51 | 4.15 | 2.79 | 2.29 | 1.46 | 4.29 | 2.65 | 2.43 | 1.36 |
| 697 | 4.02 | 2.96 | 2.17 | 1.53 | 4.17 | 2.80 | 2.28 | 1.48 | 4.31 | 2.66 | 2.42 | 1.38 |
| 698 | 4.05 | 2.98 | 2.16 | 1.54 | 4.19 | 2.82 | 2.27 | 1.49 | 4.33 | 2.68 | 2.41 | 1.39 |
| 699 | 4.08 | 2.99 | 2.16 | 1.56 | 4.22 | 2.83 | 2.27 | 1.51 | 4.36 | 2.69 | 2.41 | 1.41 |
| 700 | 4.10 | 3.01 | 2.15 | 1.58 | 4.24 | 2.84 | 2.26 | 1.53 | 4.38 | 2.70 | 2.40 | 1.42 |
| 701 | 4.13 | 3.02 | 2.14 | 1.59 | 4.26 | 2.85 | 2.25 | 1.54 | 4.40 | 2.71 | 2.40 | 1.44 |
| 702 | 4.16 | 3.03 | 2.14 | 1.61 | 4.29 | 2.87 | 2.25 | 1.56 | 4.42 | 2.72 | 2.39 | 1.45 |
| 703 | 4.19 | 3.05 | 2.13 | 1.63 | 4.31 | 2.88 | 2.24 | 1.58 | 4.44 | 2.73 | 2.39 | 1.47 |

| | | | | | | | | | | | |
|---|---|---|---|---|---|---|---|---|---|---|---|
| 704 | 4.21 | 3.06 | 2.12 | 1.65 | 4.34 | 2.89 | 2.24 | 1.59 | 4.47 | 2.74 | 2.38 | 1.48 |
| 705 | 4.24 | 3.07 | 2.12 | 1.66 | 4.36 | 2.89 | 2.23 | 1.61 | 4.49 | 2.74 | 2.38 | 1.50 |
| 706 | 4.27 | 3.08 | 2.11 | 1.68 | 4.39 | 2.90 | 2.23 | 1.63 | 4.51 | 2.75 | 2.37 | 1.52 |
| 707 | 4.30 | 3.08 | 2.11 | 1.70 | 4.42 | 2.91 | 2.22 | 1.64 | 4.54 | 2.76 | 2.37 | 1.53 |
| 708 | 4.33 | 3.09 | 2.10 | 1.72 | 4.44 | 2.92 | 2.22 | 1.66 | 4.56 | 2.76 | 2.36 | 1.55 |
| 709 | 4.36 | 3.10 | 2.09 | 1.74 | 4.47 | 2.92 | 2.21 | 1.68 | 4.59 | 2.77 | 2.36 | 1.56 |
| 710 | 4.39 | 3.10 | 2.09 | 1.76 | 4.50 | 2.93 | 2.21 | 1.70 | 4.61 | 2.77 | 2.35 | 1.58 |
| 711 | 4.42 | 3.11 | 2.08 | 1.77 | 4.52 | 2.93 | 2.20 | 1.71 | 4.63 | 2.78 | 2.35 | 1.60 |
| 712 | 4.45 | 3.11 | 2.08 | 1.79 | 4.55 | 2.93 | 2.20 | 1.73 | 4.66 | 2.78 | 2.34 | 1.62 |
| 713 | 4.48 | 3.12 | 2.08 | 1.81 | 4.58 | 2.94 | 2.19 | 1.75 | 4.68 | 2.78 | 2.34 | 1.63 |
| 714 | 4.51 | 3.12 | 2.07 | 1.83 | 4.60 | 2.94 | 2.19 | 1.77 | 4.71 | 2.78 | 2.34 | 1.65 |
| 715 | 4.54 | 3.12 | 2.07 | 1.85 | 4.63 | 2.94 | 2.19 | 1.79 | 4.73 | 2.78 | 2.33 | 1.67 |
| 716 | 4.57 | 3.12 | 2.06 | 1.87 | 4.66 | 2.94 | 2.18 | 1.81 | 4.75 | 2.78 | 2.33 | 1.68 |
| 717 | 4.60 | 3.12 | 2.06 | 1.89 | 4.68 | 2.94 | 2.18 | 1.83 | 4.78 | 2.78 | 2.33 | 1.70 |
| 718 | 4.63 | 3.11 | 2.06 | 1.91 | 4.71 | 2.94 | 2.18 | 1.85 | 4.80 | 2.78 | 2.32 | 1.72 |
| 719 | 4.66 | 3.11 | 2.05 | 1.93 | 4.74 | 2.93 | 2.17 | 1.86 | 4.83 | 2.78 | 2.32 | 1.74 |
| 720 | 4.69 | 3.11 | 2.05 | 1.95 | 4.76 | 2.93 | 2.17 | 1.88 | 4.85 | 2.78 | 2.32 | 1.76 |
| 721 | 4.72 | 3.10 | 2.05 | 1.97 | 4.79 | 2.93 | 2.17 | 1.90 | 4.87 | 2.77 | 2.32 | 1.78 |
| 722 | 4.75 | 3.10 | 2.04 | 2.00 | 4.81 | 2.92 | 2.17 | 1.92 | 4.90 | 2.77 | 2.32 | 1.79 |
| 723 | 4.78 | 3.09 | 2.04 | 2.02 | 4.84 | 2.92 | 2.17 | 1.94 | 4.92 | 2.77 | 2.31 | 1.81 |
| 724 | 4.81 | 3.08 | 2.04 | 2.04 | 4.87 | 2.91 | 2.16 | 1.96 | 4.94 | 2.76 | 2.31 | 1.83 |
| 725 | 4.83 | 3.07 | 2.04 | 2.06 | 4.89 | 2.90 | 2.16 | 1.98 | 4.97 | 2.76 | 2.31 | 1.85 |
| 726 | 4.86 | 3.06 | 2.04 | 2.08 | 4.92 | 2.90 | 2.16 | 2.00 | 4.99 | 2.75 | 2.31 | 1.87 |
| 727 | 4.89 | 3.05 | 2.04 | 2.10 | 4.94 | 2.89 | 2.16 | 2.03 | 5.01 | 2.74 | 2.31 | 1.89 |
| 728 | 4.92 | 3.04 | 2.04 | 2.12 | 4.96 | 2.88 | 2.16 | 2.05 | 5.03 | 2.73 | 2.31 | 1.91 |
| 729 | 4.94 | 3.03 | 2.03 | 2.15 | 4.99 | 2.87 | 2.16 | 2.07 | 5.05 | 2.73 | 2.31 | 1.93 |
| 730 | 4.97 | 3.02 | 2.03 | 2.17 | 5.01 | 2.86 | 2.16 | 2.09 | 5.07 | 2.72 | 2.31 | 1.95 |
| 731 | 4.99 | 3.01 | 2.03 | 2.19 | 5.03 | 2.85 | 2.16 | 2.11 | 5.10 | 2.71 | 2.31 | 1.97 |
| 732 | 5.02 | 2.99 | 2.04 | 2.21 | 5.05 | 2.84 | 2.16 | 2.13 | 5.12 | 2.70 | 2.31 | 1.99 |
| 733 | 5.04 | 2.98 | 2.04 | 2.23 | 5.08 | 2.83 | 2.16 | 2.15 | 5.14 | 2.69 | 2.31 | 2.01 |
| 734 | 5.06 | 2.96 | 2.04 | 2.26 | 5.10 | 2.81 | 2.17 | 2.17 | 5.16 | 2.68 | 2.32 | 2.03 |
| 735 | 5.09 | 2.95 | 2.04 | 2.28 | 5.12 | 2.80 | 2.17 | 2.19 | 5.17 | 2.67 | 2.32 | 2.05 |
| 736 | 5.11 | 2.93 | 2.04 | 2.30 | 5.14 | 2.79 | 2.17 | 2.21 | 5.19 | 2.66 | 2.32 | 2.07 |
| 737 | 5.13 | 2.92 | 2.04 | 2.32 | 5.16 | 2.77 | 2.17 | 2.24 | 5.21 | 2.64 | 2.32 | 2.09 |
| 738 | 5.15 | 2.90 | 2.05 | 2.35 | 5.18 | 2.76 | 2.18 | 2.26 | 5.23 | 2.63 | 2.33 | 2.11 |
| 739 | 5.17 | 2.88 | 2.05 | 2.37 | 5.20 | 2.74 | 2.18 | 2.28 | 5.25 | 2.62 | 2.33 | 2.13 |
| 740 | 5.19 | 2.86 | 2.05 | 2.39 | 5.21 | 2.73 | 2.18 | 2.30 | 5.26 | 2.61 | 2.33 | 2.15 |
| 741 | 5.21 | 2.85 | 2.06 | 2.42 | 5.23 | 2.71 | 2.19 | 2.32 | 5.28 | 2.59 | 2.34 | 2.17 |

| | | | | | | | | | | | |
|---|---|---|---|---|---|---|---|---|---|---|---|
| 742 | 5.22 | 2.83 | 2.06 | 2.44 | 5.25 | 2.70 | 2.19 | 2.34 | 5.30 | 2.58 | 2.34 | 2.19 |
| 743 | 5.24 | 2.81 | 2.06 | 2.46 | 5.26 | 2.68 | 2.19 | 2.36 | 5.31 | 2.57 | 2.34 | 2.21 |
| 744 | 5.26 | 2.79 | 2.07 | 2.49 | 5.28 | 2.67 | 2.20 | 2.39 | 5.33 | 2.55 | 2.35 | 2.23 |
| 745 | 5.27 | 2.77 | 2.07 | 2.51 | 5.29 | 2.65 | 2.20 | 2.41 | 5.34 | 2.54 | 2.36 | 2.25 |
| 746 | 5.29 | 2.75 | 2.08 | 2.53 | 5.31 | 2.63 | 2.21 | 2.43 | 5.36 | 2.52 | 2.36 | 2.27 |
| 747 | 5.30 | 2.73 | 2.08 | 2.55 | 5.32 | 2.61 | 2.22 | 2.45 | 5.37 | 2.51 | 2.37 | 2.29 |
| 748 | 5.32 | 2.71 | 2.09 | 2.58 | 5.34 | 2.60 | 2.22 | 2.47 | 5.38 | 2.49 | 2.37 | 2.31 |
| 749 | 5.33 | 2.69 | 2.10 | 2.60 | 5.35 | 2.58 | 2.23 | 2.49 | 5.40 | 2.48 | 2.38 | 2.34 |
| 750 | 5.34 | 2.67 | 2.10 | 2.62 | 5.36 | 2.56 | 2.24 | 2.52 | 5.41 | 2.46 | 2.39 | 2.36 |
| 751 | 5.35 | 2.65 | 2.11 | 2.65 | 5.37 | 2.54 | 2.24 | 2.54 | 5.42 | 2.45 | 2.40 | 2.38 |
| 752 | 5.36 | 2.63 | 2.12 | 2.67 | 5.38 | 2.52 | 2.25 | 2.56 | 5.43 | 2.43 | 2.40 | 2.40 |
| 753 | 5.37 | 2.61 | 2.13 | 2.69 | 5.39 | 2.51 | 2.26 | 2.58 | 5.44 | 2.41 | 2.41 | 2.42 |
| 754 | 5.38 | 2.58 | 2.14 | 2.71 | 5.40 | 2.49 | 2.27 | 2.60 | 5.45 | 2.40 | 2.42 | 2.44 |
| 755 | 5.39 | 2.56 | 2.15 | 2.74 | 5.41 | 2.47 | 2.28 | 2.62 | 5.46 | 2.38 | 2.43 | 2.46 |
| 756 | 5.40 | 2.54 | 2.15 | 2.76 | 5.42 | 2.45 | 2.29 | 2.64 | 5.47 | 2.36 | 2.44 | 2.48 |
| 757 | 5.41 | 2.52 | 2.16 | 2.78 | 5.43 | 2.43 | 2.30 | 2.66 | 5.48 | 2.35 | 2.45 | 2.50 |
| 758 | 5.42 | 2.50 | 2.17 | 2.80 | 5.44 | 2.41 | 2.31 | 2.68 | 5.49 | 2.33 | 2.46 | 2.52 |
| 759 | 5.42 | 2.48 | 2.18 | 2.82 | 5.44 | 2.39 | 2.32 | 2.70 | 5.49 | 2.31 | 2.47 | 2.53 |
| 760 | 5.43 | 2.46 | 2.20 | 2.85 | 5.45 | 2.38 | 2.33 | 2.72 | 5.50 | 2.30 | 2.48 | 2.55 |
| 761 | 5.43 | 2.43 | 2.21 | 2.87 | 5.46 | 2.36 | 2.34 | 2.75 | 5.51 | 2.28 | 2.49 | 2.57 |
| 762 | 5.44 | 2.41 | 2.22 | 2.89 | 5.46 | 2.34 | 2.35 | 2.77 | 5.51 | 2.26 | 2.50 | 2.59 |
| 763 | 5.44 | 2.39 | 2.23 | 2.91 | 5.47 | 2.32 | 2.36 | 2.78 | 5.52 | 2.25 | 2.51 | 2.61 |
| 764 | 5.44 | 2.37 | 2.24 | 2.93 | 5.47 | 2.30 | 2.38 | 2.80 | 5.53 | 2.23 | 2.53 | 2.63 |
| 765 | 5.45 | 2.35 | 2.26 | 2.95 | 5.48 | 2.28 | 2.39 | 2.82 | 5.53 | 2.21 | 2.54 | 2.65 |
| 766 | 5.45 | 2.33 | 2.27 | 2.97 | 5.48 | 2.26 | 2.40 | 2.84 | 5.54 | 2.20 | 2.55 | 2.66 |
| 767 | 5.45 | 2.31 | 2.28 | 2.99 | 5.48 | 2.24 | 2.41 | 2.86 | 5.54 | 2.18 | 2.57 | 2.68 |
| 768 | 5.45 | 2.29 | 2.29 | 3.01 | 5.49 | 2.23 | 2.43 | 2.88 | 5.54 | 2.16 | 2.58 | 2.70 |
| 769 | 5.45 | 2.27 | 2.31 | 3.03 | 5.49 | 2.21 | 2.44 | 2.90 | 5.55 | 2.15 | 2.59 | 2.72 |
| 770 | 5.45 | 2.25 | 2.32 | 3.05 | 5.49 | 2.19 | 2.45 | 2.92 | 5.55 | 2.13 | 2.61 | 2.73 |
| 771 | 5.45 | 2.23 | 2.34 | 3.07 | 5.49 | 2.17 | 2.47 | 2.93 | 5.55 | 2.11 | 2.62 | 2.75 |
| 772 | 5.45 | 2.21 | 2.35 | 3.09 | 5.49 | 2.15 | 2.48 | 2.95 | 5.55 | 2.10 | 2.64 | 2.77 |
| 773 | 5.45 | 2.19 | 2.37 | 3.11 | 5.49 | 2.13 | 2.50 | 2.97 | 5.56 | 2.08 | 2.65 | 2.78 |
| 774 | 5.45 | 2.17 | 2.38 | 3.13 | 5.49 | 2.12 | 2.51 | 2.99 | 5.56 | 2.06 | 2.67 | 2.80 |
| 775 | 5.45 | 2.15 | 2.40 | 3.14 | 5.49 | 2.10 | 2.53 | 3.00 | 5.56 | 2.05 | 2.68 | 2.81 |
| 776 | 5.44 | 2.13 | 2.41 | 3.16 | 5.49 | 2.08 | 2.54 | 3.02 | 5.56 | 2.03 | 2.70 | 2.83 |
| 777 | 5.44 | 2.11 | 2.43 | 3.18 | 5.49 | 2.06 | 2.56 | 3.04 | 5.56 | 2.02 | 2.71 | 2.84 |
| 778 | 5.44 | 2.09 | 2.45 | 3.20 | 5.49 | 2.05 | 2.58 | 3.05 | 5.56 | 2.00 | 2.73 | 2.86 |
| 779 | 5.43 | 2.07 | 2.46 | 3.21 | 5.48 | 2.03 | 2.59 | 3.07 | 5.56 | 1.98 | 2.75 | 2.87 |

| | | | | | | | | | | | |
|---|---|---|---|---|---|---|---|---|---|---|---|
| 780 | 5.43 | 2.05 | 2.48 | 3.23 | 5.48 | 2.01 | 2.61 | 3.08 | 5.56 | 1.97 | 2.76 | 2.89 |
| 781 | 5.42 | 2.03 | 2.50 | 3.25 | 5.48 | 1.99 | 2.63 | 3.10 | 5.56 | 1.95 | 2.78 | 2.90 |
| 782 | 5.42 | 2.01 | 2.51 | 3.26 | 5.48 | 1.98 | 2.64 | 3.11 | 5.56 | 1.94 | 2.80 | 2.91 |
| 783 | 5.41 | 2.00 | 2.53 | 3.28 | 5.47 | 1.96 | 2.66 | 3.13 | 5.55 | 1.92 | 2.81 | 2.93 |
| 784 | 5.41 | 1.98 | 2.55 | 3.29 | 5.47 | 1.94 | 2.68 | 3.14 | 5.55 | 1.91 | 2.83 | 2.94 |
| 785 | 5.40 | 1.96 | 2.57 | 3.31 | 5.46 | 1.93 | 2.69 | 3.15 | 5.55 | 1.89 | 2.85 | 2.95 |
| 786 | 5.39 | 1.94 | 2.58 | 3.32 | 5.46 | 1.91 | 2.71 | 3.17 | 5.55 | 1.88 | 2.87 | 2.96 |
| 787 | 5.39 | 1.93 | 2.60 | 3.34 | 5.45 | 1.90 | 2.73 | 3.18 | 5.54 | 1.86 | 2.88 | 2.97 |
| 788 | 5.38 | 1.91 | 2.62 | 3.35 | 5.45 | 1.88 | 2.75 | 3.19 | 5.54 | 1.85 | 2.90 | 2.98 |
| 789 | 5.37 | 1.89 | 2.64 | 3.36 | 5.44 | 1.87 | 2.76 | 3.20 | 5.54 | 1.83 | 2.92 | 2.99 |
| 790 | 5.36 | 1.88 | 2.66 | 3.37 | 5.44 | 1.85 | 2.78 | 3.21 | 5.53 | 1.82 | 2.94 | 3.00 |
| 791 | 5.36 | 1.86 | 2.67 | 3.39 | 5.43 | 1.83 | 2.80 | 3.22 | 5.53 | 1.81 | 2.96 | 3.01 |
| 792 | 5.35 | 1.84 | 2.69 | 3.40 | 5.42 | 1.82 | 2.82 | 3.24 | 5.52 | 1.79 | 2.97 | 3.02 |
| 793 | 5.34 | 1.83 | 2.71 | 3.41 | 5.42 | 1.81 | 2.84 | 3.25 | 5.52 | 1.78 | 2.99 | 3.03 |
| 794 | 5.33 | 1.81 | 2.73 | 3.42 | 5.41 | 1.79 | 2.85 | 3.26 | 5.51 | 1.77 | 3.01 | 3.04 |
| 795 | 5.32 | 1.80 | 2.75 | 3.43 | 5.40 | 1.78 | 2.87 | 3.27 | 5.51 | 1.75 | 3.03 | 3.05 |
| 796 | 5.31 | 1.78 | 2.77 | 3.44 | 5.40 | 1.76 | 2.89 | 3.28 | 5.50 | 1.74 | 3.05 | 3.06 |
| 797 | 5.30 | 1.77 | 2.78 | 3.45 | 5.39 | 1.75 | 2.91 | 3.28 | 5.50 | 1.73 | 3.06 | 3.06 |
| 798 | 5.29 | 1.75 | 2.80 | 3.47 | 5.38 | 1.73 | 2.93 | 3.29 | 5.49 | 1.71 | 3.08 | 3.07 |
| 799 | 5.28 | 1.74 | 2.82 | 3.47 | 5.37 | 1.72 | 2.94 | 3.30 | 5.49 | 1.70 | 3.10 | 3.08 |
| 800 | 5.27 | 1.72 | 2.84 | 3.48 | 5.36 | 1.71 | 2.96 | 3.31 | 5.48 | 1.69 | 3.12 | 3.09 |
| 801 | 5.26 | 1.71 | 2.86 | 3.49 | 5.35 | 1.69 | 2.98 | 3.32 | 5.47 | 1.67 | 3.13 | 3.09 |
| 802 | 5.24 | 1.70 | 2.88 | 3.50 | 5.34 | 1.68 | 3.00 | 3.33 | 5.47 | 1.66 | 3.15 | 3.10 |
| 803 | 5.23 | 1.68 | 2.89 | 3.51 | 5.34 | 1.67 | 3.02 | 3.33 | 5.46 | 1.65 | 3.17 | 3.10 |
| 804 | 5.22 | 1.67 | 2.91 | 3.52 | 5.33 | 1.66 | 3.03 | 3.34 | 5.45 | 1.64 | 3.19 | 3.11 |
| 805 | 5.21 | 1.66 | 2.93 | 3.53 | 5.32 | 1.64 | 3.05 | 3.35 | 5.44 | 1.63 | 3.20 | 3.11 |
| 806 | 5.20 | 1.64 | 2.95 | 3.53 | 5.31 | 1.63 | 3.07 | 3.35 | 5.44 | 1.61 | 3.22 | 3.12 |
| 807 | 5.18 | 1.63 | 2.97 | 3.54 | 5.30 | 1.62 | 3.09 | 3.36 | 5.43 | 1.60 | 3.24 | 3.12 |
| 808 | 5.17 | 1.62 | 2.98 | 3.55 | 5.29 | 1.61 | 3.10 | 3.36 | 5.42 | 1.59 | 3.26 | 3.12 |
| 809 | 5.16 | 1.61 | 3.00 | 3.55 | 5.27 | 1.60 | 3.12 | 3.37 | 5.41 | 1.58 | 3.27 | 3.13 |
| 810 | 5.15 | 1.59 | 3.02 | 3.56 | 5.26 | 1.58 | 3.14 | 3.37 | 5.40 | 1.57 | 3.29 | 3.13 |
| 811 | 5.13 | 1.58 | 3.04 | 3.57 | 5.25 | 1.57 | 3.15 | 3.38 | 5.39 | 1.56 | 3.31 | 3.13 |
| 812 | 5.12 | 1.57 | 3.06 | 3.57 | 5.24 | 1.56 | 3.17 | 3.38 | 5.39 | 1.55 | 3.32 | 3.14 |
| 813 | 5.10 | 1.56 | 3.07 | 3.58 | 5.23 | 1.55 | 3.19 | 3.39 | 5.38 | 1.54 | 3.34 | 3.14 |
| 814 | 5.09 | 1.55 | 3.09 | 3.58 | 5.22 | 1.54 | 3.20 | 3.39 | 5.37 | 1.53 | 3.36 | 3.14 |
| 815 | 5.08 | 1.54 | 3.11 | 3.59 | 5.21 | 1.53 | 3.22 | 3.40 | 5.36 | 1.52 | 3.37 | 3.14 |
| 816 | 5.06 | 1.53 | 3.12 | 3.59 | 5.19 | 1.52 | 3.24 | 3.40 | 5.35 | 1.51 | 3.39 | 3.15 |
| 817 | 5.05 | 1.52 | 3.14 | 3.60 | 5.18 | 1.51 | 3.25 | 3.40 | 5.34 | 1.50 | 3.40 | 3.15 |

| | | | | | | | | | | | |
|---|---|---|---|---|---|---|---|---|---|---|---|
| 818 | 5.03 | 1.51 | 3.16 | 3.60 | 5.17 | 1.50 | 3.27 | 3.41 | 5.33 | 1.49 | 3.42 | 3.15 |
| 819 | 5.02 | 1.50 | 3.17 | 3.61 | 5.16 | 1.49 | 3.29 | 3.41 | 5.32 | 1.48 | 3.43 | 3.15 |
| 820 | 5.00 | 1.49 | 3.19 | 3.61 | 5.14 | 1.48 | 3.30 | 3.41 | 5.31 | 1.47 | 3.45 | 3.15 |
| 821 | 4.99 | 1.48 | 3.21 | 3.62 | 5.13 | 1.47 | 3.32 | 3.42 | 5.30 | 1.46 | 3.46 | 3.15 |
| 822 | 4.97 | 1.47 | 3.22 | 3.62 | 5.12 | 1.46 | 3.33 | 3.42 | 5.29 | 1.45 | 3.48 | 3.15 |
| 823 | 4.95 | 1.46 | 3.24 | 3.62 | 5.10 | 1.45 | 3.35 | 3.42 | 5.28 | 1.44 | 3.49 | 3.15 |
| 824 | 4.94 | 1.45 | 3.25 | 3.63 | 5.09 | 1.44 | 3.36 | 3.42 | 5.26 | 1.43 | 3.51 | 3.15 |
| 825 | 4.92 | 1.44 | 3.27 | 3.63 | 5.08 | 1.43 | 3.38 | 3.42 | 5.25 | 1.43 | 3.52 | 3.15 |
| 826 | 4.91 | 1.43 | 3.28 | 3.63 | 5.06 | 1.43 | 3.39 | 3.43 | 5.24 | 1.42 | 3.54 | 3.16 |
| 827 | 4.89 | 1.42 | 3.30 | 3.64 | 5.05 | 1.42 | 3.41 | 3.43 | 5.23 | 1.41 | 3.55 | 3.16 |
| 828 | 4.87 | 1.41 | 3.31 | 3.64 | 5.04 | 1.41 | 3.42 | 3.43 | 5.22 | 1.40 | 3.56 | 3.15 |
| 829 | 4.85 | 1.41 | 3.33 | 3.64 | 5.02 | 1.40 | 3.44 | 3.43 | 5.21 | 1.39 | 3.58 | 3.15 |
| 830 | 4.84 | 1.40 | 3.34 | 3.65 | 5.01 | 1.39 | 3.45 | 3.43 | 5.19 | 1.39 | 3.59 | 3.15 |
| 831 | 4.82 | 1.39 | 3.36 | 3.65 | 4.99 | 1.39 | 3.46 | 3.43 | 5.18 | 1.38 | 3.60 | 3.15 |
| 832 | 4.80 | 1.38 | 3.37 | 3.65 | 4.98 | 1.38 | 3.48 | 3.44 | 5.17 | 1.37 | 3.62 | 3.15 |
| 833 | 4.78 | 1.38 | 3.39 | 3.65 | 4.96 | 1.37 | 3.49 | 3.44 | 5.16 | 1.36 | 3.63 | 3.15 |
| 834 | 4.77 | 1.37 | 3.40 | 3.66 | 4.95 | 1.36 | 3.50 | 3.44 | 5.14 | 1.36 | 3.64 | 3.15 |
| 835 | 4.75 | 1.36 | 3.41 | 3.66 | 4.93 | 1.36 | 3.52 | 3.44 | 5.13 | 1.35 | 3.65 | 3.15 |
| 836 | 4.73 | 1.35 | 3.43 | 3.66 | 4.91 | 1.35 | 3.53 | 3.44 | 5.12 | 1.34 | 3.67 | 3.15 |
| 837 | 4.71 | 1.35 | 3.44 | 3.66 | 4.90 | 1.34 | 3.54 | 3.44 | 5.11 | 1.34 | 3.68 | 3.15 |
| 838 | 4.69 | 1.34 | 3.46 | 3.66 | 4.88 | 1.34 | 3.56 | 3.44 | 5.09 | 1.33 | 3.69 | 3.15 |
| 839 | 4.67 | 1.33 | 3.47 | 3.67 | 4.87 | 1.33 | 3.57 | 3.44 | 5.08 | 1.32 | 3.70 | 3.15 |
| 840 | 4.65 | 1.33 | 3.48 | 3.67 | 4.85 | 1.32 | 3.58 | 3.44 | 5.07 | 1.32 | 3.71 | 3.15 |
| 841 | 4.63 | 1.32 | 3.50 | 3.67 | 4.83 | 1.32 | 3.59 | 3.44 | 5.05 | 1.31 | 3.72 | 3.14 |
| 842 | 4.61 | 1.32 | 3.51 | 3.67 | 4.82 | 1.31 | 3.61 | 3.44 | 5.04 | 1.31 | 3.74 | 3.14 |
| 843 | 4.59 | 1.31 | 3.52 | 3.67 | 4.80 | 1.31 | 3.62 | 3.44 | 5.02 | 1.30 | 3.75 | 3.14 |
| 844 | 4.57 | 1.31 | 3.53 | 3.68 | 4.78 | 1.30 | 3.63 | 3.44 | 5.01 | 1.30 | 3.76 | 3.14 |
| 845 | 4.55 | 1.30 | 3.55 | 3.68 | 4.76 | 1.30 | 3.64 | 3.44 | 4.99 | 1.29 | 3.77 | 3.14 |
| 846 | 4.53 | 1.30 | 3.56 | 3.68 | 4.75 | 1.29 | 3.65 | 3.45 | 4.98 | 1.29 | 3.78 | 3.14 |
| 847 | 4.51 | 1.29 | 3.57 | 3.68 | 4.73 | 1.29 | 3.66 | 3.45 | 4.97 | 1.28 | 3.79 | 3.14 |
| 848 | 4.49 | 1.29 | 3.58 | 3.68 | 4.71 | 1.28 | 3.68 | 3.45 | 4.95 | 1.28 | 3.80 | 3.14 |
| 849 | 4.47 | 1.28 | 3.60 | 3.68 | 4.69 | 1.28 | 3.69 | 3.45 | 4.94 | 1.27 | 3.81 | 3.13 |
| 850 | 4.45 | 1.28 | 3.61 | 3.69 | 4.67 | 1.27 | 3.70 | 3.45 | 4.92 | 1.27 | 3.82 | 3.13 |
| 851 | 4.43 | 1.27 | 3.62 | 3.69 | 4.66 | 1.27 | 3.71 | 3.45 | 4.90 | 1.26 | 3.83 | 3.13 |
| 852 | 4.41 | 1.27 | 3.63 | 3.69 | 4.64 | 1.26 | 3.72 | 3.45 | 4.89 | 1.26 | 3.84 | 3.13 |
| 853 | 4.38 | 1.27 | 3.64 | 3.69 | 4.62 | 1.26 | 3.73 | 3.45 | 4.87 | 1.25 | 3.85 | 3.13 |
| 854 | 4.36 | 1.26 | 3.66 | 3.69 | 4.60 | 1.26 | 3.74 | 3.45 | 4.86 | 1.25 | 3.86 | 3.13 |
| 855 | 4.34 | 1.26 | 3.67 | 3.69 | 4.58 | 1.25 | 3.75 | 3.45 | 4.84 | 1.25 | 3.87 | 3.13 |

| | | | | | | | | | | | |
|---|---|---|---|---|---|---|---|---|---|---|---|
| 856 | 4.32 | 1.26 | 3.68 | 3.70 | 4.56 | 1.25 | 3.76 | 3.45 | 4.82 | 1.24 | 3.88 | 3.13 |
| 857 | 4.29 | 1.25 | 3.69 | 3.70 | 4.54 | 1.25 | 3.77 | 3.45 | 4.81 | 1.24 | 3.88 | 3.12 |
| 858 | 4.27 | 1.25 | 3.70 | 3.70 | 4.52 | 1.24 | 3.78 | 3.45 | 4.79 | 1.24 | 3.89 | 3.12 |
| 859 | 4.25 | 1.25 | 3.71 | 3.70 | 4.50 | 1.24 | 3.79 | 3.45 | 4.77 | 1.23 | 3.90 | 3.12 |
| 860 | 4.22 | 1.25 | 3.72 | 3.70 | 4.48 | 1.24 | 3.80 | 3.45 | 4.76 | 1.23 | 3.91 | 3.12 |
| 861 | 4.20 | 1.24 | 3.73 | 3.70 | 4.46 | 1.24 | 3.81 | 3.45 | 4.74 | 1.23 | 3.92 | 3.12 |
| 862 | 4.17 | 1.24 | 3.74 | 3.71 | 4.44 | 1.23 | 3.82 | 3.45 | 4.72 | 1.23 | 3.93 | 3.12 |
| 863 | 4.15 | 1.24 | 3.76 | 3.71 | 4.42 | 1.23 | 3.83 | 3.45 | 4.70 | 1.22 | 3.94 | 3.12 |
| 864 | 4.12 | 1.24 | 3.77 | 3.71 | 4.39 | 1.23 | 3.84 | 3.45 | 4.69 | 1.22 | 3.94 | 3.12 |
| 865 | 4.10 | 1.24 | 3.78 | 3.71 | 4.37 | 1.23 | 3.85 | 3.45 | 4.67 | 1.22 | 3.95 | 3.11 |
| 866 | 4.07 | 1.24 | 3.79 | 3.71 | 4.35 | 1.23 | 3.86 | 3.45 | 4.65 | 1.22 | 3.96 | 3.11 |
| 867 | 4.05 | 1.24 | 3.80 | 3.71 | 4.33 | 1.23 | 3.87 | 3.45 | 4.63 | 1.22 | 3.97 | 3.11 |
| 868 | 4.02 | 1.24 | 3.81 | 3.71 | 4.31 | 1.22 | 3.88 | 3.45 | 4.61 | 1.21 | 3.97 | 3.11 |
| 869 | 4.00 | 1.24 | 3.82 | 3.72 | 4.28 | 1.22 | 3.89 | 3.45 | 4.59 | 1.21 | 3.98 | 3.11 |
| 870 | 3.97 | 1.23 | 3.83 | 3.72 | 4.26 | 1.22 | 3.90 | 3.45 | 4.58 | 1.21 | 3.99 | 3.11 |
| 871 | 3.94 | 1.23 | 3.84 | 3.72 | 4.24 | 1.22 | 3.90 | 3.45 | 4.56 | 1.21 | 4.00 | 3.11 |
| 872 | 3.92 | 1.24 | 3.85 | 3.72 | 4.21 | 1.22 | 3.91 | 3.45 | 4.54 | 1.21 | 4.00 | 3.11 |
| 873 | 3.89 | 1.24 | 3.86 | 3.72 | 4.19 | 1.22 | 3.92 | 3.45 | 4.52 | 1.21 | 4.01 | 3.11 |
| 874 | 3.86 | 1.24 | 3.87 | 3.72 | 4.17 | 1.22 | 3.93 | 3.45 | 4.50 | 1.21 | 4.02 | 3.11 |
| 875 | 3.83 | 1.24 | 3.88 | 3.72 | 4.14 | 1.22 | 3.94 | 3.45 | 4.48 | 1.21 | 4.03 | 3.10 |
| 876 | 3.80 | 1.24 | 3.89 | 3.73 | 4.12 | 1.22 | 3.95 | 3.45 | 4.46 | 1.21 | 4.03 | 3.10 |
| 877 | 3.77 | 1.24 | 3.90 | 3.73 | 4.09 | 1.22 | 3.96 | 3.45 | 4.43 | 1.21 | 4.04 | 3.10 |
| 878 | 3.74 | 1.24 | 3.91 | 3.73 | 4.07 | 1.23 | 3.96 | 3.45 | 4.41 | 1.21 | 4.05 | 3.10 |
| 879 | 3.71 | 1.24 | 3.92 | 3.73 | 4.04 | 1.23 | 3.97 | 3.45 | 4.39 | 1.21 | 4.05 | 3.10 |
| 880 | 3.68 | 1.25 | 3.93 | 3.73 | 4.01 | 1.23 | 3.98 | 3.45 | 4.37 | 1.21 | 4.06 | 3.10 |
| 881 | 3.65 | 1.25 | 3.94 | 3.73 | 3.99 | 1.23 | 3.99 | 3.46 | 4.35 | 1.21 | 4.07 | 3.10 |
| 882 | 3.62 | 1.25 | 3.95 | 3.74 | 3.96 | 1.23 | 4.00 | 3.46 | 4.33 | 1.22 | 4.07 | 3.10 |
| 883 | 3.59 | 1.25 | 3.96 | 3.74 | 3.93 | 1.23 | 4.01 | 3.46 | 4.30 | 1.22 | 4.08 | 3.10 |
| 884 | 3.56 | 1.26 | 3.97 | 3.74 | 3.91 | 1.24 | 4.01 | 3.46 | 4.28 | 1.22 | 4.09 | 3.10 |
| 885 | 3.53 | 1.26 | 3.98 | 3.74 | 3.88 | 1.24 | 4.02 | 3.46 | 4.26 | 1.22 | 4.09 | 3.10 |
| 886 | 3.49 | 1.26 | 3.99 | 3.74 | 3.85 | 1.24 | 4.03 | 3.46 | 4.23 | 1.22 | 4.10 | 3.10 |
| 887 | 3.46 | 1.27 | 4.00 | 3.74 | 3.82 | 1.25 | 4.04 | 3.46 | 4.21 | 1.23 | 4.10 | 3.10 |
| 888 | 3.43 | 1.27 | 4.00 | 3.74 | 3.79 | 1.25 | 4.04 | 3.46 | 4.19 | 1.23 | 4.11 | 3.10 |
| 889 | 3.39 | 1.28 | 4.01 | 3.75 | 3.76 | 1.25 | 4.05 | 3.46 | 4.16 | 1.23 | 4.12 | 3.10 |
| 890 | 3.36 | 1.28 | 4.02 | 3.75 | 3.73 | 1.26 | 4.06 | 3.46 | 4.14 | 1.24 | 4.12 | 3.10 |
| 891 | 3.32 | 1.29 | 4.03 | 3.75 | 3.70 | 1.26 | 4.07 | 3.46 | 4.11 | 1.24 | 4.13 | 3.10 |
| 892 | 3.29 | 1.29 | 4.04 | 3.75 | 3.67 | 1.27 | 4.07 | 3.46 | 4.09 | 1.25 | 4.13 | 3.10 |
| 893 | 3.25 | 1.30 | 4.05 | 3.75 | 3.64 | 1.27 | 4.08 | 3.46 | 4.06 | 1.25 | 4.14 | 3.10 |

| | | | | | | | | | | | |
|---|---|---|---|---|---|---|---|---|---|---|---|
| 894 | 3.22 | 1.31 | 4.06 | 3.75 | 3.61 | 1.28 | 4.09 | 3.46 | 4.04 | 1.25 | 4.14 | 3.10 |
| 895 | 3.18 | 1.31 | 4.07 | 3.75 | 3.58 | 1.28 | 4.09 | 3.46 | 4.01 | 1.26 | 4.15 | 3.10 |
| 896 | 3.14 | 1.32 | 4.08 | 3.75 | 3.55 | 1.29 | 4.10 | 3.46 | 3.98 | 1.26 | 4.16 | 3.10 |
| 897 | 3.10 | 1.33 | 4.08 | 3.76 | 3.51 | 1.30 | 4.11 | 3.46 | 3.96 | 1.27 | 4.16 | 3.10 |
| 898 | 3.06 | 1.33 | 4.09 | 3.76 | 3.48 | 1.30 | 4.11 | 3.47 | 3.93 | 1.28 | 4.17 | 3.10 |
| 899 | 3.03 | 1.34 | 4.10 | 3.76 | 3.45 | 1.31 | 4.12 | 3.47 | 3.90 | 1.28 | 4.17 | 3.10 |
| 900 | 2.99 | 1.35 | 4.11 | 3.76 | 3.41 | 1.32 | 4.13 | 3.47 | 3.87 | 1.29 | 4.18 | 3.10 |
| 901 | 2.95 | 1.36 | 4.12 | 3.76 | 3.38 | 1.33 | 4.13 | 3.47 | 3.84 | 1.30 | 4.18 | 3.10 |
| 902 | 2.90 | 1.37 | 4.13 | 3.76 | 3.34 | 1.33 | 4.14 | 3.47 | 3.82 | 1.30 | 4.19 | 3.10 |
| 903 | 2.86 | 1.38 | 4.13 | 3.76 | 3.31 | 1.34 | 4.15 | 3.47 | 3.79 | 1.31 | 4.19 | 3.10 |
| 904 | 2.82 | 1.39 | 4.14 | 3.77 | 3.27 | 1.35 | 4.15 | 3.47 | 3.76 | 1.32 | 4.20 | 3.10 |
| 905 | 2.78 | 1.40 | 4.15 | 3.77 | 3.23 | 1.36 | 4.16 | 3.47 | 3.73 | 1.33 | 4.20 | 3.10 |
| 906 | 2.73 | 1.41 | 4.16 | 3.77 | 3.20 | 1.37 | 4.16 | 3.47 | 3.69 | 1.34 | 4.20 | 3.10 |
| 907 | 2.69 | 1.43 | 4.16 | 3.77 | 3.16 | 1.38 | 4.17 | 3.47 | 3.66 | 1.35 | 4.21 | 3.10 |
| 908 | 2.64 | 1.44 | 4.17 | 3.77 | 3.12 | 1.39 | 4.17 | 3.48 | 3.63 | 1.36 | 4.21 | 3.10 |
| 909 | 2.60 | 1.45 | 4.18 | 3.77 | 3.08 | 1.41 | 4.18 | 3.48 | 3.60 | 1.37 | 4.22 | 3.10 |
| 910 | 2.55 | 1.46 | 4.18 | 3.78 | 3.04 | 1.42 | 4.19 | 3.48 | 3.57 | 1.38 | 4.22 | 3.10 |
| 911 | 2.50 | 1.48 | 4.19 | 3.78 | 3.00 | 1.43 | 4.19 | 3.48 | 3.53 | 1.39 | 4.23 | 3.10 |
| 912 | 2.45 | 1.49 | 4.20 | 3.78 | 2.96 | 1.44 | 4.20 | 3.48 | 3.50 | 1.40 | 4.23 | 3.10 |
| 913 | 2.40 | 1.51 | 4.21 | 3.78 | 2.91 | 1.46 | 4.20 | 3.48 | 3.46 | 1.41 | 4.23 | 3.10 |
| 914 | 2.35 | 1.53 | 4.21 | 3.78 | 2.87 | 1.47 | 4.21 | 3.48 | 3.43 | 1.43 | 4.24 | 3.11 |
| 915 | 2.30 | 1.54 | 4.22 | 3.78 | 2.83 | 1.49 | 4.21 | 3.49 | 3.39 | 1.44 | 4.24 | 3.11 |
| 916 | 2.25 | 1.56 | 4.22 | 3.79 | 2.78 | 1.50 | 4.22 | 3.49 | 3.36 | 1.46 | 4.24 | 3.11 |
| 917 | 2.20 | 1.58 | 4.23 | 3.79 | 2.74 | 1.52 | 4.22 | 3.49 | 3.32 | 1.47 | 4.25 | 3.11 |
| 918 | 2.14 | 1.60 | 4.24 | 3.79 | 2.69 | 1.54 | 4.22 | 3.49 | 3.28 | 1.49 | 4.25 | 3.11 |
| 919 | 2.09 | 1.62 | 4.24 | 3.79 | 2.64 | 1.56 | 4.23 | 3.49 | 3.25 | 1.50 | 4.25 | 3.11 |
| 920 | 2.03 | 1.64 | 4.25 | 3.79 | 2.60 | 1.58 | 4.23 | 3.50 | 3.21 | 1.52 | 4.26 | 3.11 |
| 921 | 1.97 | 1.66 | 4.25 | 3.80 | 2.55 | 1.60 | 4.24 | 3.50 | 3.17 | 1.54 | 4.26 | 3.12 |
| 922 | 1.91 | 1.69 | 4.26 | 3.80 | 2.50 | 1.62 | 4.24 | 3.50 | 3.13 | 1.55 | 4.26 | 3.12 |
| 923 | 1.86 | 1.71 | 4.26 | 3.80 | 2.45 | 1.64 | 4.24 | 3.50 | 3.09 | 1.57 | 4.26 | 3.12 |
| 924 | 1.79 | 1.73 | 4.27 | 3.81 | 2.40 | 1.66 | 4.25 | 3.51 | 3.05 | 1.59 | 4.27 | 3.12 |
| 925 | 1.73 | 1.76 | 4.27 | 3.81 | 2.34 | 1.68 | 4.25 | 3.51 | 3.00 | 1.62 | 4.27 | 3.13 |
| 926 | 1.67 | 1.79 | 4.28 | 3.81 | 2.29 | 1.71 | 4.25 | 3.51 | 2.96 | 1.64 | 4.27 | 3.13 |
| 927 | 1.61 | 1.82 | 4.28 | 3.81 | 2.24 | 1.73 | 4.26 | 3.52 | 2.92 | 1.66 | 4.27 | 3.13 |
| 928 | 1.54 | 1.85 | 4.28 | 3.82 | 2.18 | 1.76 | 4.26 | 3.52 | 2.87 | 1.68 | 4.28 | 3.14 |
| 929 | 1.47 | 1.88 | 4.29 | 3.82 | 2.12 | 1.79 | 4.26 | 3.52 | 2.83 | 1.71 | 4.28 | 3.14 |
| 930 | 1.40 | 1.91 | 4.29 | 3.83 | 2.07 | 1.82 | 4.26 | 3.53 | 2.78 | 1.73 | 4.28 | 3.14 |
| 931 | 1.34 | 1.94 | 4.30 | 3.83 | 2.01 | 1.85 | 4.27 | 3.53 | 2.74 | 1.76 | 4.28 | 3.15 |

| | | | | | | | | | | | |
|---|---|---|---|---|---|---|---|---|---|---|---|
| 932 | 1.26 | 1.98 | 4.30 | 3.83 | 1.95 | 1.88 | 4.27 | 3.54 | 2.69 | 1.79 | 4.28 | 3.15 |
| 933 | 1.19 | 2.01 | 4.30 | 3.84 | 1.89 | 1.91 | 4.27 | 3.54 | 2.64 | 1.82 | 4.28 | 3.15 |
| 934 | 1.12 | 2.05 | 4.30 | 3.84 | 1.82 | 1.95 | 4.27 | 3.55 | 2.59 | 1.85 | 4.29 | 3.16 |
| 935 | 1.04 | 2.09 | 4.31 | 3.85 | 1.76 | 1.98 | 4.27 | 3.55 | 2.54 | 1.88 | 4.29 | 3.16 |
| 936 | 0.96 | 2.13 | 4.31 | 3.85 | 1.70 | 2.02 | 4.28 | 3.56 | 2.49 | 1.92 | 4.29 | 3.17 |
| 937 | 0.88 | 2.18 | 4.31 | 3.86 | 1.63 | 2.06 | 4.28 | 3.56 | 2.44 | 1.95 | 4.29 | 3.17 |
| 938 | 0.80 | 2.22 | 4.31 | 3.87 | 1.56 | 2.10 | 4.28 | 3.57 | 2.38 | 1.99 | 4.29 | 3.18 |
| 939 | 0.72 | 2.27 | 4.31 | 3.87 | 1.49 | 2.14 | 4.28 | 3.58 | 2.33 | 2.03 | 4.29 | 3.19 |
| 940 | 0.64 | 2.32 | 4.32 | 3.88 | 1.42 | 2.19 | 4.28 | 3.58 | 2.28 | 2.06 | 4.29 | 3.19 |
| 941 | 0.55 | 2.37 | 4.32 | 3.89 | 1.35 | 2.24 | 4.28 | 3.59 | 2.22 | 2.11 | 4.29 | 3.20 |
| 942 | 0.46 | 2.43 | 4.32 | 3.89 | 1.28 | 2.28 | 4.28 | 3.60 | 2.16 | 2.15 | 4.29 | 3.20 |
| 943 | 0.37 | 2.49 | 4.32 | 3.90 | 1.20 | 2.34 | 4.28 | 3.61 | 2.10 | 2.20 | 4.29 | 3.21 |
| 944 | 0.28 | 2.55 | 4.32 | 3.91 | 1.13 | 2.39 | 4.28 | 3.62 | 2.05 | 2.24 | 4.29 | 3.22 |
| 945 | 0.19 | 2.61 | 4.32 | 3.92 | 1.05 | 2.45 | 4.28 | 3.63 | 1.99 | 2.29 | 4.29 | 3.23 |
| 946 | 0.09 | 2.68 | 4.32 | 3.93 | 0.97 | 2.51 | 4.28 | 3.64 | 1.92 | 2.35 | 4.29 | 3.24 |
| 947 | -0.01 | 2.75 | 4.32 | 3.94 | 0.89 | 2.57 | 4.28 | 3.65 | 1.86 | 2.40 | 4.30 | 3.25 |
| 948 | -0.11 | 2.82 | 4.32 | 3.95 | 0.80 | 2.63 | 4.28 | 3.66 | 1.80 | 2.46 | 4.30 | 3.26 |
| 949 | -0.21 | 2.90 | 4.32 | 3.96 | 0.72 | 2.70 | 4.28 | 3.67 | 1.74 | 2.52 | 4.30 | 3.27 |
| 950 | -0.32 | 2.98 | 4.32 | 3.97 | 0.63 | 2.77 | 4.28 | 3.68 | 1.67 | 2.58 | 4.30 | 3.28 |
| 951 | -0.42 | 3.06 | 4.32 | 3.99 | 0.55 | 2.85 | 4.28 | 3.69 | 1.60 | 2.65 | 4.30 | 3.29 |
| 952 | -0.53 | 3.15 | 4.32 | 4.00 | 0.46 | 2.93 | 4.28 | 3.71 | 1.54 | 2.72 | 4.30 | 3.30 |
| 953 | -0.65 | 3.24 | 4.32 | 4.02 | 0.36 | 3.01 | 4.28 | 3.72 | 1.47 | 2.79 | 4.30 | 3.31 |
| 954 | -0.76 | 3.34 | 4.32 | 4.03 | 0.27 | 3.10 | 4.28 | 3.74 | 1.40 | 2.86 | 4.30 | 3.32 |
| 955 | -0.88 | 3.45 | 4.32 | 4.05 | 0.18 | 3.19 | 4.28 | 3.75 | 1.33 | 2.94 | 4.30 | 3.34 |
| 956 | -1.00 | 3.56 | 4.32 | 4.06 | 0.08 | 3.29 | 4.28 | 3.77 | 1.26 | 3.03 | 4.30 | 3.35 |
| 957 | -1.12 | 3.67 | 4.32 | 4.08 | -0.02 | 3.39 | 4.28 | 3.78 | 1.19 | 3.11 | 4.30 | 3.37 |
| 958 | -1.24 | 3.80 | 4.32 | 4.10 | -0.12 | 3.50 | 4.28 | 3.80 | 1.12 | 3.21 | 4.30 | 3.38 |
| 959 | -1.37 | 3.92 | 4.32 | 4.12 | -0.22 | 3.61 | 4.29 | 3.82 | 1.05 | 3.30 | 4.30 | 3.40 |
| 960 | -1.50 | 4.06 | 4.32 | 4.14 | -0.32 | 3.73 | 4.29 | 3.84 | 0.97 | 3.41 | 4.30 | 3.41 |
| 961 | -1.63 | 4.20 | 4.32 | 4.16 | -0.43 | 3.86 | 4.29 | 3.86 | 0.90 | 3.51 | 4.31 | 3.43 |
| 962 | -1.76 | 4.36 | 4.32 | 4.19 | -0.53 | 3.99 | 4.29 | 3.88 | 0.83 | 3.63 | 4.31 | 3.45 |
| 963 | -1.90 | 4.52 | 4.32 | 4.21 | -0.64 | 4.13 | 4.29 | 3.91 | 0.75 | 3.74 | 4.31 | 3.47 |
| 964 | -2.04 | 4.69 | 4.32 | 4.24 | -0.75 | 4.28 | 4.29 | 3.93 | 0.68 | 3.87 | 4.31 | 3.49 |
| 965 | -2.17 | 4.87 | 4.32 | 4.26 | -0.85 | 4.44 | 4.30 | 3.95 | 0.61 | 4.00 | 4.32 | 3.51 |
| 966 | -2.32 | 5.06 | 4.33 | 4.29 | -0.96 | 4.61 | 4.30 | 3.98 | 0.53 | 4.14 | 4.32 | 3.53 |
| 967 | -2.46 | 5.26 | 4.33 | 4.32 | -1.07 | 4.78 | 4.30 | 4.01 | 0.46 | 4.28 | 4.33 | 3.56 |
| 968 | -2.60 | 5.48 | 4.33 | 4.35 | -1.18 | 4.97 | 4.31 | 4.04 | 0.39 | 4.43 | 4.33 | 3.58 |
| 969 | -2.74 | 5.70 | 4.34 | 4.38 | -1.29 | 5.16 | 4.31 | 4.06 | 0.32 | 4.59 | 4.34 | 3.60 |

| | | | | | | | | | | | |
|---|---|---|---|---|---|---|---|---|---|---|---|
| 970 | -2.89 | 5.95 | 4.34 | 4.42 | -1.40 | 5.37 | 4.32 | 4.10 | 0.26 | 4.76 | 4.34 | 3.63 |
| 971 | -3.03 | 6.20 | 4.35 | 4.45 | -1.50 | 5.59 | 4.33 | 4.13 | 0.20 | 4.94 | 4.35 | 3.65 |
| 972 | -3.17 | 6.47 | 4.36 | 4.49 | -1.60 | 5.82 | 4.34 | 4.16 | 0.14 | 5.13 | 4.36 | 3.68 |
| 973 | -3.31 | 6.76 | 4.37 | 4.53 | -1.70 | 6.07 | 4.35 | 4.19 | 0.08 | 5.32 | 4.37 | 3.71 |
| 974 | -3.45 | 7.07 | 4.38 | 4.57 | -1.80 | 6.33 | 4.36 | 4.23 | 0.03 | 5.53 | 4.38 | 3.74 |
| 975 | -3.58 | 7.40 | 4.39 | 4.61 | -1.89 | 6.60 | 4.37 | 4.26 | -0.02 | 5.74 | 4.40 | 3.77 |
| 976 | -3.70 | 7.74 | 4.41 | 4.65 | -1.98 | 6.89 | 4.39 | 4.30 | -0.06 | 5.97 | 4.41 | 3.80 |
| 977 | -3.82 | 8.11 | 4.42 | 4.69 | -2.05 | 7.20 | 4.41 | 4.34 | -0.09 | 6.21 | 4.42 | 3.83 |
| 978 | -3.93 | 8.50 | 4.44 | 4.74 | -2.12 | 7.53 | 4.42 | 4.38 | -0.12 | 6.46 | 4.44 | 3.86 |
| 979 | -4.03 | 8.92 | 4.46 | 4.79 | -2.18 | 7.87 | 4.45 | 4.42 | -0.13 | 6.72 | 4.46 | 3.89 |
| 980 | -4.12 | 9.35 | 4.49 | 4.83 | -2.23 | 8.23 | 4.47 | 4.46 | -0.14 | 6.99 | 4.48 | 3.93 |
| 981 | -4.19 | 9.82 | 4.51 | 4.88 | -2.26 | 8.61 | 4.49 | 4.50 | -0.13 | 7.27 | 4.50 | 3.96 |
| 982 | -4.24 | 10.31 | 4.54 | 4.93 | -2.27 | 9.00 | 4.52 | 4.55 | -0.11 | 7.56 | 4.53 | 3.99 |
| 983 | -4.27 | 10.82 | 4.57 | 4.98 | -2.27 | 9.42 | 4.55 | 4.59 | -0.07 | 7.87 | 4.56 | 4.03 |
| 984 | -4.28 | 11.37 | 4.61 | 5.03 | -2.24 | 9.86 | 4.59 | 4.63 | -0.01 | 8.18 | 4.58 | 4.06 |
| 985 | -4.25 | 11.94 | 4.65 | 5.09 | -2.19 | 10.31 | 4.62 | 4.67 | 0.06 | 8.50 | 4.62 | 4.10 |
| 986 | -4.19 | 12.54 | 4.69 | 5.14 | -2.11 | 10.78 | 4.66 | 4.72 | 0.16 | 8.84 | 4.65 | 4.13 |
| 987 | -4.09 | 13.16 | 4.74 | 5.19 | -2.00 | 11.27 | 4.71 | 4.76 | 0.27 | 9.17 | 4.69 | 4.16 |
| 988 | -3.94 | 13.80 | 4.79 | 5.24 | -1.85 | 11.77 | 4.75 | 4.80 | 0.42 | 9.52 | 4.72 | 4.19 |
| 989 | -3.75 | 14.47 | 4.85 | 5.29 | -1.67 | 12.28 | 4.80 | 4.84 | 0.58 | 9.87 | 4.77 | 4.23 |
| 990 | -3.50 | 15.15 | 4.91 | 5.34 | -1.44 | 12.80 | 4.85 | 4.88 | 0.78 | 10.21 | 4.81 | 4.26 |
| 991 | -3.19 | 15.84 | 4.97 | 5.38 | -1.16 | 13.33 | 4.91 | 4.92 | 1.01 | 10.56 | 4.86 | 4.29 |
| 992 | -2.81 | 16.54 | 5.04 | 5.43 | -0.84 | 13.85 | 4.97 | 4.95 | 1.26 | 10.90 | 4.91 | 4.31 |
| 993 | -2.37 | 17.23 | 5.11 | 5.47 | -0.46 | 14.37 | 5.03 | 4.99 | 1.55 | 11.23 | 4.96 | 4.34 |
| 994 | -1.85 | 17.91 | 5.19 | 5.51 | -0.04 | 14.87 | 5.10 | 5.02 | 1.87 | 11.56 | 5.01 | 4.36 |
| 995 | -1.25 | 18.58 | 5.27 | 5.54 | 0.45 | 15.36 | 5.17 | 5.05 | 2.23 | 11.86 | 5.07 | 4.38 |
| 996 | -0.58 | 19.21 | 5.35 | 5.58 | 0.98 | 15.81 | 5.24 | 5.07 | 2.61 | 12.15 | 5.12 | 4.40 |
| 997 | 0.17 | 19.79 | 5.44 | 5.60 | 1.57 | 16.24 | 5.31 | 5.10 | 3.02 | 12.41 | 5.18 | 4.42 |
| 998 | 0.99 | 20.32 | 5.53 | 5.63 | 2.21 | 16.62 | 5.39 | 5.11 | 3.46 | 12.64 | 5.24 | 4.43 |
| 999 | 1.88 | 20.78 | 5.62 | 5.64 | 2.89 | 16.95 | 5.47 | 5.13 | 3.92 | 12.84 | 5.31 | 4.44 |
| 1000 | 2.83 | 21.16 | 5.72 | 5.66 | 3.61 | 17.22 | 5.55 | 5.14 | 4.41 | 13.00 | 5.37 | 4.45 |
| 1001 | 3.82 | 21.45 | 5.81 | 5.67 | 4.36 | 17.43 | 5.63 | 5.14 | 4.91 | 13.13 | 5.44 | 4.45 |
| 1002 | 4.84 | 21.65 | 5.91 | 5.67 | 5.13 | 17.57 | 5.71 | 5.15 | 5.42 | 13.21 | 5.50 | 4.46 |
| 1003 | 5.88 | 21.75 | 6.01 | 5.66 | 5.91 | 17.63 | 5.79 | 5.14 | 5.94 | 13.25 | 5.57 | 4.45 |
| 1004 | 6.93 | 21.74 | 6.11 | 5.65 | 6.70 | 17.63 | 5.87 | 5.13 | 6.47 | 13.25 | 5.63 | 4.45 |
| 1005 | 7.97 | 21.63 | 6.20 | 5.64 | 7.48 | 17.55 | 5.96 | 5.12 | 6.99 | 13.21 | 5.70 | 4.44 |
| 1006 | 8.99 | 21.42 | 6.30 | 5.62 | 8.25 | 17.41 | 6.04 | 5.11 | 7.50 | 13.13 | 5.76 | 4.42 |
| 1007 | 9.96 | 21.12 | 6.40 | 5.59 | 8.99 | 17.20 | 6.12 | 5.08 | 7.99 | 13.00 | 5.83 | 4.41 |

| | | | | | | | | | | | |
|---|---|---|---|---|---|---|---|---|---|---|---|
| 1008 | 10.88 | 20.73 | 6.49 | 5.56 | 9.70 | 16.93 | 6.20 | 5.06 | 8.47 | 12.84 | 5.89 | 4.39 |
| 1009 | 11.75 | 20.27 | 6.58 | 5.52 | 10.36 | 16.60 | 6.27 | 5.03 | 8.93 | 12.65 | 5.95 | 4.36 |
| 1010 | 12.55 | 19.75 | 6.67 | 5.48 | 10.98 | 16.23 | 6.35 | 5.00 | 9.36 | 12.42 | 6.01 | 4.34 |
| 1011 | 13.27 | 19.18 | 6.75 | 5.44 | 11.56 | 15.81 | 6.42 | 4.96 | 9.77 | 12.17 | 6.07 | 4.31 |
| 1012 | 13.93 | 18.56 | 6.84 | 5.39 | 12.08 | 15.37 | 6.49 | 4.92 | 10.14 | 11.89 | 6.13 | 4.28 |
| 1013 | 14.51 | 17.92 | 6.91 | 5.33 | 12.55 | 14.90 | 6.56 | 4.88 | 10.49 | 11.60 | 6.18 | 4.25 |
| 1014 | 15.01 | 17.26 | 6.99 | 5.28 | 12.97 | 14.41 | 6.62 | 4.83 | 10.81 | 11.29 | 6.24 | 4.21 |
| 1015 | 15.45 | 16.60 | 7.06 | 5.22 | 13.34 | 13.92 | 6.68 | 4.78 | 11.09 | 10.97 | 6.29 | 4.17 |
| 1016 | 15.82 | 15.93 | 7.12 | 5.15 | 13.66 | 13.41 | 6.74 | 4.73 | 11.35 | 10.64 | 6.33 | 4.13 |
| 1017 | 16.13 | 15.27 | 7.18 | 5.09 | 13.93 | 12.91 | 6.80 | 4.68 | 11.58 | 10.31 | 6.38 | 4.09 |
| 1018 | 16.38 | 14.62 | 7.24 | 5.03 | 14.16 | 12.41 | 6.85 | 4.63 | 11.77 | 9.98 | 6.42 | 4.05 |
| 1019 | 16.58 | 13.98 | 7.29 | 4.96 | 14.35 | 11.93 | 6.89 | 4.57 | 11.95 | 9.65 | 6.46 | 4.01 |
| 1020 | 16.73 | 13.36 | 7.34 | 4.89 | 14.51 | 11.45 | 6.94 | 4.52 | 12.09 | 9.32 | 6.50 | 3.96 |
| 1021 | 16.84 | 12.77 | 7.39 | 4.82 | 14.63 | 10.98 | 6.98 | 4.46 | 12.22 | 9.00 | 6.53 | 3.92 |
| 1022 | 16.92 | 12.20 | 7.43 | 4.76 | 14.72 | 10.53 | 7.02 | 4.40 | 12.32 | 8.68 | 6.57 | 3.87 |
| 1023 | 16.96 | 11.65 | 7.47 | 4.69 | 14.78 | 10.09 | 7.05 | 4.35 | 12.41 | 8.37 | 6.60 | 3.83 |
| 1024 | 16.97 | 11.13 | 7.50 | 4.62 | 14.82 | 9.68 | 7.09 | 4.29 | 12.47 | 8.07 | 6.62 | 3.78 |
| 1025 | 16.96 | 10.63 | 7.53 | 4.55 | 14.84 | 9.28 | 7.12 | 4.23 | 12.52 | 7.78 | 6.65 | 3.74 |
| 1026 | 16.93 | 10.16 | 7.56 | 4.49 | 14.85 | 8.89 | 7.14 | 4.18 | 12.55 | 7.49 | 6.67 | 3.69 |
| 1027 | 16.88 | 9.71 | 7.58 | 4.42 | 14.83 | 8.53 | 7.17 | 4.12 | 12.58 | 7.22 | 6.70 | 3.65 |
| 1028 | 16.82 | 9.28 | 7.61 | 4.36 | 14.80 | 8.18 | 7.19 | 4.07 | 12.59 | 6.96 | 6.72 | 3.60 |
| 1029 | 16.74 | 8.88 | 7.62 | 4.29 | 14.76 | 7.84 | 7.21 | 4.01 | 12.58 | 6.71 | 6.73 | 3.56 |
| 1030 | 16.66 | 8.50 | 7.64 | 4.23 | 14.71 | 7.53 | 7.22 | 3.96 | 12.57 | 6.47 | 6.75 | 3.52 |
| 1031 | 16.56 | 8.14 | 7.65 | 4.17 | 14.65 | 7.23 | 7.24 | 3.90 | 12.56 | 6.23 | 6.76 | 3.47 |
| 1032 | 16.46 | 7.80 | 7.67 | 4.11 | 14.59 | 6.94 | 7.25 | 3.85 | 12.53 | 6.01 | 6.78 | 3.43 |
| 1033 | 16.35 | 7.48 | 7.68 | 4.05 | 14.52 | 6.67 | 7.26 | 3.80 | 12.50 | 5.80 | 6.79 | 3.39 |
| 1034 | 16.23 | 7.17 | 7.68 | 3.99 | 14.44 | 6.41 | 7.27 | 3.75 | 12.46 | 5.60 | 6.80 | 3.35 |
| 1035 | 16.11 | 6.89 | 7.69 | 3.93 | 14.36 | 6.17 | 7.28 | 3.70 | 12.42 | 5.40 | 6.81 | 3.31 |
| 1036 | 15.99 | 6.62 | 7.70 | 3.88 | 14.27 | 5.94 | 7.29 | 3.66 | 12.38 | 5.22 | 6.81 | 3.27 |
| 1037 | 15.87 | 6.36 | 7.70 | 3.83 | 14.18 | 5.72 | 7.29 | 3.61 | 12.33 | 5.04 | 6.82 | 3.23 |
| 1038 | 15.75 | 6.12 | 7.70 | 3.77 | 14.09 | 5.51 | 7.30 | 3.56 | 12.27 | 4.87 | 6.83 | 3.19 |
| 1039 | 15.62 | 5.89 | 7.70 | 3.72 | 14.00 | 5.31 | 7.30 | 3.52 | 12.22 | 4.71 | 6.83 | 3.15 |
| 1040 | 15.50 | 5.67 | 7.70 | 3.67 | 13.91 | 5.13 | 7.30 | 3.47 | 12.16 | 4.56 | 6.83 | 3.12 |
| 1041 | 15.37 | 5.46 | 7.70 | 3.63 | 13.82 | 4.95 | 7.31 | 3.43 | 12.11 | 4.41 | 6.84 | 3.08 |
| 1042 | 15.25 | 5.27 | 7.70 | 3.58 | 13.72 | 4.78 | 7.31 | 3.39 | 12.05 | 4.27 | 6.84 | 3.05 |
| 1043 | 15.13 | 5.08 | 7.70 | 3.53 | 13.63 | 4.62 | 7.31 | 3.35 | 11.99 | 4.14 | 6.84 | 3.01 |
| 1044 | 15.01 | 4.91 | 7.69 | 3.49 | 13.54 | 4.46 | 7.30 | 3.31 | 11.93 | 4.01 | 6.84 | 2.98 |
| 1045 | 14.89 | 4.74 | 7.69 | 3.45 | 13.45 | 4.32 | 7.30 | 3.27 | 11.87 | 3.89 | 6.84 | 2.95 |

| | | | | | | | | | | | |
|---|---|---|---|---|---|---|---|---|---|---|---|
| 1046 | 14.77 | 4.58 | 7.68 | 3.40 | 13.35 | 4.18 | 7.30 | 3.24 | 11.80 | 3.77 | 6.84 | 2.92 |
| 1047 | 14.65 | 4.43 | 7.68 | 3.36 | 13.26 | 4.05 | 7.30 | 3.20 | 11.74 | 3.66 | 6.84 | 2.89 |
| 1048 | 14.54 | 4.29 | 7.67 | 3.32 | 13.17 | 3.92 | 7.29 | 3.16 | 11.68 | 3.55 | 6.84 | 2.86 |
| 1049 | 14.43 | 4.15 | 7.67 | 3.29 | 13.08 | 3.80 | 7.29 | 3.13 | 11.62 | 3.45 | 6.84 | 2.83 |
| 1050 | 14.31 | 4.03 | 7.66 | 3.25 | 13.00 | 3.69 | 7.29 | 3.10 | 11.56 | 3.35 | 6.83 | 2.80 |
| 1051 | 14.21 | 3.90 | 7.65 | 3.21 | 12.91 | 3.58 | 7.28 | 3.06 | 11.50 | 3.26 | 6.83 | 2.77 |
| 1052 | 14.10 | 3.79 | 7.64 | 3.18 | 12.82 | 3.48 | 7.28 | 3.03 | 11.44 | 3.17 | 6.83 | 2.74 |
| 1053 | 13.99 | 3.67 | 7.64 | 3.14 | 12.74 | 3.38 | 7.27 | 3.00 | 11.38 | 3.09 | 6.82 | 2.72 |
| 1054 | 13.89 | 3.57 | 7.63 | 3.11 | 12.66 | 3.28 | 7.27 | 2.97 | 11.32 | 3.00 | 6.82 | 2.69 |
| 1055 | 13.79 | 3.47 | 7.62 | 3.08 | 12.58 | 3.19 | 7.26 | 2.94 | 11.26 | 2.93 | 6.82 | 2.67 |
| 1056 | 13.69 | 3.37 | 7.61 | 3.04 | 12.50 | 3.10 | 7.26 | 2.91 | 11.20 | 2.85 | 6.81 | 2.64 |
| 1057 | 13.59 | 3.28 | 7.60 | 3.01 | 12.42 | 3.02 | 7.25 | 2.88 | 11.14 | 2.78 | 6.81 | 2.62 |
| 1058 | 13.50 | 3.19 | 7.60 | 2.98 | 12.34 | 2.94 | 7.24 | 2.86 | 11.09 | 2.71 | 6.81 | 2.60 |
| 1059 | 13.41 | 3.10 | 7.59 | 2.95 | 12.27 | 2.87 | 7.24 | 2.83 | 11.03 | 2.64 | 6.80 | 2.57 |
| 1060 | 13.31 | 3.02 | 7.58 | 2.92 | 12.19 | 2.79 | 7.23 | 2.80 | 10.98 | 2.58 | 6.80 | 2.55 |
| 1061 | 13.23 | 2.94 | 7.57 | 2.90 | 12.12 | 2.72 | 7.22 | 2.78 | 10.92 | 2.52 | 6.79 | 2.53 |
| 1062 | 13.14 | 2.87 | 7.56 | 2.87 | 12.05 | 2.66 | 7.22 | 2.75 | 10.87 | 2.46 | 6.79 | 2.51 |
| 1063 | 13.05 | 2.80 | 7.55 | 2.84 | 11.98 | 2.59 | 7.21 | 2.73 | 10.82 | 2.40 | 6.78 | 2.49 |
| 1064 | 12.97 | 2.73 | 7.54 | 2.82 | 11.91 | 2.53 | 7.20 | 2.71 | 10.76 | 2.35 | 6.78 | 2.47 |
| 1065 | 12.89 | 2.66 | 7.54 | 2.79 | 11.84 | 2.47 | 7.20 | 2.68 | 10.71 | 2.29 | 6.77 | 2.45 |
| 1066 | 12.81 | 2.60 | 7.53 | 2.77 | 11.77 | 2.41 | 7.19 | 2.66 | 10.66 | 2.24 | 6.77 | 2.43 |
| 1067 | 12.73 | 2.54 | 7.52 | 2.74 | 11.71 | 2.36 | 7.18 | 2.64 | 10.61 | 2.19 | 6.76 | 2.41 |
| 1068 | 12.65 | 2.48 | 7.51 | 2.72 | 11.65 | 2.31 | 7.18 | 2.62 | 10.56 | 2.15 | 6.76 | 2.39 |
| 1069 | 12.58 | 2.42 | 7.50 | 2.69 | 11.58 | 2.26 | 7.17 | 2.59 | 10.52 | 2.10 | 6.75 | 2.37 |
| 1070 | 12.50 | 2.37 | 7.49 | 2.67 | 11.52 | 2.21 | 7.16 | 2.57 | 10.47 | 2.06 | 6.75 | 2.35 |
| 1071 | 12.43 | 2.32 | 7.48 | 2.65 | 11.46 | 2.16 | 7.16 | 2.55 | 10.42 | 2.02 | 6.74 | 2.33 |
| 1072 | 12.36 | 2.27 | 7.47 | 2.63 | 11.40 | 2.11 | 7.15 | 2.53 | 10.38 | 1.98 | 6.74 | 2.32 |
| 1073 | 12.29 | 2.22 | 7.46 | 2.61 | 11.35 | 2.07 | 7.14 | 2.51 | 10.33 | 1.94 | 6.73 | 2.30 |
| 1074 | 12.22 | 2.17 | 7.46 | 2.58 | 11.29 | 2.03 | 7.14 | 2.49 | 10.29 | 1.90 | 6.73 | 2.28 |
| 1075 | 12.16 | 2.13 | 7.45 | 2.56 | 11.23 | 1.99 | 7.13 | 2.48 | 10.25 | 1.86 | 6.72 | 2.27 |
| 1076 | 12.09 | 2.09 | 7.44 | 2.54 | 11.18 | 1.95 | 7.12 | 2.46 | 10.21 | 1.83 | 6.72 | 2.25 |
| 1077 | 12.03 | 2.04 | 7.43 | 2.52 | 11.13 | 1.91 | 7.12 | 2.44 | 10.16 | 1.79 | 6.71 | 2.24 |
| 1078 | 11.96 | 2.00 | 7.42 | 2.50 | 11.07 | 1.87 | 7.11 | 2.42 | 10.12 | 1.76 | 6.71 | 2.22 |
| 1079 | 11.90 | 1.96 | 7.41 | 2.49 | 11.02 | 1.84 | 7.10 | 2.40 | 10.08 | 1.73 | 6.70 | 2.20 |
| 1080 | 11.84 | 1.93 | 7.41 | 2.47 | 10.97 | 1.80 | 7.10 | 2.39 | 10.04 | 1.70 | 6.70 | 2.19 |
| 1081 | 11.78 | 1.89 | 7.40 | 2.45 | 10.92 | 1.77 | 7.09 | 2.37 | 10.00 | 1.67 | 6.69 | 2.18 |
| 1082 | 11.73 | 1.85 | 7.39 | 2.43 | 10.88 | 1.74 | 7.08 | 2.35 | 9.97 | 1.64 | 6.69 | 2.16 |
| 1083 | 11.67 | 1.82 | 7.38 | 2.41 | 10.83 | 1.71 | 7.08 | 2.34 | 9.93 | 1.61 | 6.68 | 2.15 |

| | | | | | | | | | | | |
|---|---|---|---|---|---|---|---|---|---|---|---|
| 1084 | 11.61 | 1.79 | 7.37 | 2.40 | 10.78 | 1.68 | 7.07 | 2.32 | 9.89 | 1.58 | 6.68 | 2.13 |
| 1085 | 11.56 | 1.76 | 7.36 | 2.38 | 10.74 | 1.65 | 7.06 | 2.31 | 9.86 | 1.56 | 6.67 | 2.12 |
| 1086 | 11.51 | 1.73 | 7.36 | 2.36 | 10.69 | 1.62 | 7.06 | 2.29 | 9.82 | 1.53 | 6.67 | 2.11 |
| 1087 | 11.45 | 1.70 | 7.35 | 2.35 | 10.65 | 1.59 | 7.05 | 2.27 | 9.79 | 1.50 | 6.66 | 2.09 |
| 1088 | 11.40 | 1.67 | 7.34 | 2.33 | 10.60 | 1.57 | 7.04 | 2.26 | 9.75 | 1.48 | 6.66 | 2.08 |
| 1089 | 11.35 | 1.64 | 7.33 | 2.31 | 10.56 | 1.54 | 7.04 | 2.25 | 9.72 | 1.46 | 6.65 | 2.07 |
| 1090 | 11.30 | 1.61 | 7.33 | 2.30 | 10.52 | 1.52 | 7.03 | 2.23 | 9.68 | 1.43 | 6.65 | 2.05 |
| 1091 | 11.26 | 1.58 | 7.32 | 2.28 | 10.48 | 1.49 | 7.02 | 2.22 | 9.65 | 1.41 | 6.64 | 2.04 |
| 1092 | 11.21 | 1.56 | 7.31 | 2.27 | 10.44 | 1.47 | 7.02 | 2.20 | 9.62 | 1.39 | 6.64 | 2.03 |
| 1093 | 11.16 | 1.53 | 7.30 | 2.25 | 10.40 | 1.45 | 7.01 | 2.19 | 9.59 | 1.37 | 6.63 | 2.02 |
| 1094 | 11.11 | 1.51 | 7.30 | 2.24 | 10.36 | 1.42 | 7.01 | 2.18 | 9.56 | 1.35 | 6.63 | 2.01 |
| 1095 | 11.07 | 1.49 | 7.29 | 2.22 | 10.32 | 1.40 | 7.00 | 2.16 | 9.53 | 1.33 | 6.62 | 1.99 |
| 1096 | 11.03 | 1.46 | 7.28 | 2.21 | 10.28 | 1.38 | 6.99 | 2.15 | 9.50 | 1.31 | 6.62 | 1.98 |
| 1097 | 10.98 | 1.44 | 7.27 | 2.20 | 10.25 | 1.36 | 6.99 | 2.14 | 9.47 | 1.29 | 6.61 | 1.97 |
| 1098 | 10.94 | 1.42 | 7.27 | 2.18 | 10.21 | 1.34 | 6.98 | 2.12 | 9.44 | 1.27 | 6.61 | 1.96 |
| 1099 | 10.90 | 1.40 | 7.26 | 2.17 | 10.17 | 1.32 | 6.98 | 2.11 | 9.41 | 1.26 | 6.61 | 1.95 |
| 1100 | 10.86 | 1.38 | 7.25 | 2.15 | 10.14 | 1.30 | 6.97 | 2.10 | 9.38 | 1.24 | 6.60 | 1.94 |
| 1101 | 10.82 | 1.36 | 7.25 | 2.14 | 10.10 | 1.28 | 6.97 | 2.09 | 9.35 | 1.22 | 6.60 | 1.93 |
| 1102 | 10.78 | 1.34 | 7.24 | 2.13 | 10.07 | 1.27 | 6.96 | 2.07 | 9.33 | 1.20 | 6.59 | 1.92 |
| 1103 | 10.74 | 1.32 | 7.23 | 2.12 | 10.04 | 1.25 | 6.95 | 2.06 | 9.30 | 1.19 | 6.59 | 1.91 |
| 1104 | 10.70 | 1.30 | 7.23 | 2.10 | 10.00 | 1.23 | 6.95 | 2.05 | 9.27 | 1.17 | 6.58 | 1.90 |
| 1105 | 10.66 | 1.28 | 7.22 | 2.09 | 9.97 | 1.22 | 6.94 | 2.04 | 9.25 | 1.16 | 6.58 | 1.89 |
| 1106 | 10.62 | 1.27 | 7.21 | 2.08 | 9.94 | 1.20 | 6.94 | 2.03 | 9.22 | 1.14 | 6.57 | 1.88 |
| 1107 | 10.59 | 1.25 | 7.20 | 2.07 | 9.91 | 1.18 | 6.93 | 2.02 | 9.19 | 1.13 | 6.57 | 1.87 |
| 1108 | 10.55 | 1.23 | 7.20 | 2.05 | 9.88 | 1.17 | 6.93 | 2.01 | 9.17 | 1.11 | 6.57 | 1.86 |
| 1109 | 10.51 | 1.22 | 7.19 | 2.04 | 9.85 | 1.15 | 6.92 | 1.99 | 9.15 | 1.10 | 6.56 | 1.85 |
| 1110 | 10.48 | 1.20 | 7.19 | 2.03 | 9.82 | 1.14 | 6.92 | 1.98 | 9.12 | 1.09 | 6.56 | 1.84 |
| 1111 | 10.44 | 1.19 | 7.18 | 2.02 | 9.79 | 1.12 | 6.91 | 1.97 | 9.10 | 1.07 | 6.55 | 1.83 |
| 1112 | 10.41 | 1.17 | 7.17 | 2.01 | 9.76 | 1.11 | 6.91 | 1.96 | 9.07 | 1.06 | 6.55 | 1.82 |
| 1113 | 10.38 | 1.16 | 7.17 | 2.00 | 9.73 | 1.10 | 6.90 | 1.95 | 9.05 | 1.05 | 6.55 | 1.81 |
| 1114 | 10.35 | 1.14 | 7.16 | 1.99 | 9.70 | 1.08 | 6.89 | 1.94 | 9.03 | 1.04 | 6.54 | 1.80 |
| 1115 | 10.31 | 1.13 | 7.15 | 1.97 | 9.68 | 1.07 | 6.89 | 1.93 | 9.01 | 1.02 | 6.54 | 1.79 |
| 1116 | 10.28 | 1.11 | 7.15 | 1.96 | 9.65 | 1.06 | 6.88 | 1.92 | 8.98 | 1.01 | 6.53 | 1.78 |
| 1117 | 10.25 | 1.10 | 7.14 | 1.95 | 9.62 | 1.05 | 6.88 | 1.91 | 8.96 | 1.00 | 6.53 | 1.77 |
| 1118 | 10.22 | 1.09 | 7.14 | 1.94 | 9.60 | 1.03 | 6.87 | 1.90 | 8.94 | 0.99 | 6.53 | 1.76 |
| 1119 | 10.19 | 1.07 | 7.13 | 1.93 | 9.57 | 1.02 | 6.87 | 1.89 | 8.92 | 0.98 | 6.52 | 1.76 |
| 1120 | 10.16 | 1.06 | 7.12 | 1.92 | 9.55 | 1.01 | 6.86 | 1.88 | 8.90 | 0.97 | 6.52 | 1.75 |
| 1121 | 10.13 | 1.05 | 7.12 | 1.91 | 9.52 | 1.00 | 6.86 | 1.87 | 8.88 | 0.96 | 6.51 | 1.74 |
| 1122 | 10.10 | 1.04 | 7.11 | 1.90 | 9.50 | 0.99 | 6.85 | 1.86 | 8.86 | 0.95 | 6.51 | 1.73 |
| 1123 | 10.07 | 1.03 | 7.11 | 1.89 | 9.47 | 0.98 | 6.85 | 1.85 | 8.84 | 0.94 | 6.51 | 1.72 |
| 1124 | 10.04 | 1.02 | 7.10 | 1.88 | 9.45 | 0.97 | 6.85 | 1.84 | 8.82 | 0.93 | 6.50 | 1.71 |
| 1125 | 10.02 | 1.00 | 7.09 | 1.87 | 9.42 | 0.96 | 6.84 | 1.83 | 8.80 | 0.92 | 6.50 | 1.71 |
| 1126 | 9.99 | 0.99 | 7.09 | 1.86 | 9.40 | 0.95 | 6.84 | 1.83 | 8.78 | 0.91 | 6.49 | 1.70 |
| 1127 | 9.96 | 0.98 | 7.08 | 1.85 | 9.38 | 0.94 | 6.83 | 1.82 | 8.76 | 0.90 | 6.49 | 1.69 |
| 1128 | 9.94 | 0.97 | 7.08 | 1.84 | 9.35 | 0.93 | 6.83 | 1.81 | 8.74 | 0.89 | 6.49 | 1.68 |
| 1129 | 9.91 | 0.96 | 7.07 | 1.83 | 9.33 | 0.92 | 6.82 | 1.80 | 8.72 | 0.88 | 6.48 | 1.67 |
| 1130 | 9.88 | 0.95 | 7.07 | 1.82 | 9.31 | 0.91 | 6.82 | 1.79 | 8.70 | 0.87 | 6.48 | 1.67 |
| 1131 | 9.86 | 0.94 | 7.06 | 1.81 | 9.29 | 0.90 | 6.81 | 1.78 | 8.69 | 0.87 | 6.48 | 1.66 |
| 1132 | 9.83 | 0.93 | 7.06 | 1.80 | 9.27 | 0.89 | 6.81 | 1.77 | 8.67 | 0.86 | 6.47 | 1.65 |
| 1133 | 9.81 | 0.92 | 7.05 | 1.80 | 9.24 | 0.88 | 6.80 | 1.77 | 8.65 | 0.85 | 6.47 | 1.64 |
| 1134 | 9.78 | 0.91 | 7.04 | 1.79 | 9.22 | 0.87 | 6.80 | 1.76 | 8.63 | 0.84 | 6.47 | 1.64 |
| 1135 | 9.76 | 0.90 | 7.04 | 1.78 | 9.20 | 0.87 | 6.79 | 1.75 | 8.62 | 0.83 | 6.46 | 1.63 |
| 1136 | 9.74 | 0.90 | 7.03 | 1.77 | 9.18 | 0.86 | 6.79 | 1.74 | 8.60 | 0.83 | 6.46 | 1.62 |
| 1137 | 9.71 | 0.89 | 7.03 | 1.76 | 9.16 | 0.85 | 6.79 | 1.73 | 8.58 | 0.82 | 6.45 | 1.62 |
| 1138 | 9.69 | 0.88 | 7.02 | 1.75 | 9.14 | 0.84 | 6.78 | 1.72 | 8.57 | 0.81 | 6.45 | 1.61 |
| 1139 | 9.67 | 0.87 | 7.02 | 1.74 | 9.12 | 0.83 | 6.78 | 1.72 | 8.55 | 0.80 | 6.45 | 1.60 |
| 1140 | 9.64 | 0.86 | 7.01 | 1.74 | 9.10 | 0.83 | 6.77 | 1.71 | 8.53 | 0.80 | 6.44 | 1.59 |
| 1141 | 9.62 | 0.85 | 7.01 | 1.73 | 9.08 | 0.82 | 6.77 | 1.70 | 8.52 | 0.79 | 6.44 | 1.59 |
| 1142 | 9.60 | 0.85 | 7.00 | 1.72 | 9.06 | 0.81 | 6.76 | 1.69 | 8.50 | 0.78 | 6.44 | 1.58 |
| 1143 | 9.58 | 0.84 | 7.00 | 1.71 | 9.05 | 0.80 | 6.76 | 1.69 | 8.49 | 0.78 | 6.43 | 1.57 |
| 1144 | 9.56 | 0.83 | 6.99 | 1.70 | 9.03 | 0.80 | 6.76 | 1.68 | 8.47 | 0.77 | 6.43 | 1.57 |
| 1145 | 9.54 | 0.82 | 6.99 | 1.69 | 9.01 | 0.79 | 6.75 | 1.67 | 8.46 | 0.76 | 6.43 | 1.56 |
| 1146 | 9.51 | 0.82 | 6.98 | 1.69 | 8.99 | 0.78 | 6.75 | 1.66 | 8.44 | 0.76 | 6.42 | 1.55 |
| 1147 | 9.49 | 0.81 | 6.98 | 1.68 | 8.97 | 0.78 | 6.74 | 1.66 | 8.43 | 0.75 | 6.42 | 1.55 |
| 1148 | 9.47 | 0.80 | 6.97 | 1.67 | 8.96 | 0.77 | 6.74 | 1.65 | 8.41 | 0.74 | 6.42 | 1.54 |
| 1149 | 9.45 | 0.80 | 6.97 | 1.66 | 8.94 | 0.76 | 6.74 | 1.64 | 8.40 | 0.74 | 6.41 | 1.53 |
| 1150 | 9.43 | 0.79 | 6.96 | 1.66 | 8.92 | 0.76 | 6.73 | 1.63 | 8.38 | 0.73 | 6.41 | 1.53 |
| 1151 | 9.41 | 0.78 | 6.96 | 1.65 | 8.90 | 0.75 | 6.73 | 1.63 | 8.37 | 0.73 | 6.41 | 1.52 |
| 1152 | 9.39 | 0.77 | 6.95 | 1.64 | 8.89 | 0.74 | 6.72 | 1.62 | 8.36 | 0.72 | 6.41 | 1.51 |
| 1153 | 9.38 | 0.77 | 6.95 | 1.63 | 8.87 | 0.74 | 6.72 | 1.61 | 8.34 | 0.71 | 6.40 | 1.51 |
| 1154 | 9.36 | 0.76 | 6.94 | 1.62 | 8.86 | 0.73 | 6.72 | 1.61 | 8.33 | 0.71 | 6.40 | 1.50 |
| 1155 | 9.34 | 0.76 | 6.94 | 1.62 | 8.84 | 0.73 | 6.71 | 1.60 | 8.32 | 0.70 | 6.40 | 1.50 |
| 1156 | 9.32 | 0.75 | 6.93 | 1.61 | 8.82 | 0.72 | 6.71 | 1.59 | 8.30 | 0.70 | 6.39 | 1.49 |
| 1157 | 9.30 | 0.74 | 6.93 | 1.60 | 8.81 | 0.72 | 6.70 | 1.58 | 8.29 | 0.69 | 6.39 | 1.48 |
| 1158 | 9.28 | 0.74 | 6.92 | 1.60 | 8.79 | 0.71 | 6.70 | 1.58 | 8.28 | 0.69 | 6.39 | 1.48 |
| 1159 | 9.27 | 0.73 | 6.92 | 1.59 | 8.78 | 0.70 | 6.70 | 1.57 | 8.26 | 0.68 | 6.38 | 1.47 |
| 1160 | 9.25 | 0.73 | 6.92 | 1.58 | 8.76 | 0.70 | 6.69 | 1.56 | 8.25 | 0.68 | 6.38 | 1.47 |
| 1161 | 9.23 | 0.72 | 6.91 | 1.57 | 8.75 | 0.69 | 6.69 | 1.56 | 8.24 | 0.67 | 6.38 | 1.46 |
| 1162 | 9.21 | 0.71 | 6.91 | 1.57 | 8.73 | 0.69 | 6.68 | 1.55 | 8.23 | 0.67 | 6.37 | 1.45 |
| 1163 | 9.20 | 0.71 | 6.90 | 1.56 | 8.72 | 0.68 | 6.68 | 1.55 | 8.21 | 0.66 | 6.37 | 1.45 |
| 1164 | 9.18 | 0.70 | 6.90 | 1.55 | 8.70 | 0.68 | 6.68 | 1.54 | 8.20 | 0.66 | 6.37 | 1.44 |
| 1165 | 9.16 | 0.70 | 6.89 | 1.55 | 8.69 | 0.67 | 6.67 | 1.53 | 8.19 | 0.65 | 6.36 | 1.44 |
| 1166 | 9.15 | 0.69 | 6.89 | 1.54 | 8.67 | 0.67 | 6.67 | 1.53 | 8.18 | 0.65 | 6.36 | 1.43 |
| 1167 | 9.13 | 0.69 | 6.88 | 1.53 | 8.66 | 0.66 | 6.67 | 1.52 | 8.17 | 0.64 | 6.36 | 1.43 |
| 1168 | 9.11 | 0.68 | 6.88 | 1.53 | 8.65 | 0.66 | 6.66 | 1.51 | 8.16 | 0.64 | 6.36 | 1.42 |
| 1169 | 9.10 | 0.68 | 6.88 | 1.52 | 8.63 | 0.65 | 6.66 | 1.51 | 8.14 | 0.64 | 6.35 | 1.41 |
| 1170 | 9.08 | 0.67 | 6.87 | 1.51 | 8.62 | 0.65 | 6.65 | 1.50 | 8.13 | 0.63 | 6.35 | 1.41 |
| 1171 | 9.07 | 0.67 | 6.87 | 1.51 | 8.60 | 0.64 | 6.65 | 1.50 | 8.12 | 0.63 | 6.35 | 1.40 |
| 1172 | 9.05 | 0.66 | 6.86 | 1.50 | 8.59 | 0.64 | 6.65 | 1.49 | 8.11 | 0.62 | 6.34 | 1.40 |
| 1173 | 9.04 | 0.66 | 6.86 | 1.50 | 8.58 | 0.64 | 6.64 | 1.48 | 8.10 | 0.62 | 6.34 | 1.39 |
| 1174 | 9.02 | 0.65 | 6.85 | 1.49 | 8.57 | 0.63 | 6.64 | 1.48 | 8.09 | 0.61 | 6.34 | 1.39 |
| 1175 | 9.01 | 0.65 | 6.85 | 1.48 | 8.55 | 0.63 | 6.64 | 1.47 | 8.08 | 0.61 | 6.34 | 1.38 |
| 1176 | 8.99 | 0.65 | 6.85 | 1.48 | 8.54 | 0.62 | 6.63 | 1.47 | 8.07 | 0.61 | 6.33 | 1.38 |
| 1177 | 8.98 | 0.64 | 6.84 | 1.47 | 8.53 | 0.62 | 6.63 | 1.46 | 8.06 | 0.60 | 6.33 | 1.37 |
| 1178 | 8.96 | 0.64 | 6.84 | 1.46 | 8.51 | 0.62 | 6.63 | 1.45 | 8.05 | 0.60 | 6.33 | 1.37 |
| 1179 | 8.95 | 0.63 | 6.83 | 1.46 | 8.50 | 0.61 | 6.62 | 1.45 | 8.04 | 0.59 | 6.32 | 1.36 |
| 1180 | 8.93 | 0.63 | 6.83 | 1.45 | 8.49 | 0.61 | 6.62 | 1.44 | 8.03 | 0.59 | 6.32 | 1.36 |
| 1181 | 8.92 | 0.62 | 6.83 | 1.45 | 8.48 | 0.60 | 6.62 | 1.44 | 8.02 | 0.59 | 6.32 | 1.35 |
| 1182 | 8.91 | 0.62 | 6.82 | 1.44 | 8.47 | 0.60 | 6.61 | 1.43 | 8.01 | 0.58 | 6.32 | 1.35 |
| 1183 | 8.89 | 0.62 | 6.82 | 1.44 | 8.45 | 0.60 | 6.61 | 1.43 | 8.00 | 0.58 | 6.31 | 1.34 |
| 1184 | 8.88 | 0.61 | 6.81 | 1.43 | 8.44 | 0.59 | 6.61 | 1.42 | 7.99 | 0.58 | 6.31 | 1.34 |
| 1185 | 8.87 | 0.61 | 6.81 | 1.42 | 8.43 | 0.59 | 6.60 | 1.41 | 7.98 | 0.57 | 6.31 | 1.33 |
| 1186 | 8.85 | 0.60 | 6.80 | 1.42 | 8.42 | 0.58 | 6.60 | 1.41 | 7.97 | 0.57 | 6.31 | 1.33 |
| 1187 | 8.84 | 0.60 | 6.80 | 1.41 | 8.41 | 0.58 | 6.60 | 1.40 | 7.96 | 0.57 | 6.30 | 1.32 |
| 1188 | 8.83 | 0.60 | 6.80 | 1.41 | 8.40 | 0.58 | 6.59 | 1.40 | 7.95 | 0.56 | 6.30 | 1.32 |
| 1189 | 8.81 | 0.59 | 6.79 | 1.40 | 8.38 | 0.57 | 6.59 | 1.39 | 7.94 | 0.56 | 6.30 | 1.31 |
| 1190 | 8.80 | 0.59 | 6.79 | 1.40 | 8.37 | 0.57 | 6.59 | 1.39 | 7.93 | 0.56 | 6.29 | 1.31 |
| 1191 | 8.79 | 0.59 | 6.79 | 1.39 | 8.36 | 0.57 | 6.58 | 1.38 | 7.92 | 0.55 | 6.29 | 1.30 |
| 1192 | 8.77 | 0.58 | 6.78 | 1.38 | 8.35 | 0.56 | 6.58 | 1.38 | 7.91 | 0.55 | 6.29 | 1.30 |
| 1193 | 8.76 | 0.58 | 6.78 | 1.38 | 8.34 | 0.56 | 6.58 | 1.37 | 7.90 | 0.55 | 6.29 | 1.29 |
| 1194 | 8.75 | 0.57 | 6.77 | 1.37 | 8.33 | 0.56 | 6.57 | 1.37 | 7.89 | 0.54 | 6.28 | 1.29 |
| 1195 | 8.74 | 0.57 | 6.77 | 1.37 | 8.32 | 0.55 | 6.57 | 1.36 | 7.88 | 0.54 | 6.28 | 1.28 |
| 1196 | 8.73 | 0.57 | 6.77 | 1.36 | 8.31 | 0.55 | 6.57 | 1.36 | 7.87 | 0.54 | 6.28 | 1.28 |
| 1197 | 8.71 | 0.56 | 6.76 | 1.36 | 8.30 | 0.55 | 6.56 | 1.35 | 7.87 | 0.53 | 6.28 | 1.27 |

| | | | | | | | | | | | |
|---|---|---|---|---|---|---|---|---|---|---|---|
| 1198 | 8.70 | 0.56 | 6.76 | 1.35 | 8.29 | 0.54 | 6.56 | 1.35 | 7.86 | 0.53 | 6.27 | 1.27 |
| 1199 | 8.69 | 0.56 | 6.75 | 1.35 | 8.28 | 0.54 | 6.56 | 1.34 | 7.85 | 0.53 | 6.27 | 1.27 |
| 1200 | 8.68 | 0.55 | 6.75 | 1.34 | 8.27 | 0.54 | 6.55 | 1.34 | 7.84 | 0.53 | 6.27 | 1.26 |
| 1201 | 8.67 | 0.55 | 6.75 | 1.34 | 8.26 | 0.54 | 6.55 | 1.33 | 7.83 | 0.52 | 6.27 | 1.26 |
| 1202 | 8.66 | 0.55 | 6.74 | 1.33 | 8.25 | 0.53 | 6.55 | 1.33 | 7.82 | 0.52 | 6.26 | 1.25 |
| 1203 | 8.64 | 0.55 | 6.74 | 1.33 | 8.24 | 0.53 | 6.54 | 1.32 | 7.82 | 0.52 | 6.26 | 1.25 |
| 1204 | 8.63 | 0.54 | 6.74 | 1.32 | 8.23 | 0.53 | 6.54 | 1.32 | 7.81 | 0.51 | 6.26 | 1.24 |
| 1205 | 8.62 | 0.54 | 6.73 | 1.32 | 8.22 | 0.52 | 6.54 | 1.31 | 7.80 | 0.51 | 6.26 | 1.24 |
| 1206 | 8.61 | 0.54 | 6.73 | 1.31 | 8.21 | 0.52 | 6.53 | 1.31 | 7.79 | 0.51 | 6.25 | 1.24 |
| 1207 | 8.60 | 0.53 | 6.73 | 1.31 | 8.20 | 0.52 | 6.53 | 1.30 | 7.78 | 0.51 | 6.25 | 1.23 |
| 1208 | 8.59 | 0.53 | 6.72 | 1.30 | 8.19 | 0.52 | 6.53 | 1.30 | 7.78 | 0.50 | 6.25 | 1.23 |
| 1209 | 8.58 | 0.53 | 6.72 | 1.30 | 8.18 | 0.51 | 6.53 | 1.29 | 7.77 | 0.50 | 6.25 | 1.22 |
| 1210 | 8.57 | 0.52 | 6.71 | 1.29 | 8.17 | 0.51 | 6.52 | 1.29 | 7.76 | 0.50 | 6.24 | 1.22 |
| 1211 | 8.56 | 0.52 | 6.71 | 1.29 | 8.16 | 0.51 | 6.52 | 1.29 | 7.75 | 0.50 | 6.24 | 1.21 |
| 1212 | 8.55 | 0.52 | 6.71 | 1.28 | 8.15 | 0.50 | 6.52 | 1.28 | 7.74 | 0.49 | 6.24 | 1.21 |
| 1213 | 8.54 | 0.52 | 6.70 | 1.28 | 8.15 | 0.50 | 6.51 | 1.28 | 7.74 | 0.49 | 6.24 | 1.21 |
| 1214 | 8.53 | 0.51 | 6.70 | 1.27 | 8.14 | 0.50 | 6.51 | 1.27 | 7.73 | 0.49 | 6.23 | 1.20 |
| 1215 | 8.52 | 0.51 | 6.70 | 1.27 | 8.13 | 0.50 | 6.51 | 1.27 | 7.72 | 0.49 | 6.23 | 1.20 |
| 1216 | 8.51 | 0.51 | 6.69 | 1.26 | 8.12 | 0.49 | 6.50 | 1.26 | 7.71 | 0.48 | 6.23 | 1.19 |
| 1217 | 8.50 | 0.51 | 6.69 | 1.26 | 8.11 | 0.49 | 6.50 | 1.26 | 7.71 | 0.48 | 6.23 | 1.19 |
| 1218 | 8.49 | 0.50 | 6.69 | 1.25 | 8.10 | 0.49 | 6.50 | 1.25 | 7.70 | 0.48 | 6.22 | 1.19 |
| 1219 | 8.48 | 0.50 | 6.68 | 1.25 | 8.09 | 0.49 | 6.50 | 1.25 | 7.69 | 0.48 | 6.22 | 1.18 |
| 1220 | 8.47 | 0.50 | 6.68 | 1.25 | 8.08 | 0.48 | 6.49 | 1.25 | 7.69 | 0.47 | 6.22 | 1.18 |
| 1221 | 8.46 | 0.50 | 6.68 | 1.24 | 8.08 | 0.48 | 6.49 | 1.24 | 7.68 | 0.47 | 6.22 | 1.17 |
| 1222 | 8.45 | 0.49 | 6.67 | 1.24 | 8.07 | 0.48 | 6.49 | 1.24 | 7.67 | 0.47 | 6.21 | 1.17 |
| 1223 | 8.44 | 0.49 | 6.67 | 1.23 | 8.06 | 0.48 | 6.48 | 1.23 | 7.66 | 0.47 | 6.21 | 1.17 |
| 1224 | 8.43 | 0.49 | 6.67 | 1.23 | 8.05 | 0.47 | 6.48 | 1.23 | 7.66 | 0.47 | 6.21 | 1.16 |
| 1225 | 8.42 | 0.49 | 6.66 | 1.22 | 8.04 | 0.47 | 6.48 | 1.22 | 7.65 | 0.46 | 6.21 | 1.16 |
| 1226 | 8.41 | 0.48 | 6.66 | 1.22 | 8.04 | 0.47 | 6.48 | 1.22 | 7.64 | 0.46 | 6.20 | 1.15 |
| 1227 | 8.40 | 0.48 | 6.66 | 1.22 | 8.03 | 0.47 | 6.47 | 1.22 | 7.64 | 0.46 | 6.20 | 1.15 |
| 1228 | 8.39 | 0.48 | 6.65 | 1.21 | 8.02 | 0.47 | 6.47 | 1.21 | 7.63 | 0.46 | 6.20 | 1.15 |
| 1229 | 8.38 | 0.48 | 6.65 | 1.21 | 8.01 | 0.46 | 6.47 | 1.21 | 7.62 | 0.45 | 6.20 | 1.14 |
| 1230 | 8.37 | 0.47 | 6.65 | 1.20 | 8.00 | 0.46 | 6.46 | 1.20 | 7.62 | 0.45 | 6.20 | 1.14 |
| 1231 | 8.37 | 0.47 | 6.64 | 1.20 | 8.00 | 0.46 | 6.46 | 1.20 | 7.61 | 0.45 | 6.19 | 1.14 |
| 1232 | 8.36 | 0.47 | 6.64 | 1.19 | 7.99 | 0.46 | 6.46 | 1.20 | 7.60 | 0.45 | 6.19 | 1.13 |
| 1233 | 8.35 | 0.47 | 6.64 | 1.19 | 7.98 | 0.46 | 6.46 | 1.19 | 7.60 | 0.45 | 6.19 | 1.13 |
| 1234 | 8.34 | 0.46 | 6.63 | 1.19 | 7.97 | 0.45 | 6.45 | 1.19 | 7.59 | 0.44 | 6.19 | 1.12 |
| 1235 | 8.33 | 0.46 | 6.63 | 1.18 | 7.97 | 0.45 | 6.45 | 1.18 | 7.59 | 0.44 | 6.18 | 1.12 |
| 1236 | 8.32 | 0.46 | 6.63 | 1.18 | 7.96 | 0.45 | 6.45 | 1.18 | 7.58 | 0.44 | 6.18 | 1.12 |
| 1237 | 8.31 | 0.46 | 6.62 | 1.17 | 7.95 | 0.45 | 6.45 | 1.18 | 7.57 | 0.44 | 6.18 | 1.11 |
| 1238 | 8.31 | 0.46 | 6.62 | 1.17 | 7.94 | 0.45 | 6.44 | 1.17 | 7.57 | 0.44 | 6.18 | 1.11 |
| 1239 | 8.30 | 0.45 | 6.62 | 1.17 | 7.94 | 0.44 | 6.44 | 1.17 | 7.56 | 0.44 | 6.18 | 1.11 |
| 1240 | 8.29 | 0.45 | 6.61 | 1.16 | 7.93 | 0.44 | 6.44 | 1.16 | 7.55 | 0.43 | 6.17 | 1.10 |
| 1241 | 8.28 | 0.45 | 6.61 | 1.16 | 7.92 | 0.44 | 6.43 | 1.16 | 7.55 | 0.43 | 6.17 | 1.10 |
| 1242 | 8.27 | 0.45 | 6.61 | 1.15 | 7.92 | 0.44 | 6.43 | 1.16 | 7.54 | 0.43 | 6.17 | 1.10 |
| 1243 | 8.26 | 0.45 | 6.61 | 1.15 | 7.91 | 0.44 | 6.43 | 1.15 | 7.54 | 0.43 | 6.17 | 1.09 |
| 1244 | 8.26 | 0.44 | 6.60 | 1.15 | 7.90 | 0.43 | 6.43 | 1.15 | 7.53 | 0.43 | 6.16 | 1.09 |
| 1245 | 8.25 | 0.44 | 6.60 | 1.14 | 7.89 | 0.43 | 6.42 | 1.15 | 7.53 | 0.42 | 6.16 | 1.09 |
| 1246 | 8.24 | 0.44 | 6.60 | 1.14 | 7.89 | 0.43 | 6.42 | 1.14 | 7.52 | 0.42 | 6.16 | 1.08 |
| 1247 | 8.23 | 0.44 | 6.59 | 1.13 | 7.88 | 0.43 | 6.42 | 1.14 | 7.51 | 0.42 | 6.16 | 1.08 |
| 1248 | 8.23 | 0.44 | 6.59 | 1.13 | 7.87 | 0.43 | 6.42 | 1.13 | 7.51 | 0.42 | 6.16 | 1.08 |
| 1249 | 8.22 | 0.43 | 6.59 | 1.13 | 7.87 | 0.42 | 6.41 | 1.13 | 7.50 | 0.42 | 6.15 | 1.07 |
| 1250 | 8.21 | 0.43 | 6.58 | 1.12 | 7.86 | 0.42 | 6.41 | 1.13 | 7.50 | 0.42 | 6.15 | 1.07 |
| 1251 | 8.20 | 0.43 | 6.58 | 1.12 | 7.85 | 0.42 | 6.41 | 1.12 | 7.49 | 0.41 | 6.15 | 1.07 |
| 1252 | 8.20 | 0.43 | 6.58 | 1.12 | 7.85 | 0.42 | 6.41 | 1.12 | 7.49 | 0.41 | 6.15 | 1.06 |
| 1253 | 8.19 | 0.43 | 6.57 | 1.11 | 7.84 | 0.42 | 6.40 | 1.12 | 7.48 | 0.41 | 6.14 | 1.06 |
| 1254 | 8.18 | 0.43 | 6.57 | 1.11 | 7.83 | 0.42 | 6.40 | 1.11 | 7.47 | 0.41 | 6.14 | 1.06 |
| 1255 | 8.17 | 0.42 | 6.57 | 1.11 | 7.83 | 0.41 | 6.40 | 1.11 | 7.47 | 0.41 | 6.14 | 1.05 |
| 1256 | 8.17 | 0.42 | 6.57 | 1.10 | 7.82 | 0.41 | 6.40 | 1.11 | 7.46 | 0.41 | 6.14 | 1.05 |
| 1257 | 8.16 | 0.42 | 6.56 | 1.10 | 7.82 | 0.41 | 6.39 | 1.10 | 7.46 | 0.40 | 6.14 | 1.05 |
| 1258 | 8.15 | 0.42 | 6.56 | 1.09 | 7.81 | 0.41 | 6.39 | 1.10 | 7.45 | 0.40 | 6.13 | 1.04 |
| 1259 | 8.14 | 0.42 | 6.56 | 1.09 | 7.80 | 0.41 | 6.39 | 1.10 | 7.45 | 0.40 | 6.13 | 1.04 |
| 1260 | 8.14 | 0.42 | 6.55 | 1.09 | 7.80 | 0.41 | 6.39 | 1.09 | 7.44 | 0.40 | 6.13 | 1.04 |
| 1261 | 8.13 | 0.41 | 6.55 | 1.08 | 7.79 | 0.40 | 6.38 | 1.09 | 7.44 | 0.40 | 6.13 | 1.03 |
| 1262 | 8.12 | 0.41 | 6.55 | 1.08 | 7.78 | 0.40 | 6.38 | 1.09 | 7.43 | 0.40 | 6.13 | 1.03 |
| 1263 | 8.12 | 0.41 | 6.55 | 1.08 | 7.78 | 0.40 | 6.38 | 1.08 | 7.43 | 0.39 | 6.12 | 1.03 |
| 1264 | 8.11 | 0.41 | 6.54 | 1.07 | 7.77 | 0.40 | 6.38 | 1.08 | 7.42 | 0.39 | 6.12 | 1.03 |
| 1265 | 8.10 | 0.41 | 6.54 | 1.07 | 7.77 | 0.40 | 6.37 | 1.08 | 7.42 | 0.39 | 6.12 | 1.02 |
| 1266 | 8.10 | 0.41 | 6.54 | 1.07 | 7.76 | 0.40 | 6.37 | 1.07 | 7.41 | 0.39 | 6.12 | 1.02 |
| 1267 | 8.09 | 0.40 | 6.53 | 1.06 | 7.75 | 0.40 | 6.37 | 1.07 | 7.41 | 0.39 | 6.12 | 1.02 |
| 1268 | 8.08 | 0.40 | 6.53 | 1.06 | 7.75 | 0.39 | 6.37 | 1.07 | 7.40 | 0.39 | 6.11 | 1.01 |
| 1269 | 8.08 | 0.40 | 6.53 | 1.06 | 7.74 | 0.39 | 6.36 | 1.06 | 7.40 | 0.39 | 6.11 | 1.01 |
| 1270 | 8.07 | 0.40 | 6.53 | 1.05 | 7.74 | 0.39 | 6.36 | 1.06 | 7.39 | 0.38 | 6.11 | 1.01 |
| 1271 | 8.06 | 0.40 | 6.52 | 1.05 | 7.73 | 0.39 | 6.36 | 1.06 | 7.39 | 0.38 | 6.11 | 1.00 |
| 1272 | 8.06 | 0.40 | 6.52 | 1.05 | 7.73 | 0.39 | 6.36 | 1.05 | 7.38 | 0.38 | 6.11 | 1.00 |
| 1273 | 8.05 | 0.39 | 6.52 | 1.04 | 7.72 | 0.39 | 6.35 | 1.05 | 7.38 | 0.38 | 6.10 | 1.00 |

| | | | | | | | | | | | |
|---|---|---|---|---|---|---|---|---|---|---|---|
| 1274 | 8.04 | 0.39 | 6.51 | 1.04 | 7.71 | 0.38 | 6.35 | 1.05 | 7.37 | 0.38 | 6.10 | 1.00 |
| 1275 | 8.04 | 0.39 | 6.51 | 1.04 | 7.71 | 0.38 | 6.35 | 1.04 | 7.37 | 0.38 | 6.10 | 0.99 |
| 1276 | 8.03 | 0.39 | 6.51 | 1.03 | 7.70 | 0.38 | 6.35 | 1.04 | 7.36 | 0.38 | 6.10 | 0.99 |
| 1277 | 8.02 | 0.39 | 6.51 | 1.03 | 7.70 | 0.38 | 6.34 | 1.04 | 7.36 | 0.38 | 6.10 | 0.99 |
| 1278 | 8.02 | 0.39 | 6.50 | 1.03 | 7.69 | 0.38 | 6.34 | 1.03 | 7.35 | 0.37 | 6.09 | 0.98 |
| 1279 | 8.01 | 0.39 | 6.50 | 1.02 | 7.69 | 0.38 | 6.34 | 1.03 | 7.35 | 0.37 | 6.09 | 0.98 |
| 1280 | 8.01 | 0.38 | 6.50 | 1.02 | 7.68 | 0.38 | 6.34 | 1.03 | 7.34 | 0.37 | 6.09 | 0.98 |
| 1281 | 8.00 | 0.38 | 6.50 | 1.02 | 7.68 | 0.38 | 6.33 | 1.03 | 7.34 | 0.37 | 6.09 | 0.98 |
| 1282 | 7.99 | 0.38 | 6.49 | 1.02 | 7.67 | 0.37 | 6.33 | 1.02 | 7.34 | 0.37 | 6.09 | 0.97 |
| 1283 | 7.99 | 0.38 | 6.49 | 1.01 | 7.67 | 0.37 | 6.33 | 1.02 | 7.33 | 0.37 | 6.08 | 0.97 |
| 1284 | 7.98 | 0.38 | 6.49 | 1.01 | 7.66 | 0.37 | 6.33 | 1.02 | 7.33 | 0.37 | 6.08 | 0.97 |
| 1285 | 7.97 | 0.38 | 6.49 | 1.01 | 7.65 | 0.37 | 6.33 | 1.01 | 7.32 | 0.36 | 6.08 | 0.96 |
| 1286 | 7.97 | 0.38 | 6.48 | 1.00 | 7.65 | 0.37 | 6.32 | 1.01 | 7.32 | 0.36 | 6.08 | 0.96 |
| 1287 | 7.96 | 0.37 | 6.48 | 1.00 | 7.64 | 0.37 | 6.32 | 1.01 | 7.31 | 0.36 | 6.08 | 0.96 |
| 1288 | 7.96 | 0.37 | 6.48 | 1.00 | 7.64 | 0.37 | 6.32 | 1.00 | 7.31 | 0.36 | 6.07 | 0.96 |
| 1289 | 7.95 | 0.37 | 6.47 | 0.99 | 7.63 | 0.36 | 6.32 | 1.00 | 7.30 | 0.36 | 6.07 | 0.95 |
| 1290 | 7.95 | 0.37 | 6.47 | 0.99 | 7.63 | 0.36 | 6.31 | 1.00 | 7.30 | 0.36 | 6.07 | 0.95 |
| 1291 | 7.94 | 0.37 | 6.47 | 0.99 | 7.62 | 0.36 | 6.31 | 1.00 | 7.30 | 0.36 | 6.07 | 0.95 |
| 1292 | 7.93 | 0.37 | 6.47 | 0.99 | 7.62 | 0.36 | 6.31 | 0.99 | 7.29 | 0.36 | 6.07 | 0.95 |
| 1293 | 7.93 | 0.37 | 6.46 | 0.98 | 7.61 | 0.36 | 6.31 | 0.99 | 7.29 | 0.36 | 6.06 | 0.94 |
| 1294 | 7.92 | 0.37 | 6.46 | 0.98 | 7.61 | 0.36 | 6.31 | 0.99 | 7.28 | 0.35 | 6.06 | 0.94 |
| 1295 | 7.92 | 0.36 | 6.46 | 0.98 | 7.60 | 0.36 | 6.30 | 0.98 | 7.28 | 0.35 | 6.06 | 0.94 |
| 1296 | 7.91 | 0.36 | 6.46 | 0.97 | 7.60 | 0.36 | 6.30 | 0.98 | 7.28 | 0.35 | 6.06 | 0.94 |
| 1297 | 7.91 | 0.36 | 6.45 | 0.97 | 7.59 | 0.36 | 6.30 | 0.98 | 7.27 | 0.35 | 6.06 | 0.93 |
| 1298 | 7.90 | 0.36 | 6.45 | 0.97 | 7.59 | 0.35 | 6.30 | 0.98 | 7.27 | 0.35 | 6.06 | 0.93 |
| 1299 | 7.90 | 0.36 | 6.45 | 0.97 | 7.59 | 0.35 | 6.29 | 0.97 | 7.26 | 0.35 | 6.05 | 0.93 |
| 1300 | 7.89 | 0.36 | 6.45 | 0.96 | 7.58 | 0.35 | 6.29 | 0.97 | 7.26 | 0.35 | 6.05 | 0.93 |
| 1301 | 7.88 | 0.36 | 6.44 | 0.96 | 7.58 | 0.35 | 6.29 | 0.97 | 7.25 | 0.35 | 6.05 | 0.92 |
| 1302 | 7.88 | 0.36 | 6.44 | 0.96 | 7.57 | 0.35 | 6.29 | 0.97 | 7.25 | 0.35 | 6.05 | 0.92 |
| 1303 | 7.87 | 0.35 | 6.44 | 0.95 | 7.57 | 0.35 | 6.29 | 0.96 | 7.25 | 0.34 | 6.05 | 0.92 |
| 1304 | 7.87 | 0.35 | 6.44 | 0.95 | 7.56 | 0.35 | 6.28 | 0.96 | 7.24 | 0.34 | 6.04 | 0.92 |
| 1305 | 7.86 | 0.35 | 6.43 | 0.95 | 7.56 | 0.35 | 6.28 | 0.96 | 7.24 | 0.34 | 6.04 | 0.91 |
| 1306 | 7.86 | 0.35 | 6.43 | 0.95 | 7.55 | 0.34 | 6.28 | 0.95 | 7.23 | 0.34 | 6.04 | 0.91 |
| 1307 | 7.85 | 0.35 | 6.43 | 0.94 | 7.55 | 0.34 | 6.28 | 0.95 | 7.23 | 0.34 | 6.04 | 0.91 |
| 1308 | 7.85 | 0.35 | 6.43 | 0.94 | 7.54 | 0.34 | 6.27 | 0.95 | 7.23 | 0.34 | 6.04 | 0.91 |
| 1309 | 7.84 | 0.35 | 6.42 | 0.94 | 7.54 | 0.34 | 6.27 | 0.95 | 7.22 | 0.34 | 6.04 | 0.90 |
| 1310 | 7.84 | 0.35 | 6.42 | 0.94 | 7.53 | 0.34 | 6.27 | 0.94 | 7.22 | 0.34 | 6.03 | 0.90 |
| 1311 | 7.83 | 0.35 | 6.42 | 0.93 | 7.53 | 0.34 | 6.27 | 0.94 | 7.22 | 0.34 | 6.03 | 0.90 |
| 1312 | 7.83 | 0.34 | 6.42 | 0.93 | 7.53 | 0.34 | 6.27 | 0.94 | 7.21 | 0.33 | 6.03 | 0.90 |
| 1313 | 7.82 | 0.34 | 6.42 | 0.93 | 7.52 | 0.34 | 6.26 | 0.94 | 7.21 | 0.33 | 6.03 | 0.89 |
| 1314 | 7.82 | 0.34 | 6.41 | 0.92 | 7.52 | 0.34 | 6.26 | 0.93 | 7.20 | 0.33 | 6.03 | 0.89 |
| 1315 | 7.81 | 0.34 | 6.41 | 0.92 | 7.51 | 0.34 | 6.26 | 0.93 | 7.20 | 0.33 | 6.02 | 0.89 |
| 1316 | 7.81 | 0.34 | 6.41 | 0.92 | 7.51 | 0.33 | 6.26 | 0.93 | 7.20 | 0.33 | 6.02 | 0.89 |
| 1317 | 7.80 | 0.34 | 6.41 | 0.92 | 7.50 | 0.33 | 6.26 | 0.93 | 7.19 | 0.33 | 6.02 | 0.88 |
| 1318 | 7.80 | 0.34 | 6.40 | 0.91 | 7.50 | 0.33 | 6.25 | 0.92 | 7.19 | 0.33 | 6.02 | 0.88 |
| 1319 | 7.79 | 0.34 | 6.40 | 0.91 | 7.50 | 0.33 | 6.25 | 0.92 | 7.19 | 0.33 | 6.02 | 0.88 |
| 1320 | 7.79 | 0.34 | 6.40 | 0.91 | 7.49 | 0.33 | 6.25 | 0.92 | 7.18 | 0.33 | 6.02 | 0.88 |
| 1321 | 7.78 | 0.33 | 6.40 | 0.91 | 7.49 | 0.33 | 6.25 | 0.92 | 7.18 | 0.33 | 6.01 | 0.88 |
| 1322 | 7.78 | 0.33 | 6.39 | 0.90 | 7.48 | 0.33 | 6.25 | 0.91 | 7.18 | 0.32 | 6.01 | 0.87 |
| 1323 | 7.77 | 0.33 | 6.39 | 0.90 | 7.48 | 0.33 | 6.24 | 0.91 | 7.17 | 0.32 | 6.01 | 0.87 |
| 1324 | 7.77 | 0.33 | 6.39 | 0.90 | 7.47 | 0.33 | 6.24 | 0.91 | 7.17 | 0.32 | 6.01 | 0.87 |
| 1325 | 7.76 | 0.33 | 6.39 | 0.90 | 7.47 | 0.33 | 6.24 | 0.91 | 7.17 | 0.32 | 6.01 | 0.87 |
| 1326 | 7.76 | 0.33 | 6.38 | 0.89 | 7.47 | 0.32 | 6.24 | 0.90 | 7.16 | 0.32 | 6.01 | 0.86 |
| 1327 | 7.76 | 0.33 | 6.38 | 0.89 | 7.46 | 0.32 | 6.24 | 0.90 | 7.16 | 0.32 | 6.00 | 0.86 |
| 1328 | 7.75 | 0.33 | 6.38 | 0.89 | 7.46 | 0.32 | 6.23 | 0.90 | 7.15 | 0.32 | 6.00 | 0.86 |
| 1329 | 7.75 | 0.33 | 6.38 | 0.89 | 7.45 | 0.32 | 6.23 | 0.90 | 7.15 | 0.32 | 6.00 | 0.86 |
| 1330 | 7.74 | 0.33 | 6.38 | 0.89 | 7.45 | 0.32 | 6.23 | 0.89 | 7.15 | 0.32 | 6.00 | 0.86 |
| 1331 | 7.74 | 0.32 | 6.37 | 0.88 | 7.45 | 0.32 | 6.23 | 0.89 | 7.14 | 0.32 | 6.00 | 0.85 |
| 1332 | 7.73 | 0.32 | 6.37 | 0.88 | 7.44 | 0.32 | 6.23 | 0.89 | 7.14 | 0.32 | 6.00 | 0.85 |
| 1333 | 7.73 | 0.32 | 6.37 | 0.88 | 7.44 | 0.32 | 6.22 | 0.89 | 7.14 | 0.31 | 5.99 | 0.85 |
| 1334 | 7.72 | 0.32 | 6.37 | 0.88 | 7.43 | 0.32 | 6.22 | 0.89 | 7.13 | 0.31 | 5.99 | 0.85 |
| 1335 | 7.72 | 0.32 | 6.36 | 0.87 | 7.43 | 0.32 | 6.22 | 0.88 | 7.13 | 0.31 | 5.99 | 0.84 |
| 1336 | 7.72 | 0.32 | 6.36 | 0.87 | 7.43 | 0.31 | 6.22 | 0.88 | 7.13 | 0.31 | 5.99 | 0.84 |
| 1337 | 7.71 | 0.32 | 6.36 | 0.87 | 7.42 | 0.31 | 6.22 | 0.88 | 7.12 | 0.31 | 5.99 | 0.84 |
| 1338 | 7.71 | 0.32 | 6.36 | 0.87 | 7.42 | 0.31 | 6.21 | 0.88 | 7.12 | 0.31 | 5.99 | 0.84 |
| 1339 | 7.70 | 0.32 | 6.36 | 0.86 | 7.42 | 0.31 | 6.21 | 0.87 | 7.12 | 0.31 | 5.98 | 0.84 |
| 1340 | 7.70 | 0.32 | 6.35 | 0.86 | 7.41 | 0.31 | 6.21 | 0.87 | 7.12 | 0.31 | 5.98 | 0.83 |
| 1341 | 7.69 | 0.32 | 6.35 | 0.86 | 7.41 | 0.31 | 6.21 | 0.87 | 7.11 | 0.31 | 5.98 | 0.83 |
| 1342 | 7.69 | 0.31 | 6.35 | 0.86 | 7.40 | 0.31 | 6.21 | 0.87 | 7.11 | 0.31 | 5.98 | 0.83 |
| 1343 | 7.69 | 0.31 | 6.35 | 0.86 | 7.40 | 0.31 | 6.20 | 0.87 | 7.11 | 0.31 | 5.98 | 0.83 |
| 1344 | 7.68 | 0.31 | 6.34 | 0.85 | 7.40 | 0.31 | 6.20 | 0.86 | 7.10 | 0.31 | 5.98 | 0.83 |
| 1345 | 7.68 | 0.31 | 6.34 | 0.85 | 7.39 | 0.31 | 6.20 | 0.86 | 7.10 | 0.30 | 5.97 | 0.82 |
| 1346 | 7.67 | 0.31 | 6.34 | 0.85 | 7.39 | 0.31 | 6.20 | 0.86 | 7.10 | 0.30 | 5.97 | 0.82 |
| 1347 | 7.67 | 0.31 | 6.34 | 0.85 | 7.39 | 0.31 | 6.20 | 0.86 | 7.09 | 0.30 | 5.97 | 0.82 |
| 1348 | 7.66 | 0.31 | 6.34 | 0.84 | 7.38 | 0.30 | 6.19 | 0.85 | 7.09 | 0.30 | 5.97 | 0.82 |
| 1349 | 7.66 | 0.31 | 6.33 | 0.84 | 7.38 | 0.30 | 6.19 | 0.85 | 7.09 | 0.30 | 5.97 | 0.82 |
| 1350 | 7.66 | 0.31 | 6.33 | 0.84 | 7.38 | 0.30 | 6.19 | 0.85 | 7.08 | 0.30 | 5.97 | 0.81 |
| 1351 | 7.65 | 0.31 | 6.33 | 0.84 | 7.37 | 0.30 | 6.19 | 0.85 | 7.08 | 0.30 | 5.96 | 0.81 |
| 1352 | 7.65 | 0.31 | 6.33 | 0.84 | 7.37 | 0.30 | 6.19 | 0.85 | 7.08 | 0.30 | 5.96 | 0.81 |
| 1353 | 7.64 | 0.30 | 6.33 | 0.83 | 7.37 | 0.30 | 6.19 | 0.84 | 7.07 | 0.30 | 5.96 | 0.81 |
| 1354 | 7.64 | 0.30 | 6.32 | 0.83 | 7.36 | 0.30 | 6.18 | 0.84 | 7.07 | 0.30 | 5.96 | 0.81 |
| 1355 | 7.64 | 0.30 | 6.32 | 0.83 | 7.36 | 0.30 | 6.18 | 0.84 | 7.07 | 0.30 | 5.96 | 0.80 |
| 1356 | 7.63 | 0.30 | 6.32 | 0.83 | 7.35 | 0.30 | 6.18 | 0.84 | 7.07 | 0.30 | 5.96 | 0.80 |
| 1357 | 7.63 | 0.30 | 6.32 | 0.82 | 7.35 | 0.30 | 6.18 | 0.84 | 7.06 | 0.29 | 5.96 | 0.80 |
| 1358 | 7.63 | 0.30 | 6.32 | 0.82 | 7.35 | 0.30 | 6.18 | 0.83 | 7.06 | 0.29 | 5.95 | 0.80 |
| 1359 | 7.62 | 0.30 | 6.31 | 0.82 | 7.34 | 0.30 | 6.17 | 0.83 | 7.06 | 0.29 | 5.95 | 0.80 |
| 1360 | 7.62 | 0.30 | 6.31 | 0.82 | 7.34 | 0.30 | 6.17 | 0.83 | 7.05 | 0.29 | 5.95 | 0.79 |
| 1361 | 7.61 | 0.30 | 6.31 | 0.82 | 7.34 | 0.29 | 6.17 | 0.83 | 7.05 | 0.29 | 5.95 | 0.79 |
| 1362 | 7.61 | 0.30 | 6.31 | 0.81 | 7.33 | 0.29 | 6.17 | 0.83 | 7.05 | 0.29 | 5.95 | 0.79 |
| 1363 | 7.61 | 0.30 | 6.31 | 0.81 | 7.33 | 0.29 | 6.17 | 0.82 | 7.05 | 0.29 | 5.95 | 0.79 |
| 1364 | 7.60 | 0.30 | 6.30 | 0.81 | 7.33 | 0.29 | 6.17 | 0.82 | 7.04 | 0.29 | 5.94 | 0.79 |
| 1365 | 7.60 | 0.30 | 6.30 | 0.81 | 7.32 | 0.29 | 6.16 | 0.82 | 7.04 | 0.29 | 5.94 | 0.78 |
| 1366 | 7.59 | 0.29 | 6.30 | 0.81 | 7.32 | 0.29 | 6.16 | 0.82 | 7.04 | 0.29 | 5.94 | 0.78 |
| 1367 | 7.59 | 0.29 | 6.30 | 0.80 | 7.32 | 0.29 | 6.16 | 0.82 | 7.03 | 0.29 | 5.94 | 0.78 |
| 1368 | 7.59 | 0.29 | 6.30 | 0.80 | 7.31 | 0.29 | 6.16 | 0.81 | 7.03 | 0.29 | 5.94 | 0.78 |
| 1369 | 7.58 | 0.29 | 6.29 | 0.80 | 7.31 | 0.29 | 6.16 | 0.81 | 7.03 | 0.29 | 5.94 | 0.78 |
| 1370 | 7.58 | 0.29 | 6.29 | 0.80 | 7.31 | 0.29 | 6.16 | 0.81 | 7.03 | 0.29 | 5.94 | 0.78 |
| 1371 | 7.58 | 0.29 | 6.29 | 0.80 | 7.31 | 0.29 | 6.15 | 0.81 | 7.02 | 0.28 | 5.93 | 0.77 |
| 1372 | 7.57 | 0.29 | 6.29 | 0.79 | 7.30 | 0.29 | 6.15 | 0.81 | 7.02 | 0.28 | 5.93 | 0.77 |
| 1373 | 7.57 | 0.29 | 6.29 | 0.79 | 7.30 | 0.29 | 6.15 | 0.80 | 7.02 | 0.28 | 5.93 | 0.77 |
| 1374 | 7.57 | 0.29 | 6.28 | 0.79 | 7.30 | 0.28 | 6.15 | 0.80 | 7.02 | 0.28 | 5.93 | 0.77 |
| 1375 | 7.56 | 0.29 | 6.28 | 0.79 | 7.29 | 0.28 | 6.15 | 0.80 | 7.01 | 0.28 | 5.93 | 0.77 |
| 1376 | 7.56 | 0.29 | 6.28 | 0.79 | 7.29 | 0.28 | 6.14 | 0.80 | 7.01 | 0.28 | 5.93 | 0.76 |
| 1377 | 7.56 | 0.29 | 6.28 | 0.78 | 7.29 | 0.28 | 6.14 | 0.80 | 7.01 | 0.28 | 5.93 | 0.76 |
| 1378 | 7.55 | 0.29 | 6.28 | 0.78 | 7.28 | 0.28 | 6.14 | 0.79 | 7.00 | 0.28 | 5.92 | 0.76 |
| 1379 | 7.55 | 0.28 | 6.27 | 0.78 | 7.28 | 0.28 | 6.14 | 0.79 | 7.00 | 0.28 | 5.92 | 0.76 |
| 1380 | 7.54 | 0.28 | 6.27 | 0.78 | 7.28 | 0.28 | 6.14 | 0.79 | 7.00 | 0.28 | 5.92 | 0.76 |
| 1381 | 7.54 | 0.28 | 6.27 | 0.78 | 7.27 | 0.28 | 6.14 | 0.79 | 7.00 | 0.28 | 5.92 | 0.76 |
| 1382 | 7.54 | 0.28 | 6.27 | 0.78 | 7.27 | 0.28 | 6.13 | 0.79 | 6.99 | 0.28 | 5.92 | 0.75 |
| 1383 | 7.53 | 0.28 | 6.27 | 0.77 | 7.27 | 0.28 | 6.13 | 0.79 | 6.99 | 0.28 | 5.92 | 0.75 |
| 1384 | 7.53 | 0.28 | 6.26 | 0.77 | 7.26 | 0.28 | 6.13 | 0.78 | 6.99 | 0.28 | 5.92 | 0.75 |
| 1385 | 7.53 | 0.28 | 6.26 | 0.77 | 7.26 | 0.28 | 6.13 | 0.78 | 6.99 | 0.28 | 5.91 | 0.75 |
| 1386 | 7.52 | 0.28 | 6.26 | 0.77 | 7.26 | 0.28 | 6.13 | 0.78 | 6.98 | 0.27 | 5.91 | 0.75 |
| 1387 | 7.52 | 0.28 | 6.26 | 0.77 | 7.26 | 0.28 | 6.13 | 0.78 | 6.98 | 0.27 | 5.91 | 0.75 |

| | | | | | | | | | | | |
|---|---|---|---|---|---|---|---|---|---|---|---|
| 1388 | 7.52 | 0.28 | 6.26 | 0.76 | 7.25 | 0.28 | 6.12 | 0.78 | 6.98 | 0.27 | 5.91 | 0.74 |
| 1389 | 7.51 | 0.28 | 6.26 | 0.76 | 7.25 | 0.27 | 6.12 | 0.77 | 6.98 | 0.27 | 5.91 | 0.74 |
| 1390 | 7.51 | 0.28 | 6.25 | 0.76 | 7.25 | 0.27 | 6.12 | 0.77 | 6.97 | 0.27 | 5.91 | 0.74 |
| 1391 | 7.51 | 0.28 | 6.25 | 0.76 | 7.24 | 0.27 | 6.12 | 0.77 | 6.97 | 0.27 | 5.91 | 0.74 |
| 1392 | 7.50 | 0.28 | 6.25 | 0.76 | 7.24 | 0.27 | 6.12 | 0.77 | 6.97 | 0.27 | 5.90 | 0.74 |
| 1393 | 7.50 | 0.28 | 6.25 | 0.76 | 7.24 | 0.27 | 6.12 | 0.77 | 6.97 | 0.27 | 5.90 | 0.74 |
| 1394 | 7.50 | 0.27 | 6.25 | 0.75 | 7.24 | 0.27 | 6.11 | 0.77 | 6.96 | 0.27 | 5.90 | 0.73 |
| 1395 | 7.49 | 0.27 | 6.24 | 0.75 | 7.23 | 0.27 | 6.11 | 0.76 | 6.96 | 0.27 | 5.90 | 0.73 |
| 1396 | 7.49 | 0.27 | 6.24 | 0.75 | 7.23 | 0.27 | 6.11 | 0.76 | 6.96 | 0.27 | 5.90 | 0.73 |
| 1397 | 7.49 | 0.27 | 6.24 | 0.75 | 7.23 | 0.27 | 6.11 | 0.76 | 6.96 | 0.27 | 5.90 | 0.73 |
| 1398 | 7.49 | 0.27 | 6.24 | 0.75 | 7.22 | 0.27 | 6.11 | 0.76 | 6.95 | 0.27 | 5.90 | 0.73 |
| 1399 | 7.48 | 0.27 | 6.24 | 0.75 | 7.22 | 0.27 | 6.11 | 0.76 | 6.95 | 0.27 | 5.89 | 0.73 |
| 1400 | 7.48 | 0.27 | 6.24 | 0.74 | 7.22 | 0.27 | 6.11 | 0.76 | 6.95 | 0.27 | 5.89 | 0.72 |
| 1401 | 7.48 | 0.27 | 6.23 | 0.74 | 7.22 | 0.27 | 6.10 | 0.75 | 6.95 | 0.27 | 5.89 | 0.72 |
| 1402 | 7.47 | 0.27 | 6.23 | 0.74 | 7.21 | 0.27 | 6.10 | 0.75 | 6.94 | 0.26 | 5.89 | 0.72 |
| 1403 | 7.47 | 0.27 | 6.23 | 0.74 | 7.21 | 0.27 | 6.10 | 0.75 | 6.94 | 0.26 | 5.89 | 0.72 |
| 1404 | 7.47 | 0.27 | 6.23 | 0.74 | 7.21 | 0.27 | 6.10 | 0.75 | 6.94 | 0.26 | 5.89 | 0.72 |
| 1405 | 7.46 | 0.27 | 6.23 | 0.74 | 7.21 | 0.26 | 6.10 | 0.75 | 6.94 | 0.26 | 5.89 | 0.72 |
| 1406 | 7.46 | 0.27 | 6.22 | 0.73 | 7.20 | 0.26 | 6.10 | 0.74 | 6.94 | 0.26 | 5.88 | 0.71 |
| 1407 | 7.46 | 0.27 | 6.22 | 0.73 | 7.20 | 0.26 | 6.09 | 0.74 | 6.93 | 0.26 | 5.88 | 0.71 |
| 1408 | 7.45 | 0.27 | 6.22 | 0.73 | 7.20 | 0.26 | 6.09 | 0.74 | 6.93 | 0.26 | 5.88 | 0.71 |
| 1409 | 7.45 | 0.26 | 6.22 | 0.73 | 7.19 | 0.26 | 6.09 | 0.74 | 6.93 | 0.26 | 5.88 | 0.71 |
| 1410 | 7.45 | 0.26 | 6.22 | 0.73 | 7.19 | 0.26 | 6.09 | 0.74 | 6.93 | 0.26 | 5.88 | 0.71 |
| 1411 | 7.45 | 0.26 | 6.22 | 0.73 | 7.19 | 0.26 | 6.09 | 0.74 | 6.92 | 0.26 | 5.88 | 0.71 |
| 1412 | 7.44 | 0.26 | 6.21 | 0.72 | 7.19 | 0.26 | 6.09 | 0.74 | 6.92 | 0.26 | 5.88 | 0.71 |
| 1413 | 7.44 | 0.26 | 6.21 | 0.72 | 7.18 | 0.26 | 6.08 | 0.73 | 6.92 | 0.26 | 5.88 | 0.70 |
| 1414 | 7.44 | 0.26 | 6.21 | 0.72 | 7.18 | 0.26 | 6.08 | 0.73 | 6.92 | 0.26 | 5.87 | 0.70 |
| 1415 | 7.43 | 0.26 | 6.21 | 0.72 | 7.18 | 0.26 | 6.08 | 0.73 | 6.92 | 0.26 | 5.87 | 0.70 |
| 1416 | 7.43 | 0.26 | 6.21 | 0.72 | 7.18 | 0.26 | 6.08 | 0.73 | 6.91 | 0.26 | 5.87 | 0.70 |
| 1417 | 7.43 | 0.26 | 6.21 | 0.72 | 7.17 | 0.26 | 6.08 | 0.73 | 6.91 | 0.26 | 5.87 | 0.70 |
| 1418 | 7.42 | 0.26 | 6.20 | 0.71 | 7.17 | 0.26 | 6.08 | 0.73 | 6.91 | 0.26 | 5.87 | 0.70 |
| 1419 | 7.42 | 0.26 | 6.20 | 0.71 | 7.17 | 0.26 | 6.08 | 0.72 | 6.91 | 0.25 | 5.87 | 0.70 |
| 1420 | 7.42 | 0.26 | 6.20 | 0.71 | 7.17 | 0.26 | 6.07 | 0.72 | 6.90 | 0.25 | 5.87 | 0.69 |
| 1421 | 7.42 | 0.26 | 6.20 | 0.71 | 7.16 | 0.26 | 6.07 | 0.72 | 6.90 | 0.25 | 5.86 | 0.69 |
| 1422 | 7.41 | 0.26 | 6.20 | 0.71 | 7.16 | 0.25 | 6.07 | 0.72 | 6.90 | 0.25 | 5.86 | 0.69 |
| 1423 | 7.41 | 0.26 | 6.20 | 0.71 | 7.16 | 0.25 | 6.07 | 0.72 | 6.90 | 0.25 | 5.86 | 0.69 |
| 1424 | 7.41 | 0.26 | 6.19 | 0.70 | 7.16 | 0.25 | 6.07 | 0.72 | 6.90 | 0.25 | 5.86 | 0.69 |
| 1425 | 7.40 | 0.26 | 6.19 | 0.70 | 7.15 | 0.25 | 6.07 | 0.71 | 6.89 | 0.25 | 5.86 | 0.69 |

| | | | | | | | | | | | |
|---|---|---|---|---|---|---|---|---|---|---|---|
| 1426 | 7.40 | 0.26 | 6.19 | 0.70 | 7.15 | 0.25 | 6.07 | 0.71 | 6.89 | 0.25 | 5.86 | 0.68 |
| 1427 | 7.40 | 0.25 | 6.19 | 0.70 | 7.15 | 0.25 | 6.06 | 0.71 | 6.89 | 0.25 | 5.86 | 0.68 |
| 1428 | 7.40 | 0.25 | 6.19 | 0.70 | 7.15 | 0.25 | 6.06 | 0.71 | 6.89 | 0.25 | 5.86 | 0.68 |
| 1429 | 7.39 | 0.25 | 6.19 | 0.70 | 7.14 | 0.25 | 6.06 | 0.71 | 6.88 | 0.25 | 5.85 | 0.68 |
| 1430 | 7.39 | 0.25 | 6.18 | 0.70 | 7.14 | 0.25 | 6.06 | 0.71 | 6.88 | 0.25 | 5.85 | 0.68 |
| 1431 | 7.39 | 0.25 | 6.18 | 0.69 | 7.14 | 0.25 | 6.06 | 0.71 | 6.88 | 0.25 | 5.85 | 0.68 |
| 1432 | 7.39 | 0.25 | 6.18 | 0.69 | 7.14 | 0.25 | 6.06 | 0.70 | 6.88 | 0.25 | 5.85 | 0.68 |
| 1433 | 7.38 | 0.25 | 6.18 | 0.69 | 7.13 | 0.25 | 6.05 | 0.70 | 6.88 | 0.25 | 5.85 | 0.68 |
| 1434 | 7.38 | 0.25 | 6.18 | 0.69 | 7.13 | 0.25 | 6.05 | 0.70 | 6.87 | 0.25 | 5.85 | 0.67 |
| 1435 | 7.38 | 0.25 | 6.18 | 0.69 | 7.13 | 0.25 | 6.05 | 0.70 | 6.87 | 0.25 | 5.85 | 0.67 |
| 1436 | 7.37 | 0.25 | 6.17 | 0.69 | 7.13 | 0.25 | 6.05 | 0.70 | 6.87 | 0.25 | 5.85 | 0.67 |
| 1437 | 7.37 | 0.25 | 6.17 | 0.69 | 7.12 | 0.25 | 6.05 | 0.70 | 6.87 | 0.25 | 5.84 | 0.67 |
| 1438 | 7.37 | 0.25 | 6.17 | 0.68 | 7.12 | 0.25 | 6.05 | 0.70 | 6.87 | 0.25 | 5.84 | 0.67 |
| 1439 | 7.37 | 0.25 | 6.17 | 0.68 | 7.12 | 0.25 | 6.05 | 0.69 | 6.86 | 0.24 | 5.84 | 0.67 |
| 1440 | 7.36 | 0.25 | 6.17 | 0.68 | 7.12 | 0.25 | 6.04 | 0.69 | 6.86 | 0.24 | 5.84 | 0.67 |
| 1441 | 7.36 | 0.25 | 6.17 | 0.68 | 7.12 | 0.24 | 6.04 | 0.69 | 6.86 | 0.24 | 5.84 | 0.66 |
| 1442 | 7.36 | 0.25 | 6.16 | 0.68 | 7.11 | 0.24 | 6.04 | 0.69 | 6.86 | 0.24 | 5.84 | 0.66 |
| 1443 | 7.36 | 0.25 | 6.16 | 0.68 | 7.11 | 0.24 | 6.04 | 0.69 | 6.86 | 0.24 | 5.84 | 0.66 |
| 1444 | 7.35 | 0.25 | 6.16 | 0.67 | 7.11 | 0.24 | 6.04 | 0.69 | 6.85 | 0.24 | 5.84 | 0.66 |
| 1445 | 7.35 | 0.25 | 6.16 | 0.67 | 7.11 | 0.24 | 6.04 | 0.69 | 6.85 | 0.24 | 5.83 | 0.66 |
| 1446 | 7.35 | 0.24 | 6.16 | 0.67 | 7.10 | 0.24 | 6.04 | 0.68 | 6.85 | 0.24 | 5.83 | 0.66 |
| 1447 | 7.35 | 0.24 | 6.16 | 0.67 | 7.10 | 0.24 | 6.03 | 0.68 | 6.85 | 0.24 | 5.83 | 0.66 |
| 1448 | 7.34 | 0.24 | 6.16 | 0.67 | 7.10 | 0.24 | 6.03 | 0.68 | 6.85 | 0.24 | 5.83 | 0.65 |
| 1449 | 7.34 | 0.24 | 6.15 | 0.67 | 7.10 | 0.24 | 6.03 | 0.68 | 6.84 | 0.24 | 5.83 | 0.65 |
| 1450 | 7.34 | 0.24 | 6.15 | 0.67 | 7.10 | 0.24 | 6.03 | 0.68 | 6.84 | 0.24 | 5.83 | 0.65 |
| 1451 | 7.34 | 0.24 | 6.15 | 0.67 | 7.09 | 0.24 | 6.03 | 0.68 | 6.84 | 0.24 | 5.83 | 0.65 |
| 1452 | 7.33 | 0.24 | 6.15 | 0.66 | 7.09 | 0.24 | 6.03 | 0.68 | 6.84 | 0.24 | 5.83 | 0.65 |
| 1453 | 7.33 | 0.24 | 6.15 | 0.66 | 7.09 | 0.24 | 6.03 | 0.67 | 6.84 | 0.24 | 5.83 | 0.65 |
| 1454 | 7.33 | 0.24 | 6.15 | 0.66 | 7.09 | 0.24 | 6.03 | 0.67 | 6.84 | 0.24 | 5.82 | 0.65 |
| 1455 | 7.33 | 0.24 | 6.14 | 0.66 | 7.08 | 0.24 | 6.02 | 0.67 | 6.83 | 0.24 | 5.82 | 0.65 |
| 1456 | 7.32 | 0.24 | 6.14 | 0.66 | 7.08 | 0.24 | 6.02 | 0.67 | 6.83 | 0.24 | 5.82 | 0.64 |
| 1457 | 7.32 | 0.24 | 6.14 | 0.66 | 7.08 | 0.24 | 6.02 | 0.67 | 6.83 | 0.24 | 5.82 | 0.64 |
| 1458 | 7.32 | 0.24 | 6.14 | 0.66 | 7.08 | 0.24 | 6.02 | 0.67 | 6.83 | 0.24 | 5.82 | 0.64 |
| 1459 | 7.32 | 0.24 | 6.14 | 0.65 | 7.08 | 0.24 | 6.02 | 0.67 | 6.83 | 0.24 | 5.82 | 0.64 |
| 1460 | 7.31 | 0.24 | 6.14 | 0.65 | 7.07 | 0.24 | 6.02 | 0.66 | 6.82 | 0.23 | 5.82 | 0.64 |
| 1461 | 7.31 | 0.24 | 6.14 | 0.65 | 7.07 | 0.24 | 6.02 | 0.66 | 6.82 | 0.23 | 5.82 | 0.64 |
| 1462 | 7.31 | 0.24 | 6.13 | 0.65 | 7.07 | 0.23 | 6.01 | 0.66 | 6.82 | 0.23 | 5.81 | 0.64 |
| 1463 | 7.31 | 0.24 | 6.13 | 0.65 | 7.07 | 0.23 | 6.01 | 0.66 | 6.82 | 0.23 | 5.81 | 0.64 |
| 1464 | 7.30 | 0.24 | 6.13 | 0.65 | 7.07 | 0.23 | 6.01 | 0.66 | 6.82 | 0.23 | 5.81 | 0.63 |
| 1465 | 7.30 | 0.24 | 6.13 | 0.65 | 7.06 | 0.23 | 6.01 | 0.66 | 6.82 | 0.23 | 5.81 | 0.63 |
| 1466 | 7.30 | 0.23 | 6.13 | 0.65 | 7.06 | 0.23 | 6.01 | 0.66 | 6.81 | 0.23 | 5.81 | 0.63 |
| 1467 | 7.30 | 0.23 | 6.13 | 0.64 | 7.06 | 0.23 | 6.01 | 0.66 | 6.81 | 0.23 | 5.81 | 0.63 |
| 1468 | 7.29 | 0.23 | 6.13 | 0.64 | 7.06 | 0.23 | 6.01 | 0.65 | 6.81 | 0.23 | 5.81 | 0.63 |
| 1469 | 7.29 | 0.23 | 6.12 | 0.64 | 7.05 | 0.23 | 6.01 | 0.65 | 6.81 | 0.23 | 5.81 | 0.63 |
| 1470 | 7.29 | 0.23 | 6.12 | 0.64 | 7.05 | 0.23 | 6.00 | 0.65 | 6.81 | 0.23 | 5.81 | 0.63 |
| 1471 | 7.29 | 0.23 | 6.12 | 0.64 | 7.05 | 0.23 | 6.00 | 0.65 | 6.80 | 0.23 | 5.80 | 0.63 |
| 1472 | 7.29 | 0.23 | 6.12 | 0.64 | 7.05 | 0.23 | 6.00 | 0.65 | 6.80 | 0.23 | 5.80 | 0.62 |
| 1473 | 7.28 | 0.23 | 6.12 | 0.64 | 7.05 | 0.23 | 6.00 | 0.65 | 6.80 | 0.23 | 5.80 | 0.62 |
| 1474 | 7.28 | 0.23 | 6.12 | 0.63 | 7.04 | 0.23 | 6.00 | 0.65 | 6.80 | 0.23 | 5.80 | 0.62 |
| 1475 | 7.28 | 0.23 | 6.12 | 0.63 | 7.04 | 0.23 | 6.00 | 0.65 | 6.80 | 0.23 | 5.80 | 0.62 |
| 1476 | 7.28 | 0.23 | 6.11 | 0.63 | 7.04 | 0.23 | 6.00 | 0.64 | 6.80 | 0.23 | 5.80 | 0.62 |
| 1477 | 7.27 | 0.23 | 6.11 | 0.63 | 7.04 | 0.23 | 5.99 | 0.64 | 6.79 | 0.23 | 5.80 | 0.62 |
| 1478 | 7.27 | 0.23 | 6.11 | 0.63 | 7.04 | 0.23 | 5.99 | 0.64 | 6.79 | 0.23 | 5.80 | 0.62 |
| 1479 | 7.27 | 0.23 | 6.11 | 0.63 | 7.03 | 0.23 | 5.99 | 0.64 | 6.79 | 0.23 | 5.80 | 0.62 |
| 1480 | 7.27 | 0.23 | 6.11 | 0.63 | 7.03 | 0.23 | 5.99 | 0.64 | 6.79 | 0.23 | 5.79 | 0.62 |
| 1481 | 7.27 | 0.23 | 6.11 | 0.63 | 7.03 | 0.23 | 5.99 | 0.64 | 6.79 | 0.23 | 5.79 | 0.61 |
| 1482 | 7.26 | 0.23 | 6.11 | 0.62 | 7.03 | 0.23 | 5.99 | 0.64 | 6.79 | 0.23 | 5.79 | 0.61 |
| 1483 | 7.26 | 0.23 | 6.10 | 0.62 | 7.03 | 0.23 | 5.99 | 0.64 | 6.78 | 0.22 | 5.79 | 0.61 |
| 1484 | 7.26 | 0.23 | 6.10 | 0.62 | 7.02 | 0.23 | 5.99 | 0.63 | 6.78 | 0.22 | 5.79 | 0.61 |
| 1485 | 7.26 | 0.23 | 6.10 | 0.62 | 7.02 | 0.22 | 5.98 | 0.63 | 6.78 | 0.22 | 5.79 | 0.61 |
| 1486 | 7.25 | 0.23 | 6.10 | 0.62 | 7.02 | 0.22 | 5.98 | 0.63 | 6.78 | 0.22 | 5.79 | 0.61 |
| 1487 | 7.25 | 0.23 | 6.10 | 0.62 | 7.02 | 0.22 | 5.98 | 0.63 | 6.78 | 0.22 | 5.79 | 0.61 |
| 1488 | 7.25 | 0.23 | 6.10 | 0.62 | 7.02 | 0.22 | 5.98 | 0.63 | 6.78 | 0.22 | 5.79 | 0.61 |
| 1489 | 7.25 | 0.22 | 6.10 | 0.62 | 7.02 | 0.22 | 5.98 | 0.63 | 6.77 | 0.22 | 5.78 | 0.60 |
| 1490 | 7.25 | 0.22 | 6.09 | 0.62 | 7.01 | 0.22 | 5.98 | 0.63 | 6.77 | 0.22 | 5.78 | 0.60 |
| 1491 | 7.24 | 0.22 | 6.09 | 0.61 | 7.01 | 0.22 | 5.98 | 0.63 | 6.77 | 0.22 | 5.78 | 0.60 |
| 1492 | 7.24 | 0.22 | 6.09 | 0.61 | 7.01 | 0.22 | 5.98 | 0.62 | 6.77 | 0.22 | 5.78 | 0.60 |
| 1493 | 7.24 | 0.22 | 6.09 | 0.61 | 7.01 | 0.22 | 5.97 | 0.62 | 6.77 | 0.22 | 5.78 | 0.60 |
| 1494 | 7.24 | 0.22 | 6.09 | 0.61 | 7.01 | 0.22 | 5.97 | 0.62 | 6.77 | 0.22 | 5.78 | 0.60 |
| 1495 | 7.24 | 0.22 | 6.09 | 0.61 | 7.00 | 0.22 | 5.97 | 0.62 | 6.76 | 0.22 | 5.78 | 0.60 |
| 1496 | 7.23 | 0.22 | 6.09 | 0.61 | 7.00 | 0.22 | 5.97 | 0.62 | 6.76 | 0.22 | 5.78 | 0.60 |
| 1497 | 7.23 | 0.22 | 6.09 | 0.61 | 7.00 | 0.22 | 5.97 | 0.62 | 6.76 | 0.22 | 5.78 | 0.60 |
| 1498 | 7.23 | 0.22 | 6.08 | 0.61 | 7.00 | 0.22 | 5.97 | 0.62 | 6.76 | 0.22 | 5.77 | 0.59 |
| 1499 | 7.23 | 0.22 | 6.08 | 0.60 | 7.00 | 0.22 | 5.97 | 0.62 | 6.76 | 0.22 | 5.77 | 0.59 |
| 1500 | 7.23 | 0.22 | 6.08 | 0.60 | 6.99 | 0.22 | 5.97 | 0.62 | 6.76 | 0.22 | 5.77 | 0.59 |